\documentclass[fleqn,usenatbib,useAMS]{Papers}
\usepackage{newtxtext,newtxmath}
\usepackage[T1]{fontenc}
\usepackage{subfloat}
\usepackage{makecell}
\usepackage[authoryear]{natbib}
\usepackage{float}
\usepackage{graphicx}	
\usepackage{amsmath}	

\title[Mean flow, velocity dispersion, and energy transfer in dark matter halos]{The mean flow, velocity dispersion, energy transfer and evolution of rotating and growing dark matter halos}

\author[Z. Xu]{
Zhijie (Jay) Xu,$^{1}$\thanks{E-mail: zhijie.xu@pnnl.gov; zhijiexu@hotmail.com}
\\
$^{1}$Physical and Computational Sciences Directorate, Pacific Northwest National Laboratory; Richland, WA 99352, USA\\
}

\date{Accepted XXX. Received YYY; in original form ZZZ}

\pubyear{2022}

\begin{document}
\label{firstpage}
\pagerange{\pageref{firstpage}--\pageref{lastpage}}
\maketitle

\begin{abstract}
By decomposing velocity dispersion into non-spin and spin-induced, mean flow and dispersion are analytically solved for axisymmetric rotating and growing halos. The polar flow can be neglected and azimuthal flow is directly related to dispersion. The fictitious ("Reynolds") stress acts on mean flow to enable energy transfer from mean flow to random motion and maximize system entropy. For large halos (high peak height $\nu$ at early stage of halo life) with constant concentration, there exists a self-similar radial flow (outward in core and inward in outer region). Halo mass, size and specific angular momentum increase linearly with time via fast mass accretion. Halo core spins faster than outer region. Large halos rotate with an angular velocity proportional to Hubble parameter and spin-induced dispersion is dominant. All specific energies (radial/rotational/kinetic/potential) are time-invariant. Both halo spin ($\sim$0.031) and anisotropic parameters can be analytically derived. For "small" halos with stable core and slow mass accretion (low peak height $\nu$ at late stage of halo life), radial flow vanishes. Small halos rotate with constant angular velocity and non-spin axial dispersion is dominant. Small halos are more spherical in shape, incompressible, and isotropic. Radial and azimuthal dispersion are comparable and greater than polar dispersion. Due to finite spin, kinetic energy is not equipartitioned with the greatest energy along azimuthal direction. Different from normal matter, small halos are hotter with faster spin. Halo relaxation (evolution) from early to late stage involves continuous variation of shape, density, mean flow, momentum, and energy. During relaxation, halo isotopically "stretches" with conserved specific rotational kinetic energy, increasing concentration and momentum of inertial. Halo "stretching" leads to decreasing angular velocity, increasing angular momentum and spin parameter.
\end{abstract}

\begin{keywords}
\vspace*{-10pt}
Dark matter halo; N-body simulations; Theoretical models
\end{keywords}

\begingroup
\let\clearpage\relax
\tableofcontents
\endgroup

\section{Introduction}
\label{sec:1}
The large-scale structure formation and evolution can be rigorously studied based on the self-gravitating collisionless fluid dynamics (SG-CFD) that deals with the motion of collisionless dark matter under its own gravity. While SG-CFD and hydrodynamic turbulence are different in many aspects, both contain the same essential ingredients (randomness, nonlinearity, and multiscale nature) and share many similarities with each other. 

Turbulence is ubiquitous in nature and might be the last unresolved problems in classical physics. More specifically, homogeneous isotropic incompressible turbulence has been well-studied for many decades and of important relevance to SG-CFD. The classical picture of turbulence is a eddy-mediated cascade process, where large eddies feed smaller eddies, which feed even smaller eddies, and so on to the smallest scale when viscous dissipation becomes dominant, i.e. a direct (kinetic) energy cascade \citep{Richardson:1922-Weather-Prediction-by-Numerica}. A key question for turbulence is "how the kinetic energy is transferred from the mean flow to turbulence, cascaded through scales, and destroyed by viscosity?" Or equivalently, how the turbulence initiates, propagates, and dies out. 

The energy cascade in turbulence starts with the kinetic energy obtained from mean flow by the largest eddies through Reynolds stress (arising from velocity fluctuation) acting on the mean flow. This kinetic energy is further cascaded successively to smaller and smaller eddies until viscosity dominates. The first quantitative description of energy cascade was proposed based on the similarity principles back to 1941 \citep{Kolmogoroff:1941-The-local-structure-of-turbule,Kolmogoroff:1941-Dissipation-of-energy-in-the-l}. The Reynolds stress arising from velocity fluctuation acts as a conduit to continuously draw kinetic energy from mean flow to sustain the energy cascade \citep{Andersson:2012-Computational-Fluid-Dynamics-f}. For high Reynolds flow (or vanishing viscosity), "vortex stretching" is responsible for the energy transfer from mean flow and energy cascade down the scales \citep{Taylor:1932-The-transport-of-vorticity-and,Taylor:1938-Production-and-dissipation-of-}. The shear stress induced lengthening of vortices along the direction of vorticity vector implies a thinning of vortices in the perpendicular direction \citep{Xu:2021-Inverse-and-direct-cascade-of-}. This intensifies the vorticity and leads to a rising kinetic energy due to the conservation of angular momentum. With vortices teased out into thinner and thinner filaments, kinetic energy is passed down to smaller and smaller scales and finally dissipated by molecular viscosity. 

While direct energy cascade is a dominant feature for 3D turbulence, the 2D turbulence exhibits an inverse energy cascade predicted in the late 1960s \citep{Kraichnan:1967-Inertial-Ranges-in-2-Dimension}. The fully developed 2D turbulence has both a direct cascade of enstrophy ($\omega ^{2} $) from large to small scales and an inverse cascade of kinetic energy from small to large scales \citep{Xu:2021-Inverse-and-direct-cascade-of-}. The enstrophy is passed down to smaller scales until destroyed by viscosity, while kinetic energy is passed up and destroyed on the largest scale. While vortex stretching cannot operate in a 2D turbulence, the area-conserved teasing and twisting make vortex patches thinner and longer. This facilitates a combined direct cascade of enstrophy and inverse cascade of kinetic energy. 

Just like vortex (the building block of turbulance) facilitates the energy/enstrophy cascade in 2D and 3D turbulence, halo plays a fundamental role in SG-CFD for dark matter flow. To maximize system entropy, halos and halo groups of different size are necessary to form due to the long-range interaction nature of SG-CFD \citep{Xu:2021-The-maximum-entropy-distributi,Xu:2021-Mass-functions-of-dark-matter-}. Halo structure is a major manifestation of the nonlinear gravitational collapse and building blocks of large-scale structures \citep{Neyman:1952-A-Theory-of-the-Spatial-Distri,Cooray:2002-Halo-models-of-large-scale-str}. The halo-mediated inverse mass cascade is a key feature of dark matter flow \citep{Xu:2021-Inverse-mass-cascade-mass-function}: "Little halos have big halos, That feed on their mass; And big halos have greater halos, And so on to growth". 

There exists a broad spectrum of halo sizes. Halos pass their mass onto larger and larger halos, until mass growth becomes dominant over mass propagation. The effects of mass cascade on halo mass function have been previously studied with new mass function formulated without resorting to any specific spherical or ellipsoid collapse models \citep{Xu:2021-Inverse-mass-cascade-mass-function}. The effects of mass cascade on halo deformation, energy, size and density profile are also discussed in detail \citep{Xu:2021-Inverse-mass-cascade-halo-density,Xu:2022-The-evolution-of-energy--momen}. Along with the halo-mediated mass cascade, kinetic energy (or potential energy) is simultaneously inversely (directly) cascaded with energy transfer rate proportional to the rate of mass transfer \citep{Xu:2021-Inverse-and-direct-cascade-of-}. The mass and energy cascades facilitate the development of statistical theory for dark matter flow \citep{Xu:2022-The-statistical-theory-of-2nd,Xu:2022-The-statistical-theory-of-3rd,Xu:2022-Two-thirds-law-for-pairwise-ve} with important applications for predicting dark matter particle mass and properties \citep{Xu:2022-Postulating-dark-matter-partic}, interpreting the MOND (modified Newtonian dynamics) theory \citep{Xu:2022-The-origin-of-MOND-acceleratio}, and developing the baryonic-to-halo mass relation \citep{Xu:2022-The-baryonic-to-halo-mass-rela}. However, how halos facilitate the energy transfer and cascade in SG-CFD is not completely understood. 

While "vortex stretching" (the shape change of vortex) is responsible for energy transfer and cascade in turbulence, the shape change of halo seems not sufficiently strong to be responsible for the energy cascade in SG-CFD \citep{Xu:2021-Inverse-and-direct-cascade-of-}. To better understand the role of halos in energy cascade, a complete knowledge of the mean flow, velocity dispersion, and the evolution of rotating and growing halos are required. Existing study mostly focus on the non-rotating spherical halos with vanishing radial flow \citep{Hoeft:2004-Velocity-dispersion-profiles-i,Binney:1987--Galactic-Dynamics}. Solutions for non-rotating growing halos with a nonzero radial flow were recently studied \citep{Xu:2021-Inverse-mass-cascade-halo-density}. While vortex is volume/mass conserved for incompressible flow, halos are much more complex and dynamic objects that are constantly growing, spinning, shape-changing, with a nonuniform density profile, and usually not volume- or mass-conserved. The purpose of this paper is to explore relevant solutions and evolution of rotating and growing halos and the role of halos in energy transfer and cascade in SG-CFD. 

The rest of paper is organized as follows: Section \ref{sec:2} introduces the simulation and numerical data used for this work. Section \ref{sec:3} presents solutions for the mean flow and velocity dispersions of an axisymmetric rotating and growing halo (the building block of SG-CFD) at their early and late stage of life. The momentum and energy solutions of rotating and growing halos are presented in Section \ref{sec:4}. The energy transfer between mean flow and random motion in halos is discussed in Section \ref{sec:5}, along with the halo evolution from early to late stage in Section \ref{sec:6}. A halo stretching mechanism (counterpart of vortex stretching) is proposed and studied extensively along with the energy and momentum evolution.

\section{N-body simulations and numerical data}
\label{sec:2}
The numerical data for this work is publicly available and generated from the \textit{N}-body simulations carried out by the Virgo consortium, an international collaboration that aims to perform large \textit{N}-body simulations of the formation of large-scale structures. A comprehensive description of the simulation data can be found in \citep{Frenk:2000-Public-Release-of-N-body-simul,Jenkins:1998-Evolution-of-structure-in-cold}. The same set of simulation data has been widely used in a number of different studies from clustering statistics \citep{Jenkins:1998-Evolution-of-structure-in-cold} to the formation of halo clusters in large scale environments \citep{Colberg:1999-Linking-cluster-formation-to-l}, and testing models for halo abundance and mass functions \citep{Sheth:2001-Ellipsoidal-collapse-and-an-im}. Some key parameters of N-body simulations are listed in Table \ref{tab:1}.

Two relevant datasets from this N-boby simulation, i.e. halo-based and correlation-based statistics of dark matter flow, can be found at Zenodo.org  \citep{Xu:2022-Dark_matter-flow-dataset-part1, Xu:2022-Dark_matter-flow-dataset-part2}, along with the accompanying presentation slides, "A comparative study of dark matter flow \& hydrodynamic turbulence and its applications" \citep{Xu:2022-Dark_matter-flow-and-hydrodynamic-turbulence-presentation}. All data files are also available on GitHub \citep{Xu:Dark_matter_flow_dataset_2022_all_files}.

\begin{table}
\caption{Numerical parameters of N-body simulation}
\begin{tabular}{p{0.25in}p{0.05in}p{0.05in}p{0.05in}p{0.05in}p{0.05in}p{0.4in}p{0.1in}p{0.4in}p{0.4in}} 
\hline 
Run & $\Omega_{0}$ & $\Lambda$ & $h$ & $\Gamma$ & $\sigma _{8}$ & \makecell{L\\(Mpc/h)} & $N$ & \makecell{$m_{p}$\\$M_{\odot}/h$} & \makecell{$l_{soft}$\\(Kpc/h)} \\ 
\hline 
SCDM1 & 1.0 & 0.0 & 0.5 & 0.5 & 0.51 & \centering 239.5 & $256^{3}$ & 2.27$\times 10^{11}$ & \makecell{\centering 36} \\ 
\hline 
\end{tabular}
\label{tab:1}
\end{table}

\section{Solutions for rotating and growing halos}
\label{sec:3}
\subsection{Continuity and momentum equations and azimuthal flow}
\label{sec:3.1}
Jeans' equation and solutions for spherical, stationary, and non-rotating halos can be found in many literature \citep{Hoeft:2004-Velocity-dispersion-profiles-i,Binney:1987--Galactic-Dynamics}. Solutions for spherical, growing, and non-rotating halos were also studied, where the effect of nonzero radial flow on halo density is formulated \citep{Xu:2021-Inverse-mass-cascade-halo-density}.
\begin{figure}
\includegraphics*[width=\columnwidth]{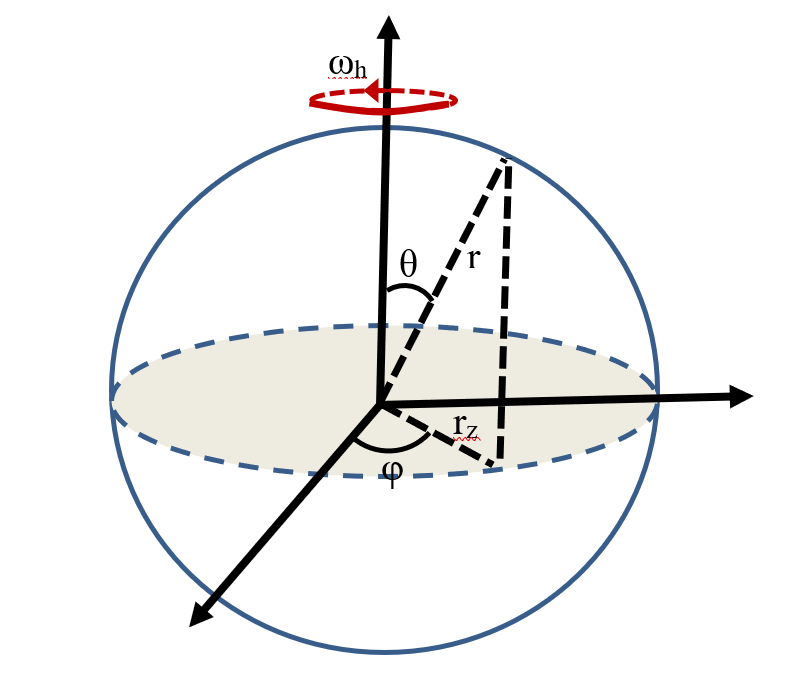}
\caption{Spherical coordinates $\left(r,\theta ,\varphi \right)$ for a halo with angular velocity $\omega _{h} \left(t\right)$, where \textit{r} is the radius, $\theta $ is the polar angle between radial vector \textbf{r} and axis of rotation, and $\varphi $ is the azimuthal angle in plane perpendicular to that axis. Distance to that axis is $r_{z} =r\sin \theta$.}
\label{fig:1}
\end{figure}
Here we consider an even more general case, i.e. spherical, growing, and rotating halos with a given angular velocity $\omega _{h} \left(t\right)$. Halos grow with a time-varying halo mass $m_{h} =m_{h} \left(t\right)$ and scale radius $r_{s} =r_{s} \left(t\right)$ due to the inverse mass cascade and mass accretion. Halo size (the virial radius) $r_{h} =c\left(t\right)r_{s} \left(t\right)$, where \textit{c} is the concentration parameter. As shown in Fig. ~\ref{fig:1}, spherical coordinates $\left(r,\theta ,\varphi \right)$ are introduced, where \textit{r} is the radius, $\theta $ is the polar angle between the radial vector \textbf{\textit{r}} and the axis of rotation, and $\varphi $ is the azimuthal angle in plane perpendicular to the axis of rotation. The distance to that axis reads $r_{z} =r\sin \theta $. 

The starting point of our formulation is the continuity equation in spherical coordinates,
\begin{equation} 
\label{ZEqnNum212331} 
\frac{\partial \rho _{h} }{\partial t} +\frac{1}{r^{2} } \frac{\partial \left(r^{2} \rho _{h} u_{r} \right)}{\partial r} +\underbrace{\frac{1}{r\sin \theta }\left(\frac{\partial \left(\rho _{h} u_{\theta } \sin \theta \right)}{\partial \theta } + \frac{\partial \left(\rho _{h} u_{\varphi } \right)}{\partial \varphi }\right) }_{1}=0,
\end{equation} 
where $\rho _{h} \equiv \rho _{h} \left(r,t\right)$ is the halo density. Mean flow along three coordinates are introduced as the radial flow $u_{r} $, polar flow (meridional flow) $u_{\theta } $, and azimuthal flow (zonal flow) $u_{\varphi } $. By considering the axisymmetry about axis of rotation, the mean azimuthal flow $u_{\varphi } =u_{\varphi } \left(r,\theta ,t\right)$ should be independent of the azimuthal angle $\varphi $. The polar flow $u_{\theta } =u_{\theta } \left(r,\theta ,t\right)$ is also independent of $\varphi $ with symmetry $u_{\theta } \left(r,\theta ,t\right)=-u_{\theta } \left(r,\pi -\theta ,t\right)$ such that $u_{\theta } \left(r,{\pi /2} ,t\right)=0$.

\begin{figure}
\includegraphics*[width=\columnwidth]{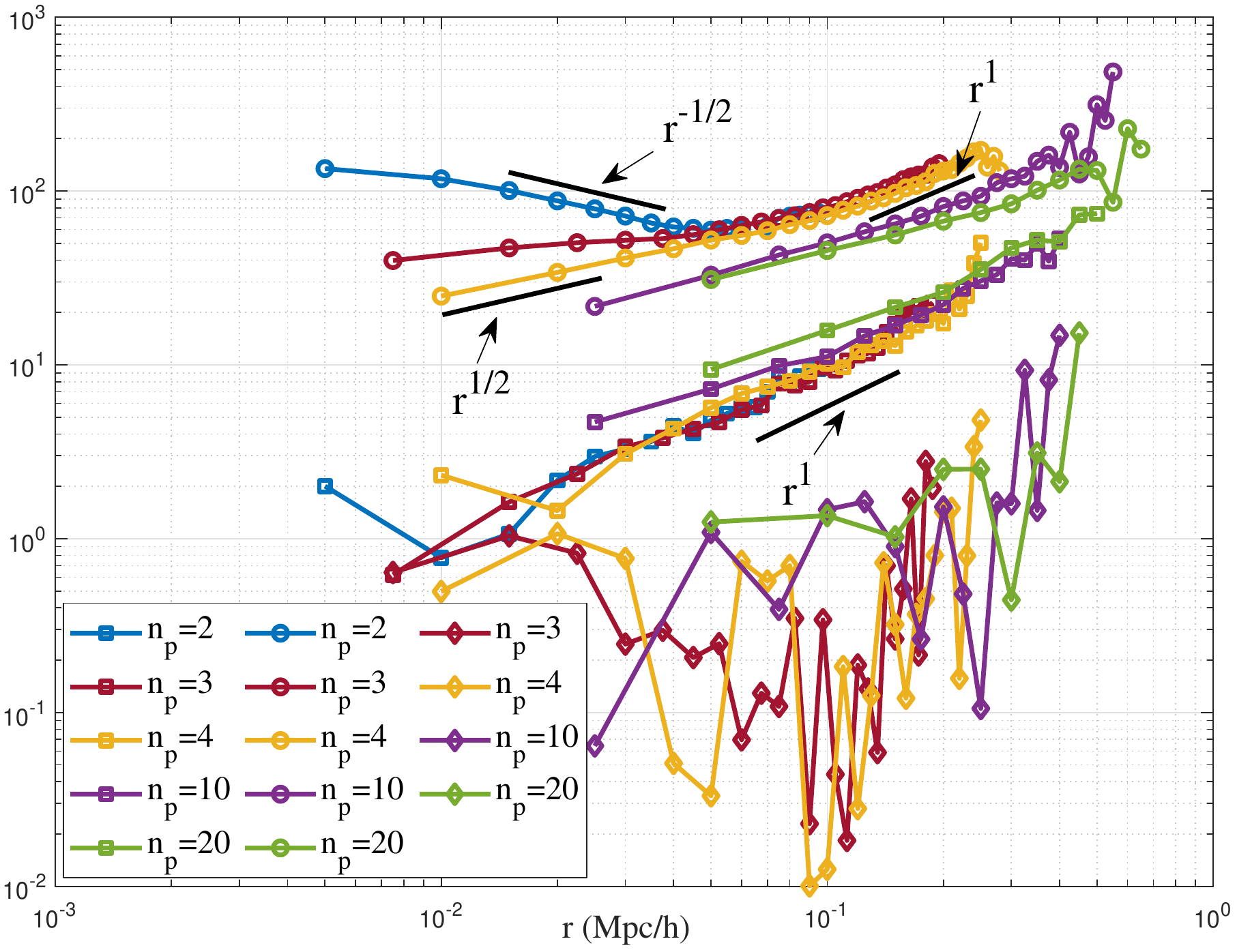}
\caption{The variation of spherical and group averaged mean (peculiar) radial ($-u_{rp} $: 'square'), azimuthal ($u_{\varphi } $:`circles'), and polar ($u_{\theta } $: `diamond') flow (unit: km/s) with radius \textit{r} for halo groups of size $n_{p} $=2, 3, 4, 10 and 20 at \textit{z}=0. For $n_{p} =2$, planar motion leads to $u_{\theta } =0$. The azimuthal flow is predicted to be $u_{\varphi } \sim r^{{-1/2} } $ for $n_{p} =2$and gradually shifts to $u_{\varphi } \sim r^{{1/2} } $ for inner region and approaching $u_{\varphi } \sim r$ for outer region. For all size of halos in figure, the peculiar radial flow $u_{rp} \approx -Hr$ from stable cluster hypothesis \citep{Xu:2021-A-non-radial-two-body-collapse}. The mean polar flow is negligible, i.e. $u_{\theta } \approx 0$ almost everywhere.}
\label{fig:2}
\end{figure}

 Observations of flow on rotating sphere strongly suggest that as the rotation rate increases, the azimuthal flow (zonal flow) will become dominant and the polar flow (meridional flow) $u_{\theta } $ may be neglected ($u_{\theta } \approx 0$) (also discussed in Fig. \ref{fig:2}). The original continuity Eq. \eqref{ZEqnNum212331} reduces to  
\begin{equation}
\label{ZEqnNum321851} 
\frac{\partial \rho _{h} }{\partial t} +\frac{1}{r^{2} } \frac{\partial \left(r^{2} \rho _{h} u_{r} \right)}{\partial r} =0,          
\end{equation} 
where the density $\rho _{h} =\rho _{h} \left(r,t\right)$ and the radial flow $u_{r} =u_{r} \left(r,t\right)$ are functions of \textit{r} and \textit{t} only. Equation \eqref{ZEqnNum321851} has been extensively studied in our previous work \citep{Xu:2021-Inverse-mass-cascade-mass-function} and used to solve for the mean radial flow $u_{r} =u_{r} \left(r,t\right)$ for a given halo density $\rho _{h} $. In current model, (in-plane) flow in concentric spherical shells is incompressible (term 1 in Eq. \eqref{ZEqnNum212331} vanishes). However, radial flow (out-of-plane) is not incompressible with $u_{r} \ne 0$. The special case is an isothermal density profile where $u_{r} =0$ such that the mean flow of entire halo is incompressible everywhere.

The full momentum equations (Jeans' equation) along three spherical coordinates read
\begin{equation}
\label{ZEqnNum140120} 
\begin{split}
&\frac{\partial u_{r} }{\partial t} +u_{r} \frac{\partial u_{r} }{\partial r} +\frac{u_{\theta } }{r} \frac{\partial u_{r} }{\partial \theta } +\frac{u_{\varphi } }{r\sin \theta } \frac{\partial u_{r} }{\partial \varphi } -\frac{u_{\theta }^{2} +u_{\varphi }^{2} }{r}\\&=-\frac{\partial \phi _{r} }{\partial r} +\frac{\sigma _{\theta \theta }^{2} +\sigma _{\varphi \varphi }^{2} }{r}\\ 
&-\frac{1}{\rho _{h} } \left[\frac{1}{r^{2} } \frac{\partial \left(r^{2} \rho _{h} \sigma _{rr}^{2} \right)}{\partial r} +\frac{1}{r\sin \theta } \left(\frac{\partial \left(\rho _{h} \sigma _{\theta r}^{2} \sin \theta \right)}{\partial \theta } + \frac{\partial \left(\rho _{h} \sigma _{\varphi r}^{2} \right)}{\partial \varphi }\right) \right],
\end{split}
\end{equation} 
\begin{equation}
\label{eq:4} 
\begin{split}
&\frac{\partial u_{\theta } }{\partial t} +u_{r} \frac{\partial u_{\theta } }{\partial r} +\frac{u_{\theta } }{r} \frac{\partial u_{\theta } }{\partial \theta } +\frac{u_{\varphi } }{r\sin \theta } \frac{\partial u_{\theta } }{\partial \varphi } +\frac{u_{r} u_{\theta } }{r} -\frac{u_{\varphi }^{2} \cot \theta }{r}\\
&=-\frac{1}{r} \frac{\partial \phi _{r} }{\partial \theta } +\frac{\sigma _{\varphi \varphi }^{2} \cot \theta -\sigma _{\theta r}^{2} +\sigma _{r\theta }^{2} }{r}\\ 
&-\frac{1}{\rho _{h} } \left[\frac{1}{r^{3} } \frac{\partial \left(r^{3} \rho _{h} \sigma _{r\theta }^{2} \right)}{\partial r} +\frac{1}{r\sin \theta } \left(\frac{\partial \left(\rho _{h} \sigma _{\theta \theta }^{2} \sin \theta \right)}{\partial \theta } + \frac{\partial \left(\rho _{h} \sigma _{\varphi \theta }^{2} \right)}{\partial \varphi }\right) \right],
\end{split}
\end{equation} 
and
\begin{equation} 
\label{ZEqnNum886213} 
\begin{split}
&\frac{\partial u_{\varphi } }{\partial t} +u_{r} \frac{\partial u_{\varphi } }{\partial r} +\frac{u_{\theta } }{r} \frac{\partial u_{\varphi } }{\partial \theta } +\frac{u_{\varphi } }{r\sin \theta } \frac{\partial u_{\varphi } }{\partial \varphi } +\frac{u_{r} u_{\varphi } }{r} +\frac{u_{\varphi } u_{\theta } \cot \theta }{r}\\
&=-\frac{1}{r\sin \theta } \frac{\partial \phi _{r} }{\partial \varphi } +\frac{\sigma _{r\varphi }^{2} -\sigma _{\varphi r}^{2} -\sigma _{\varphi \theta }^{2} \cot \theta }{r}\\  
&-\frac{1}{\rho _{h} } \left[\frac{1}{r^{3} } \frac{\partial \left(r^{3} \rho _{h} \sigma _{r\varphi }^{2} \right)}{\partial r} +\frac{1}{r\sin \theta } \left(\frac{\partial \left(\rho _{h} \sigma _{\theta \varphi }^{2} \sin \theta \right)}{\partial \theta } + \frac{\partial \left(\rho _{h} \sigma _{\varphi \varphi }^{2} \right)}{\partial \varphi } \right)\right], 
\end{split}
\end{equation} 
where the gravitational potential $\phi _{r} $ is related to halo density via the halo mass $m_{r} =m_{r} \left(r,t\right)$ within a shell of radius \textit{r},
\begin{equation}
\frac{\partial \phi _{r} }{\partial r} =\frac{Gm_{r} \left(r,t\right)}{r^{2} } \quad \textrm{and} \quad \rho _{h} =\frac{1}{4\pi r^{2} } \frac{\partial m_{r} \left(r,t\right)}{\partial r}.      
\label{ZEqnNum582884}
\end{equation}
\noindent By assuming vanishing off-diagonal velocity dispersions and the fact that all variables should be independent of the azimuthal angle $\varphi $ due to axisymmetry, i.e. 
\begin{equation} 
\label{eq:7} 
\sigma _{rr}^{2} =\sigma _{rr}^{2} \left(r,\theta ,t\right),  \sigma _{\theta \theta }^{2} =\sigma _{\theta \theta }^{2} \left(r,\theta ,t\right), \sigma _{\varphi \varphi }^{2} =\sigma _{\varphi \varphi }^{2} \left(r,\theta ,t\right),    
\end{equation} 
and
\begin{equation} 
\label{eq:8} 
\sigma _{r\theta }^{2} =0, \sigma _{r\varphi }^{2} =0, \sigma _{\varphi \theta }^{2} =0,       
\end{equation} 
momentum equations (Eq. \eqref{ZEqnNum140120}-\eqref{ZEqnNum886213}) can be significantly reduced to
\begin{equation} 
\label{ZEqnNum535400} 
\begin{split}
\frac{\partial u_{r} }{\partial t} +u_{r} \frac{\partial u_{r} }{\partial r} &+\frac{1}{\rho _{h} } \frac{\partial \left(\rho _{h} \sigma _{rr}^{2} \right)}{\partial r}\\
&+\frac{2}{r} \sigma _{rr}^{2} \underbrace{\left(1-\frac{\sigma _{\theta \theta }^{2} +\sigma _{\varphi \varphi }^{2} +u_{\varphi }^{2} }{2\sigma _{rr}^{2} } \right)}_{1}+\frac{\partial \phi _{r} }{\partial r} =0,
\end{split}
\end{equation} 
\begin{equation} 
\label{ZEqnNum618520} 
u_{\varphi }^{2} =\sigma _{\theta \theta }^{2} -\sigma _{\varphi \varphi }^{2} +\frac{\sin \theta }{\cos \theta } \frac{\partial \sigma _{\theta \theta }^{2} }{\partial \theta } ,         
\end{equation} 
\begin{equation} 
\label{ZEqnNum720202} 
\frac{\partial u_{\varphi } }{\partial t} +u_{r} \frac{\partial u_{\varphi } }{\partial r} +\frac{u_{r} u_{\varphi } }{r} =0.
\end{equation} 

The mean azimuthal flow $u_{\varphi }^{2} $ is directly related to in-plane velocity dispersions $\sigma _{\theta \theta }^{2}$ and $\sigma _{\varphi \varphi }^{2}$ in Eq. \eqref{ZEqnNum618520}. The azimuthal flow $u_{\varphi }^{} $ can be solved from Eq. \eqref{ZEqnNum720202} if $u_{r} $ is known. Note that an exact definition of the halo anisotropic parameter $\beta _{h1} $ should be (term 1 in Eq. \eqref{ZEqnNum535400})
\begin{equation} 
\label{ZEqnNum865565} 
\beta _{h1} =1-\frac{\sigma _{\theta \theta }^{2} +\sigma _{\varphi \varphi }^{2} +u_{\varphi }^{2} }{2\sigma _{rr}^{2} } , 
\end{equation} 
where the effect of azimuthal flow due to halo spin should be included. However, $u_{\varphi }^{2} $ might be relatively small compared to in-plane velocity dispersions $\sigma _{\theta \theta }^{2} $ and $\sigma _{\varphi \varphi }^{2} $ for massive halos with large velocity dispersion such that $u_{\varphi }^{2} $ can be neglected. If the azimuthal flow $u_{\varphi }^{} $ can be neglected, the anisotropic parameter $\beta _{h1} $ reduces to the standard definition in literature,
\begin{equation} 
\label{ZEqnNum597498} 
\beta _{h} =1-\frac{\sigma _{\theta \theta }^{2} +\sigma _{\varphi \varphi }^{2} }{2\sigma _{rr}^{2} } .          
\end{equation} 
Clearly, the two definitions are only consistent with each other for massive or large halos, where azimuthal flow $u_{\varphi }^{2} $ can be neglected when compared to in-plane velocity dispersions. However, small halos spin much faster than large halos at the same redshift \citep[see][Fig. 15]{Xu:2021-Inverse-and-direct-cascade-of-} and the effect of $u_{\varphi }^{2} $ can be strong. Two definitions are different with $\beta _{h1} \approx 0$ and $\beta _{h} >0$ for small and fast spinning halos. We will discuss and compare two definitions in Fig. \ref{fig:9}.

We will close this section by presenting the mean flow from \textit{N}-body simulations. For every halo identified in the system, the axis of rotation can be determined first by calculating the halo angular momentum vector $\boldsymbol{\mathrm{H}}_{h}$ \citep[see][Eq. (56)]{Xu:2021-Inverse-and-direct-cascade-of-}. All halos are positioned and aligned by the axis of rotation as shown in Fig. \ref{fig:1} such that $u_{\phi } >0$ is always true. The mean flow of every particle in halo can be obtained by projecting its peculiar velocity along three spherical coordinates. The statistics is then taken over all particles in the same spherical shell (spherical averaging) and for all halos in the same group (group averaging) to increase signal noise ratio. Groups of small halos have enough halos for reliable statistics, while groups of large halos may not have sufficient number of halos, where the average can be taken over multiple halo groups of similar sizes of a given range. 

Figure \ref{fig:2} plots the variation of the mean (peculiar) radial ($u_{rp} =u_{r} -Hr$ in square symbols), azimuthal flow ($u_{\varphi}$ in circles), and polar flow ($u_{\theta}$ in diamond symbols) with radius \textit{r} for halo groups of different sizes at \textit{z}=0. For $n_{p} =2$, planar motion leads to a vanishing polar flow $u_{\theta } =0$. The azimuthal flow is predicted to be $u_{\varphi } \sim r^{{-1/2} } $ for $n_{p} =2$ (predicted by two-body collapse model (TBCM) \citep[Eq. (103)]{Xu:2021-A-non-radial-two-body-collapse}) and gradually shifts to $u_{\varphi } \sim r^{{1/2} } $ for larger halos. For all halos in figure, the radial flow $u_{rp} \approx -Hr$ (from the stable cluster hypothesis that can be demonstrated by TBCM \citep{Xu:2021-A-non-radial-two-body-collapse}) can be a good approximation. The mean polar flow $u_{\theta}$ is negligible when compared to the mean radial and azimuthal flow, i.e. $u_{\theta } \approx 0$ almost everywhere. 

\begin{figure}
\includegraphics*[width=\columnwidth]{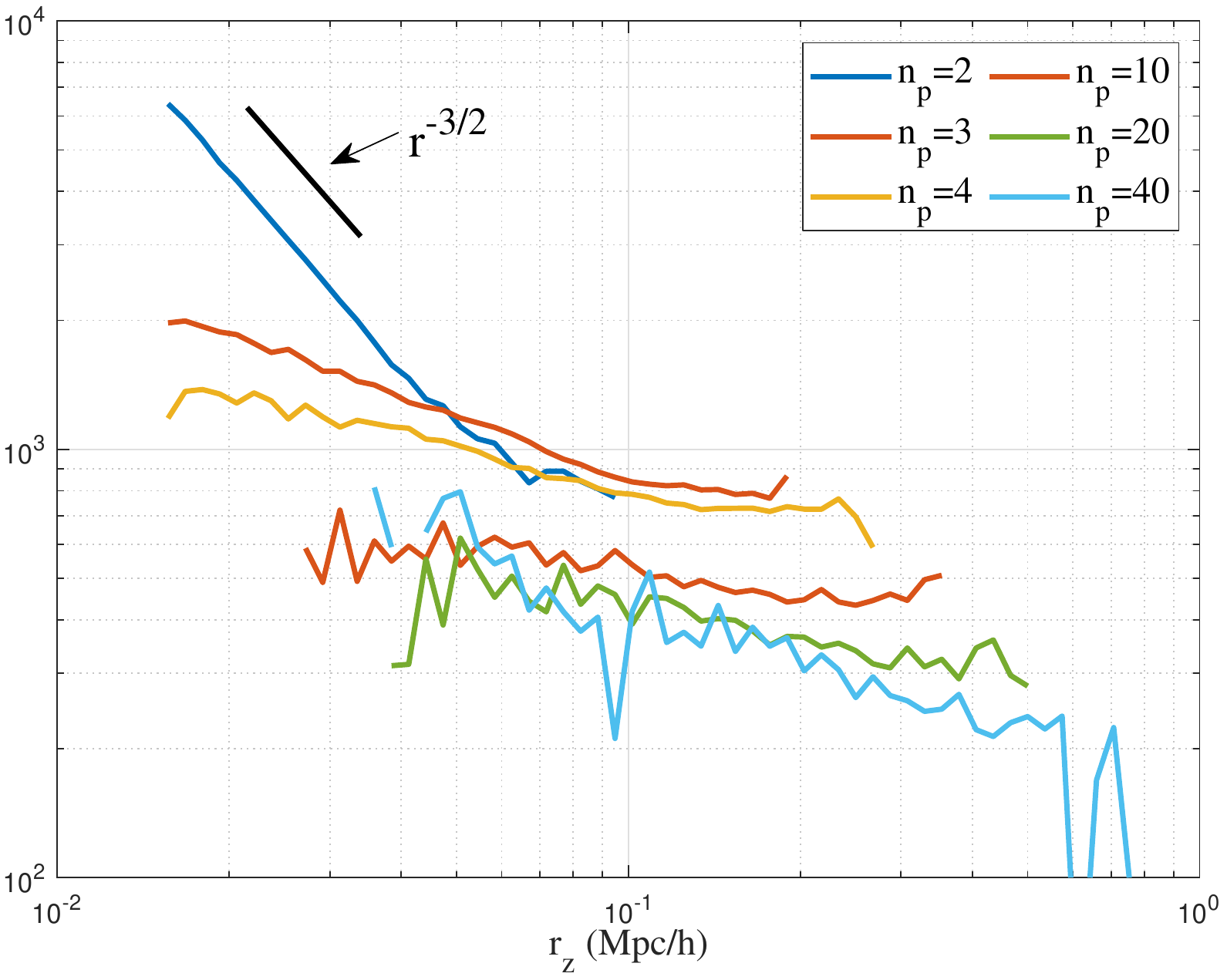}
\caption{The variation of angular velocity $\omega _{r} \left(r_{z} \right)={u_{\varphi } /r_{z} }$ (unit:$km/s/(Mpc/h)$) with $r_{z}$ (distance from axis of rotation) for halo groups of size $n_{p}$ = 2, 3, 4, 10, 20 and 40. For $n_{p} =2$, the angular velocity is predicted to be $\omega _{r} \approx 13r^{{-3/2} } $ \citep[see][Eq. (103)]{Xu:2021-A-non-radial-two-body-collapse}. Angular velocity $\omega _{r}$ decreases with halo size. For a given size, $\omega _{r}$ decreases with distance $r_{z}$ and approaches a constant $\omega _{r}$ in outer region of halos. Halo core spins faster than outer region.}
\label{fig:3}
\end{figure}

Figure \ref{fig:3} plots the variation of angular velocity $\omega _{r} \left(r_{z} \right)={u_{\varphi } /r_{z} } $ about the axis of rotation with $r_{z} $ (the distance to axis of rotation) for halo groups of different sizes. Again, for $n_{p} =2$, the angular velocity is predicted to be $\omega _{r} \sim r_{z} {}^{{-3/2} } $ \citep[see][Eq. (103)]{Xu:2021-A-non-radial-two-body-collapse}. Large halo spins slower with $\omega _{r}$ decreases with $r_{z}$ and approaches a constant $\omega _{r}$ in the outer region. The variation of effective angular velocity of entire halo ($\omega _{h}$) with halo size $n_p$ and redshift $z$ is presented in our previous work of inverse energy cascade \citep[see][Fig. 15]{Xu:2021-Inverse-and-direct-cascade-of-}.

\subsection{Evolution of halo momentum and energy}
\label{sec:3.2}
The evolution of halo momentum and energy can be studied exactly by the continuity and momentum equations. The first example is to multiply the continuity equation (Eq. \eqref{ZEqnNum321851}) and momentum Eq. \eqref{ZEqnNum535400} with $u_{r} $ and $\rho _{h} $ respectively and add them together that leads to an equation for the evolution of radial momentum $\rho _{h} u_{r} $,
\begin{equation} 
\label{ZEqnNum898930} 
\frac{\partial \rho _{h} u_{r} }{\partial t} +\frac{1}{r^{2} } \frac{\partial \rho _{h} u_{r}^{2} r^{2} }{\partial r} +\frac{\partial \left(\rho _{h} \sigma _{rr}^{2} \right)}{\partial r} +2\frac{\rho _{h} }{r} \beta _{h1} \sigma _{rr}^{2} +\rho _{h} \frac{\partial \phi _{r} }{\partial r} =0.     
\end{equation} 
The integration of Eq. \eqref{ZEqnNum898930} over the entire halo by applying $\int _{0}^{r_{h} }2\pi r^{2} \int _{0}^{\pi }\left(\bullet \right)\sin \theta d\theta dr$ to both sides of Eq. \eqref{ZEqnNum898930} leads to 
\begin{equation} 
\label{eq:15} 
\begin{split}
&\frac{\partial \bar{L}_{h} }{\partial t} +4\pi r_{h}^{2} \rho _{h} \left(r_{h} \right)u_{r} \left(r_{h} \right)\left[u_{r} \left(r_{h} \right)-\frac{\partial r_{h} }{\partial t} \right]+\\
&2\pi r_{h}^{2} \rho _{h} \left(r_{h} \right)\int _{0}^{\pi }\sigma _{rr}^{2} \left(r_{h} ,\theta \right) \sin \theta d\theta+\int _{0}^{r_{h} }4\pi r^{2} \rho _{h} \frac{\partial \phi _{r} }{\partial r} dr \\
&-\frac{1}{2} \int _{0}^{r_{h} }4\pi r\rho _{h} \int _{0}^{\pi }\left(\sigma _{\varphi \varphi }^{2} +\sigma _{\theta \theta }^{2} +u_{\varphi }^{2} \right) \sin \theta d\theta  dr=0, 
\end{split}
\end{equation} 
where the (zeroth order) halo radial momentum is defined as
\begin{equation} 
\label{eq:16} 
\bar{L}_{h} \left(a\right)=\int _{0}^{r_{h} }u_{r}^{} \left(r,a\right)4\pi r^{2} \rho _{h} \left(r,a\right)dr .\end{equation} 
The integration of Eq. \eqref{ZEqnNum898930} over the entire halo by applying $\int _{0}^{r_{h} }2\pi r^{2} \int _{0}^{\pi }\left(\bullet \right)r\sin \theta d\theta dr  $ leads to a complete virial theorem for rotating and growing halos,
\begin{equation} 
\label{eq:17}
\begin{split}
&\underbrace{\frac{\partial \bar{G}_{h} }{\partial t} }_{1}+\underbrace{4\pi r_{h}^{3} \rho _{h} \left(r_{h} \right)u_{r} \left(r_{h} \right)\left[u_{r} \left(r_{h} \right)-\frac{\partial r_{h} }{\partial t} \right]}_{2}\\
&+\underbrace{2\pi r_{h}^{3} \rho _{h} \left(r_{h} \right)\int _{0}^{\pi }\sigma _{rr}^{2} \left(r_{h} ,\theta \right) \sin \theta d\theta }_{3}\\
&-\underbrace{\int _{0}^{r_{h} }4\pi r^{2} \rho _{h} u_{r}^{2} dr }_{4}-\underbrace{\frac{1}{2} \int _{0}^{r_{h} }4\pi r^{2} \rho _{h} \left(\int _{0}^{\pi }u_{\varphi }^{2}  \sin \theta d\theta \right) dr}_{5}\\ 
&-\underbrace{\frac{1}{2} \int _{0}^{r_{h} }4\pi r^{2} \rho _{h} \left[\int _{0}^{\pi }\left(\sigma _{rr}^{2} +\sigma _{\theta \theta }^{2} +\sigma _{\varphi \varphi }^{2} \right) \sin \theta d\theta \right] dr}_{6}\\
&+\underbrace{\int _{0}^{r_{h} }4\pi r^{3} \rho _{h} \frac{\partial \phi _{r} }{\partial r} dr }_{7}=0,  
\end{split}
\end{equation} 
where halo virial quantity (first order radial momentum) is defined as 
\begin{equation} 
\label{eq:18} 
\bar{G}_{h} \left(a\right)=\int _{0}^{r_{h} }u_{r} \left(r,a\right)4\pi r^{3} \rho _{h} \left(r,a\right)dr .        
\end{equation} 
Term 2 is the surface energy due to radial flow and mass accretion at halo surface and term 3 is the surface energy due to radial velocity dispersion. Term 4 is for halo radial kinetic energy and term 5 is for the halo rotational kinetic energy, both of which are from mean flow of halo (coherent motion). Term 6 is for the kinetic energy due to the random motion. Term 7 is for the halo potential energy. The similar equation has been extensively studied \citep[see][Eq. (75)]{Xu:2021-Inverse-mass-cascade-mass-function} for an isotropic, growing, and non-rotating halo, where term 5 is not present. For virialized, non-rotating, and non-growing halos, $u_{r} =0$ and ${\partial r_{h} /\partial t} =0$ such that terms 2, 4, and 5 are not preent.

The second example is for the radial kinetic energy. Multiplying Eqs. \eqref{ZEqnNum535400} and \eqref{ZEqnNum898930} with $\rho _{h} u_{r} $ and $u_{r} $ respectively and adding them together leads to the evolution of radial kinetic energy $\left(\rho _{h} u_{r}^{2} \right)$,
\begin{equation} 
\label{ZEqnNum185081} 
\begin{split}
&\underbrace{\frac{\partial \left(\rho _{h} u_{r}^{2} \right)}{\partial t} }_{derivative}+\underbrace{\frac{1}{r^{2} } \frac{\partial \left[\left(\rho _{h} u_{r}^{2} \right)u_{r} r^{2} \right]}{\partial r} }_{advection}\\
&+\underbrace{4\beta _{h1} \frac{u_{r} }{r} \rho _{h} \sigma _{rr}^{2} }_{P_{1} }+\underbrace{2u_{r} \frac{\partial \left(\rho _{h} \sigma _{rr}^{2} \right)}{\partial r} }_{P_{3} }+\underbrace{2u_{r} \rho _{h} \frac{\partial \phi _{r} }{\partial r} }_{P_{2} }=0,
\end{split}
\end{equation} 
where the first term is the time derivative of radial kinetic energy. The second term is the advection in radial direction. The last three terms are the production of radial kinetic energy including two contributions, i.e. $P_{1} $ and $P_{3} $ from velocity radial dispersion $\sigma_{rr}^2$ and $P_{2}$ from the gravitational interaction. With $P_{2} +P_{3} \approx 0$ (gravitational force balances the pressure gradient), there is a net energy transfer between the radial mean flow $\rho _{h} u_{r}^{2} $ and random motion $\rho _{h} \sigma _{rr}^{2} $ (term $P_{1} $). The direction of transfer depends on the sign of $u_{r} $. Using the radial momentum equation Eq. \eqref{ZEqnNum535400}, we have the identity
\begin{equation} 
\label{eq:20} 
\frac{\partial \left(\rho _{h} u_{r}^{2} \right)}{\partial t} +\frac{1}{r^{2} } \frac{\partial \left[\left(\rho _{h} u_{r}^{2} \right)u_{r} r^{2} \right]}{\partial r} -2\rho _{h} u_{r}^{} \left(\frac{\partial u_{r} }{\partial t} +u_{r} \frac{\partial u_{r} }{\partial r} \right)=0,       
\end{equation} 
which will be further used to study two contributions ($P_{1} +P_{3} $ and $P_{2}$) for the production of radial kinetic energy (see Eq. \eqref{ZEqnNum644755}). 

 Integrating $\left(\rho _{h} u_{r}^{2} \right)$ in Eq. \eqref{ZEqnNum185081} leads to the total halo radial kinetic energy $\bar{K}_{r} $,
\begin{equation} 
\label{ZEqnNum362154} 
\bar{K}_{r} \left(a\right)=\frac{1}{2} \int _{0}^{r_{h} }u_{r}^{2} \left(r,a\right)4\pi r^{2} \rho _{h} \left(r,a\right)dr . 
\end{equation} 
Integrating Eq. \eqref{ZEqnNum185081} over the entire halo by applying $\int _{0}^{r_{h} }2\pi r^{2} \int _{0}^{\pi }\left(\bullet \right)\sin \theta d\theta dr  $ leads to the evolution of halo radial kinetic energy,
\begin{equation} 
\label{eq:22}
\begin{split}
&\frac{\partial \bar{K}_{r} }{\partial t} +2\pi r_{h}^{2} \rho _{h} \left(r_{h} \right)u_{r} \left(r_{h} \right)\left[u_{r} \left(r_{h} \right)\left(u_{r} \left(r_{h} \right)-\frac{\partial r_{h} }{\partial t} \right)\right.\\
&\left.+\int _{0}^{\pi }\sigma _{rr}^{2} \left(r_{h} ,\theta \right) \sin \theta d\theta \right]+\int _{0}^{r_{h} }4\pi r^{2} \rho _{h} u_{r} \frac{\partial \phi _{h} }{\partial r} dr\\ 
&-\frac{1}{2} \int _{0}^{r_{h} }4\pi r\rho _{h} u_{r} \int _{0}^{\pi }\left(\sigma _{\theta \theta }^{2} +\sigma _{\varphi \varphi }^{2} +u_{\varphi }^{2} \right) \sin \theta d\theta  dr\\
&-\int _{0}^{r_{h} }2\pi r^{2} \rho _{h} \frac{\partial u_{r} }{\partial r} \int _{0}^{\pi }\sigma _{rr}^{2}  \sin \theta d\theta  dr=0.
\end{split}
\end{equation} 

The third example is for halo angular momentum. Multiplying the continuity equation (Eq. \eqref{ZEqnNum321851}) and Eq. \eqref{ZEqnNum720202} with $u_{\varphi } $ and $\rho _{h} $ respectively and adding them together leads to the equation for the evolution of $\rho _{h} u_{\varphi } $ that is relevant to the angular momentum,
\begin{equation} 
\label{ZEqnNum701076} 
\frac{\partial \left(\rho _{h} u_{\varphi } \right)}{\partial t} +\frac{1}{r^{2} } \frac{\partial \left[\left(\rho _{h} u_{\varphi } \right)u_{r} r^{2} \right]}{\partial r} +\frac{u_{r} }{r} \left(\rho _{h} u_{\varphi } \right)=0.      
\end{equation} 
Multiplying all terms with $r_{z} =r\sin \theta $ and integrating Eq. \eqref{ZEqnNum701076} over the entire halo, i.e. applying the integration
\begin{equation} 
\label{ZEqnNum961855} 
\begin{split}
\int _{0}^{r_{h} }\int _{0}^{\pi }&\int _{0}^{2\pi }\left[\left(\bullet \right)r\sin \theta \right]r^{2} \sin \theta d\varphi d\theta dr\\
&=\int _{0}^{r_{h} }2\pi r^{2} \int _{0}^{\pi }\left(\bullet \right)r\sin ^{2} \theta d\theta dr,
\end{split}
\end{equation} 
 leads to the time variation of halo angular momentum $\bar{H}_{h} $
\begin{equation} 
\label{ZEqnNum861207} 
\frac{\partial \bar{H}_{h} }{\partial t} =2\pi r_{h}^{3} \rho _{h} \left(r_{h} \right)\int _{0}^{\pi }u_{\varphi } \left(r_{h} ,\theta \right) \sin {}^{2} \theta d\theta \left(\frac{\partial r_{h} }{\partial t} -u_{r} \left(r_{h} \right)\right),     
\end{equation} 
where angular momentum $\bar{H}_{h}$ is defined as
\begin{equation} 
\label{eq:26} 
\bar{H}_{h} =\int _{0}^{r_{h} }2\pi r^{3} \rho _{h} \left(r\right)\left(\int _{0}^{\pi }u_{\varphi } \sin ^{2} \theta d\theta  \right) dr.       
\end{equation} 
Note that integration of the first term in Eq. \eqref{ZEqnNum701076} can be separated into two contributions using the Leibniz's rule (the integration limit $r_{h} =r_{h} \left(t\right)$ is a function of \textit{t}), 
\begin{equation} 
\label{eq:27} 
\begin{split}
&\int _{0}^{r_{h} \left(t\right)}2\pi r^{2} \int _{0}^{\pi }\frac{\partial \left(\rho _{h} u_{\varphi } \right)}{\partial t} r\sin ^{2} \theta d\theta dr  \\
&=\frac{\partial \bar{H}_{h} }{\partial t} -2\pi r_{h}^{3} \rho _{h} \left(r_{h} \right)\frac{\partial r_{h} }{\partial t} \int _{0}^{\pi }u_{\varphi } \left(r_{h} ,\theta \right) \sin {}^{2} \theta d\theta.
\end{split}
\end{equation} 
Here we demonstrate that the change of halo momentum comes only from the halo growth and radial flow at halo surface (infall of matter) (Eq. \eqref{ZEqnNum861207}). Mean radial and azimuthal flow in halos do not contribute to the change of halo angular momentum. Since ${\partial r_{h} /\partial t} >0$ and $u_{r} \left(r_{h} \right) < 0$ for a growing halo, the angular momentum $\bar{H}_{h}$ should be always increasing with time for growing halos. The halo angular momentum is conserved only if ${\partial r_{h} /\partial t} =u_{r}(r_{h})=0$.

The Tidal Torque Theory relates the origin and evolution of angular momentum to the gravitational tidal torques from the environment in which halos form \citep{Peebles:1969-Origin-of-the-Angular-Momentum,White:1984-Angular-Momentum-Growth-in-Pro}. The Tidal Torque Theory (TTT) predicts a linear increase of $\bar{H}_{h} $ with time \textit{t} for a halo with a fixed given mass. Most of the halo angular momentum is obtained from the misalignment between the tidal shear field and halo shape. However, a growing halo may obtain its momentum through continuous mass acquisition (see Eq. \eqref{ZEqnNum861207}). Similar ideas were also discussed before \citep{Vitvitska:2002-The-origin-of-angular-momentum}. Mass accretion leads to a linear increase of the specific angular momentum $H_{h} \sim t$ (or total angular momentum $\bar{H}_{h} \sim t^{2}$) at the early stage of halos (Table \ref{tab:3}). 

The final example is the halo rotational kinetic energy. Multiplying Eqs. \eqref{ZEqnNum720202} and \eqref{ZEqnNum701076} with $\rho _{h} u_{\varphi } $ and $u_{\varphi } $ respectively and adding them together leads to the evolution for term $\left(\rho _{h} u_{\varphi }^{2} \right)$, 
\begin{equation} 
\label{ZEqnNum749450} 
\underbrace{\frac{\partial \left(\rho _{h} u_{\varphi }^{2} \right)}{\partial t} }_{derivative}+\underbrace{\frac{1}{r^{2} } \frac{\partial \left[\left(\rho _{h} u_{\varphi }^{2} \right)u_{r} r^{2} \right]}{\partial r} }_{advection}+\underbrace{2\rho _{h} u_{\varphi }^{2} \frac{u_{r} }{r} }_{production}=0.      
\end{equation} 
Since $u_{r} >0$ in the halo core region and $u_{r} <0$ in the halo outer region for fast growing halos \citep[see][Fig. 2]{Xu:2021-Inverse-mass-cascade-halo-density}, the rotational kinetic energy is consumed in the halo core region and generated in outer region. 

In hydrodynamic turbulence, Reynolds stress arising from velocity fluctuation continuously transfers kinetic energy from mean flow to turbulence and sustain the energy cascade. Note that $u_{\varphi }^{2} $ is closely related to the in-plane velocity dispersion (Eq. \eqref{ZEqnNum618520}), the production term in Eqs. \eqref{ZEqnNum185081} and \eqref{ZEqnNum749450} describe the energy transfer between the mean flow and random motion (turbulence) in halos. The fictitious stresses $\rho _{h} \sigma _{rr}^{2}$ and $\rho _{h} u_{\varphi }^{2} $ (equivalent to the "Reynolds stress") acts on the gradient of mean flow (${u_{r} /r} $) to facilitate the energy transfer between mean flow and random motion. 

While the energy transfer in turbulence is always one-way from mean flow to random motion, the energy transfer is two-way in halos of dark matter flow, where energy can be drawn from random motion to mean flow in outer region ($u_{r} <0$ in Eq. \eqref{ZEqnNum749450}) or from mean flow to random motion in core region ($u_{r} >0$), depending on the local sign of $u_{r} $. However, for entire halo, there is a net transfer from mean flow to random flow (see Table \ref{tab:4}).

 Just like the radial kinetic energy in Eq. \eqref{ZEqnNum362154}, halo rotational kinetic energy is defined as, 
\begin{equation} 
\label{eq:29} 
\bar{K}_{a} =\frac{1}{2} \int _{0}^{r_{h} }2\pi r^{2} \int _{0}^{\pi }\left(\rho _{h} u_{\varphi }^{2} \right)\sin \theta d\theta dr  .        
\end{equation} 
Integrating Eq. \eqref{ZEqnNum749450} with $\int _{0}^{r_{h} }2\pi r^{2} \int _{0}^{\pi }{1/2} \left(\bullet \right)\sin \theta d\theta dr  $ leads to the evolution of the total rotational kinetic energy for entire halo,
\begin{equation} 
\label{ZEqnNum463094} 
\begin{split}
\frac{\partial \bar{K}_{a} }{\partial t} &=\underbrace{\pi r_{h}^{2} \rho _{h} \left(r_{h} \right)\int _{0}^{\pi }u_{\varphi }^{2} \left(r_{h} ,\theta \right) \sin \theta d\theta \left(\frac{\partial r_{h} }{\partial t} -u_{r} \left(r_{h} \right)\right)}_{1}\\
&-\underbrace{\int _{0}^{r_{h} }2\pi r^{2} \frac{u_{r} }{r} \rho _{h} \left(\int _{0}^{\pi }u_{\varphi }^{2}  \sin \theta d\theta \right) dr}_{2},
\end{split}
\end{equation} 
where the rotational kinetic energy can be changed due to halo growth and radial flow (term 1 in Eq. \eqref{ZEqnNum463094}) on surface and the energy transfer with the random motion in bulk of halo (term 2 in Eq. \eqref{ZEqnNum463094}). By contrast, angular momentum can only be changed due to the surface term (see Eq. \eqref{ZEqnNum861207}). 

 A complete understanding of the evolution and transfer of radial and rotational kinetic energies will require solutions of mean flow and velocity dispersions. Obviously Eqs. \eqref{ZEqnNum321851}, \eqref{ZEqnNum535400}, \eqref{ZEqnNum618520}, and \eqref{ZEqnNum720202} is not a closed system. Additional assumptions are required to obtain complete solutions of the mean flow and velocity dispersions, which will be discussed in the next section. 

\subsection{General solutions for axisymmetric rotating\&growing halos}
\label{sec:3.3}
We now turn to the axisymmetric solutions of a rotating and growing spherical halo with a non-zero angular velocity. In principle, such halos can be characterized by four time-varying parameters, i.e. the halo mass $m_{h} \left(t\right)$, the angular velocity $\omega _{h} \left(t\right)$, concentration parameter $c\left(t\right)$ and scale radius $r_{s} \left(t\right)$. The halo size (virial radius) is $r_{h} \left(t\right)=c\left(t\right)r_{s} \left(t\right)$. A reduced spatial-temporal variable \textit{x} is introduced \citep[see][Eq. (60)]{Xu:2021-Inverse-mass-cascade-halo-density},
\begin{equation} 
\label{eq:31} 
x\left(r,t\right)=\frac{r}{r_{s} \left(t\right)} =\frac{c\left(t\right)r}{r_{h} \left(t\right)} .         
\end{equation} 
The time and spatial derivatives with respect to \textit{t} and \textit{r} can be derived in terms of the reduced variable \textit{x} using the chain rule,
\begin{equation}
\frac{\partial }{\partial t} =\frac{\partial }{\partial x} \frac{\partial x}{\partial t} =-\frac{x}{t} \frac{\partial \ln r_{s} }{\partial \ln t} \frac{\partial }{\partial x} \quad \textrm{and} \quad \frac{\partial }{\partial r} =\frac{\partial }{\partial x} \frac{\partial x}{\partial r} =\frac{1}{r_{s} } \frac{\partial }{\partial x}.    
\label{ZEqnNum253473}
\end{equation}
\noindent A unknown function $F\left(x\right)$ is introduced such that halo density $\rho _{h} $ and the mass $m_{r} $ enclosed in the radius \textit{r }can all be expressed in terms of function $F\left(x\right)$, 
\begin{equation}
\rho _{h} \left(r,t\right)=\frac{m_{h} \left(t\right)}{4\pi r_{s}^{3} } \frac{F^{'} \left(x\right)}{x^{2} F\left(c\right)} \quad \textrm{and} \quad m_{r} \left(r,t\right)=m_{h} \left(t\right)\frac{F\left(x\right)}{F\left(c\right)}.   
\label{ZEqnNum990243}
\end{equation}
\noindent The total mass of a virialized halo is expected to be proportional to the background density $\bar{\rho }_{0} $ at present epoch,  
\begin{equation} 
\label{ZEqnNum811003} 
m_{h}^{} \left(t\right)=\frac{4}{3} \pi r_{h}^{3} \Delta _{c} \bar{\rho }_{0} a^{-3}  ,         
\end{equation} 
where the critical ratio $\Delta _{c} =18\pi ^{2} $ can be obtained from a spherical collapse model or a two-body collapse model \citep[see][Eq. (89)]{Xu:2021-A-non-radial-two-body-collapse} for a matter dominant universe. The circular velocity at the surface of a halo and at any given radius \textit{r} can be defined as,
\begin{equation}
\begin{split}
&v_{cir}^{2} \left(a\right)=\frac{Gm_{h} \left(a\right)}{r_{h} \left(a\right)} =\frac{4\pi ^{2} r_{h}^{2} }{t^{2} } =\left(3\pi Hr_{h} \right)^{2} \\
&\textrm{and} \\
&v_{c}^{2} \left(r,a\right)=\frac{Gm_{r} \left(r,a\right)}{r} =\frac{cF\left(x\right)}{F\left(c\right)x} v_{cir}^{2}.
\end{split}
\label{ZEqnNum477002}
\end{equation}
\noindent A relation between $c\left(t\right)$, $r_{s} \left(t\right)$, and $m_{h} \left(t\right)$ is found from Eq. \eqref{ZEqnNum811003},
\begin{equation} 
\label{ZEqnNum536381} 
\frac{\partial \ln c}{\partial \ln t} +\frac{\partial \ln r_{s} }{\partial \ln t} =\frac{1}{3} \frac{\partial \ln m_{h} }{\partial \ln t} +\frac{2}{3} .        
\end{equation} 

We will focus on the solutions for two limiting situations in terms of a reduced amplitude parameter (peak height) of density fluctuation \citep{Despali:2014-Some-like-it-triaxial--the-uni},
\begin{equation} 
\label{eq:37} 
\nu ={\delta _{cr} /\sigma \left(m_{h} ,z\right)},  
\end{equation} 
where $\delta _{cr} \approx 1.68$ is the critical overdensity from spherical collapse model and $\sigma \left(m_{h} ,z\right)$ is the rms (root mean square) fluctuation of the smoothed density field. Halos at their early stage with fast mass accretion have their angular momentum increasing with time. The mass accretion and increase of angular momentum will gradually slower down with halos evolving toward the late stage of their life. 

At the same redshift, large halos tend to have a higher $\nu $ and small halos have a lower $\nu $. From this point on, "large" halos refer to the halos at early stage of its life with fast mass accretion (high $\nu $) and a growing core such that the concentration $c\left(t\right)$ is relatively time-invariant and the halo mass $m_{h} \left(t\right)\sim t$ from inverse mass cascade \citep[see][Fig. 7]{Xu:2021-Inverse-mass-cascade-mass-function}. From Eq. \eqref{ZEqnNum536381}, we should have
\begin{equation}
r_{h} \left(t\right)\sim t \quad \textrm{and} \quad r_{s} \left(t\right)\sim t.
\label{ZEqnNum681799}
\end{equation}

"Small" halos refer to low $\nu $ halos at the late stage of halo life with slow mass accretion and a stable core, where the scale radius $r_{s} \left(t\right)$ and the halo core mass (mass enclosed within $r_{s} $) $m_{r} \left(r_{s} ,t\right)$ are all relatively time-invariant such that (from Eqs. \eqref{ZEqnNum990243} and \eqref{ZEqnNum536381})
\begin{equation}
\frac{m_{r} \left(r_{s} ,t\right)}{m_{h} \left(t\right)} =\frac{F\left(1\right)}{F\left(c\right)} =C_{F} \quad \textrm{and} \quad c^{3} \sim \frac{F\left(c\right)}{F\left(1\right)} t^{2} =\frac{t^{2} }{C_{F} }.   
\label{ZEqnNum404231}
\end{equation}
\noindent Here $C_{F}$ is the ratio of core mass to halo mass and concentration $c\sim t^{{2/3} } \sim a$ for small halos with halo mass increases slowly with $m_{h}(t) \propto F(c)$. This simple relation is consistent with concentration models in \citep{Bullock:2001-A-universal-angular-momentum-p,Wechsler:2002-Concentrations-of-dark-halos-f}. 

The complete solution of the mean radial flow $u_{r} $ can be obtained by solving the continuity equation (Eq. \eqref{ZEqnNum321851}) for a given unknown function $F(x)$ \citep[see][Eq. (23)]{Xu:2021-Inverse-mass-cascade-halo-density},
\begin{equation} 
\label{ZEqnNum773554}
\begin{split}
&u_{r} \left(r\right)=u_{h} \frac{r_{s} }{t}\\ 
&\textrm{and the normalized radial flow}\\
&u_{h} \left(x\right)=x\frac{\partial \ln r_{s} }{\partial \ln t} +\left(\frac{\partial \ln F\left(c\right)}{\partial \ln t} -\frac{\partial \ln m_{h} }{\partial \ln t} \right)\frac{F\left(x\right)}{F^{'} \left(x\right)}. 
\end{split}
\end{equation} 
Obviously, $u_{h} \left(x\right)=0$ for small halos with a stable core (using Eq. \eqref{ZEqnNum404231} with constant $r_{s} $, halo mass $m_{h} \propto F(c)$). While for large halos (using Eq. \eqref{ZEqnNum681799} with a constant concentration \textit{c}),
\begin{equation} 
\label{ZEqnNum659883} 
u_{h} \left(x\right)=x-\frac{F\left(x\right)}{F^{'} \left(x\right)} .          
\end{equation} 

To derive full solutions for mean flow and velocity dispersions, the first assumption we made here is to use the separation of variables to express the mean azimuthal flow $u_{\varphi }$ as 
\begin{equation} 
\label{ZEqnNum376341} 
u_{\varphi } \left(r,\theta ,t\right)=\omega _{h} \left(t\right)r_{s}^{} \left(t\right)F_{\varphi } \left(x\right)K_{\varphi } \left(\theta \right), 
\end{equation} 
where $F_{\varphi } \left(x\right)$ and $K_{\varphi } \left(\theta \right)$ are the radial and angular functions for $u_{\varphi } $, respectively. The azimuthal flow $u_{\varphi } $ is expected to be proportional to the effective halo angular velocity $\omega _{h} $. The exact solution of $F_{\varphi } \left(x\right)$ can be derived from the momentum equation for $u_{\varphi } $ (Eq. \eqref{ZEqnNum720202}) with help of chain rule from Eq. \eqref{ZEqnNum253473},
\begin{equation} 
\label{ZEqnNum622731} 
\frac{\partial \ln F_{\varphi } }{\partial \ln x} =\frac{u_{h} \left(x\right)+x\left(\frac{\partial \ln \omega _{h} }{\partial \ln t} +\frac{\partial \ln r_{s} }{\partial \ln t} \right)}{x\frac{\partial \ln r_{s} }{\partial \ln t} -u_{h} \left(x\right)} .       
\end{equation} 

Velocity dispersions are expected to be isotropic for non-rotating halos with a spherical symmetry. The halo spin ($\omega _{h} \ne 0$) breaks the spherical symmetry and leads to the anisotropy in velocity dispersion. For spherical halos with a finite angular velocity $\omega _{h} $, velocity dispersions are only isotropic along the axis of rotation ($r_{z} =0$ or $\theta =0$ such that $u_{\varphi } =0$ on that axis), 
\begin{equation} 
\label{eq:44} 
\begin{split}
\sigma _{rr}^{2} \left(r,\theta =0,t\right)&=\sigma _{\theta \theta }^{2} \left(r,\theta =0,t\right)\\
&=\sigma _{\varphi \varphi }^{2} \left(r,\theta =0,t\right)=\sigma _{r0}^{2} \left(r,t\right),
\end{split}
\end{equation} 
where $\sigma _{r0}^{2} \left(r,t\right)$ is the axial velocity dispersion along the axis of rotation. With spin causing the velocity dispersion anisotropy, velocity dispersions can be a function of azimuthal flow $u_{\varphi }^{2} $. 

The second assumption is to express velocity dispersions as functions of the azimuthal flow $u_{\varphi }^{2} $. The first order approximation for three dispersions should read
\begin{equation}
\label{ZEqnNum528783} 
\sigma _{\theta \theta }^{2} \left(r,\theta ,t\right)=\underbrace{\sigma _{r0}^{2} \left(r,t\right)}_{1}+\underbrace{\alpha _{\varphi } \left(r,t\right)u_{\varphi }^{2} \left(r,\theta ,t\right)}_{2},       
\end{equation} 
\begin{equation}
\label{ZEqnNum154745} 
\sigma _{\varphi \varphi }^{2} \left(r,\theta ,t\right)=\sigma _{r0}^{2} \left(r,t\right)+\beta _{\varphi } \left(r,t\right)u_{\varphi }^{2} \left(r,\theta ,t\right),       
\end{equation} 
\begin{equation} 
\label{ZEqnNum892394} 
\sigma _{rr}^{2} \left(r,\theta ,t\right)=\sigma _{r0}^{2} \left(r,t\right)+\gamma _{\varphi } \left(r,t\right)u_{\varphi }^{2} \left(r,\theta ,t\right),       
\end{equation} 
where expansion coefficients $\alpha _{\varphi } $, $\beta _{\varphi } $ and $\gamma _{\varphi } $ will be determined later. This approximation decomposes the velocity dispersions into a non-spin induced axial dispersion (term 1) and a spin-induced dispersion (term 2). Substitution of Eqs. \eqref{ZEqnNum528783} and \eqref{ZEqnNum154745} into the momentum equation in polar direction (Eq. \eqref{ZEqnNum618520}) leads to the solution for angular function $K_{\varphi }^{} \left(\theta \right)$,
\begin{equation} 
\label{eq:48} 
\frac{\partial \ln u_{\varphi } }{\partial \ln \sin \theta } =\frac{\partial \ln K_{\varphi } }{\partial \ln \sin \theta } =\frac{1+\beta _{\varphi } -\alpha _{\varphi } }{2\alpha _{\varphi } } .        
\end{equation} 
With expression of $u_{\varphi } $ in Eq. \eqref{ZEqnNum376341},  the angular function $K_{\varphi }^{} \left(\theta \right)$ is 
\begin{equation}
K_{\varphi } \left(\theta \right)=\left(\sin \theta \right)^{\alpha _{\theta } } \quad \textrm{and} \quad \alpha _{\theta } =\frac{1+\beta _{\varphi } -\alpha _{\varphi } }{2\alpha _{\varphi }}.     
\label{ZEqnNum624087}
\end{equation}

Next, substitution of velocity dispersions (Eqs. \eqref{ZEqnNum528783}-\eqref{ZEqnNum892394}) into the momentum equation in radial direction (Eq. \eqref{ZEqnNum535400}) leads to two separate equations, i.e. an equation for the isotropic velocity dispersion $\sigma _{r0}^{2} $ (term 1 in Eq. \eqref{ZEqnNum528783}),  
\begin{equation} 
\label{ZEqnNum662332} 
\frac{\partial u_{r} }{\partial t} +u_{r} \frac{\partial u_{r} }{\partial r} +\frac{1}{\rho _{h} } \frac{\partial \left(\rho _{h} \sigma _{r0}^{2} \right)}{\partial r} +\frac{\partial \phi _{r} }{\partial r} +F_{a} \left(r,t\right)=0,      
\end{equation} 
and an equation for anisotropic velocity dispersions via coefficients $\alpha _{\varphi } $, $\beta _{\varphi } $ and $\gamma _{\varphi } $ (term 2 in Eqs. \eqref{ZEqnNum528783}-\eqref{ZEqnNum892394}), 
\begin{equation} 
\label{ZEqnNum451086} 
\frac{\partial \ln \gamma _{\varphi } }{\partial \ln x} +2\frac{\partial \ln u_{\varphi } }{\partial \ln x} +\frac{\partial \ln \rho _{h} }{\partial \ln x} +2-2\alpha _{a} =\frac{rF_{a} \left(r,t\right)}{\gamma _{\varphi } u_{\varphi }^{2} } .     
\end{equation} 
Here $\alpha _{a} $ is a dimensionless coefficient for the effect of anisotropy on the radial velocity dispersion through functions $\alpha _{\varphi } $, $\beta _{\varphi } $, and $\gamma _{\varphi } $, 
\begin{equation} 
\label{ZEqnNum594242} 
\alpha _{a} ={\left(\alpha _{\varphi } +\beta _{\varphi } +1\right)/2\gamma _{\varphi } } ,         
\end{equation} 
where $\alpha _{a} $ can be related to the anisotropic parameter $\beta _{h1} $. The new and the old (standard) anisotropy parameters defined in Eqs. \eqref{ZEqnNum865565} and \eqref{ZEqnNum597498} can be expressed in terms of the coefficients $\alpha _{\varphi } $, $\beta _{\varphi } $ and $\gamma _{\varphi } $ as,
\begin{equation} 
\label{eq:53} 
\beta _{h1} =\frac{1-{\left(1+\alpha _{\varphi } +\beta _{\varphi } \right)/\left(2\gamma _{\varphi } \right)} }{1+{\sigma _{r0}^{2} /\left(\gamma _{\varphi } u_{\varphi }^{2} \right)} } =\frac{1-\alpha _{a} }{1+{\sigma _{r0}^{2} /\left(\gamma _{\varphi } u_{\varphi }^{2} \right)} }  
\end{equation} 
and 
\begin{equation} 
\label{ZEqnNum486654} 
\beta _{h} =\frac{1-{\left(\alpha _{\varphi } +\beta _{\varphi } \right)/\left(2\gamma _{\varphi } \right)} }{1+{\sigma _{r0}^{2} /\left(\gamma _{\varphi } u_{\varphi }^{2} \right)} } .         
\end{equation} 
The coupling function $F_{a} \left(r,t\right)$ (with a unit of acceleration) reflects the coupling between term 1 and term 2 in Eq. \eqref{ZEqnNum528783}, i.e. how velocity dispersion $\gamma _{\varphi } u_{\varphi }^{2} $ due to halo spin and the axial dispersion $\sigma _{r0}^{2} $ are coupled. Two terms are decoupled if and only if $F_{a} \left(r,t\right)=0$. 

 The radial velocity dispersion $\sigma _{r}^{2} \left(r,t\right)$ for a non-rotating isotropic spherical growing halo ($\omega _{h}^{} =0$ and $\beta _{h1}^{} =0$ in Eq. \eqref{ZEqnNum535400}) has been extensively studied previously \citep{Xu:2021-Inverse-mass-cascade-halo-density}, where
\begin{equation} 
\label{ZEqnNum706880} 
\frac{\partial u_{r} }{\partial t} +u_{r} \frac{\partial u_{r} }{\partial r} +\frac{1}{\rho _{h} } \frac{\partial \left(\rho _{h} \sigma _{r}^{2} \right)}{\partial r} +\frac{\partial \phi _{r} }{\partial r} =0.       
\end{equation} 
The logarithmic slope of pressure can be obtained from Eq. \eqref{ZEqnNum706880} \citep[see][Eq. (73)]{Xu:2021-Inverse-mass-cascade-halo-density},
\begin{equation} 
\label{ZEqnNum479172} 
\frac{\partial \ln \left(\rho _{h} \sigma _{r}^{2} \right)}{\partial \ln x} =\frac{v_{cir}^{2} }{\sigma _{r}^{2} } \left(\frac{x^{2} -xu_{h} }{4\pi ^{2} c^{2} } \frac{\partial u_{h} }{\partial x} -\frac{v_{c}^{2} }{v_{cir}^{2} } \right).  
\end{equation} 
Obviously, $\sigma _{r0}^{2} =\sigma _{r}^{2} $ and Eq. \eqref{ZEqnNum662332} reduces to Eq. \eqref{ZEqnNum706880} only if the coupling term $F_{a} \left(r,t\right)=0$. Comparison of Eq. \eqref{ZEqnNum706880} with \eqref{ZEqnNum662332} leads to a relation between two dispersions
\begin{equation}
\label{ZEqnNum401346} 
\frac{\partial \ln \left[\rho _{h} \left(\sigma _{r}^{2} -\sigma _{r0}^{2} \right)\right]}{\partial \ln r} =\frac{rF_{a} \left(r,t\right)}{\left(\sigma _{r}^{2} -\sigma _{r0}^{2} \right)} ,        
\end{equation} 
where the coupling term $F_{a} \left(r,t\right)$ contributes to the difference between $\sigma _{r}^{2} $ of an isotropic non-rotating halo and the axial dispersion $\sigma _{r0}^{2} $ of a rotating halo. The relation between the other two radial dispersions is obtained by subtracting Eq. \eqref{ZEqnNum706880} from Eq. \eqref{ZEqnNum535400},
\begin{equation} 
\label{eq:58} 
\frac{\partial \ln \left[\rho _{h} \left(\sigma _{rr}^{2} -\sigma _{r}^{2} \right)\right]}{\partial \ln r} =-\frac{2\beta _{h1} \sigma _{rr}^{2} }{\left(\sigma _{rr}^{2} -\sigma _{r}^{2} \right)} ,        
\end{equation} 
where $\beta _{h1} $ is the new anisotropic parameter defined in Eq. \eqref{ZEqnNum865565}. However, $\sigma _{rr}^{2} $ does not necessarily equal $\sigma _{r}^{2} $ even for $\beta _{h1} =0$ because of the additional dependence of $\sigma _{rr}^{2} $ on $u_{\varphi }^{2} $ in Eq. \eqref{ZEqnNum892394}. 

Finally, the difference between radial velocity dispersion $\sigma _{rr}^{2} $ and axial dispersion $\sigma _{r0}^{2} $ reads 
\begin{equation} 
\label{eq:59} 
\frac{\partial \ln \left[\rho _{h} \left(\sigma _{rr}^{2} -\sigma _{r0}^{2} \right)\right]}{\partial \ln r} =\frac{rF_{a} \left(r,t\right)-2\beta _{h1} \sigma _{rr}^{2} }{\left(\sigma _{rr}^{2} -\sigma _{r0}^{2} \right)} ,       
\end{equation} 
which is consistent with Eq. \eqref{ZEqnNum451086} and includes two contributions from $F_{a} \left(r,t\right)$ and $\beta _{h1} $, respectively. 
 
\subsection{Solutions for small halos at late stage (low peak height \texorpdfstring{$\nu$}{})}
\label{sec:3.4}
\begin{figure}
\includegraphics*[width=\columnwidth]{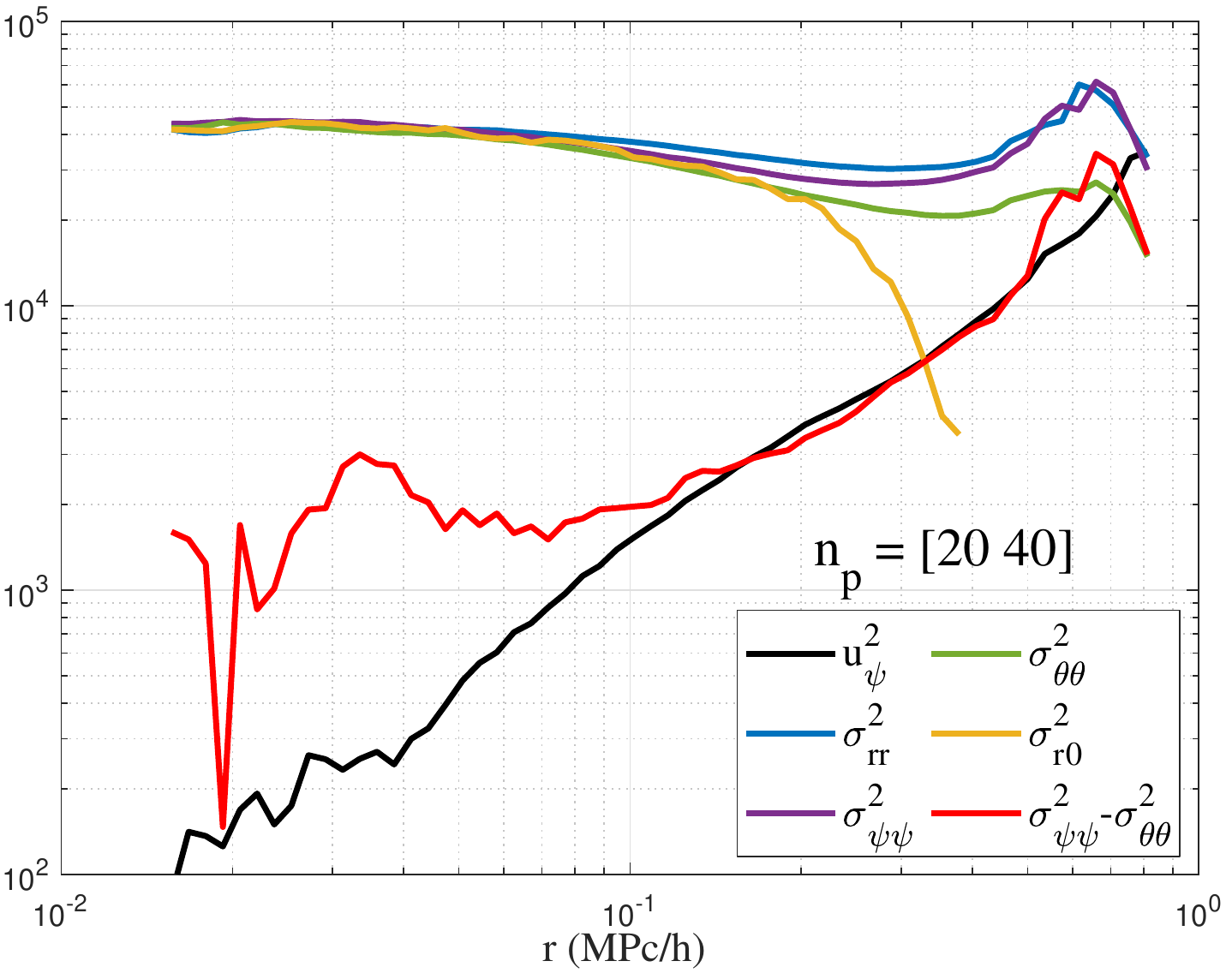}
\caption{The variation of velocity dispersions and mean azimuthal flow $u_{\varphi }^{2} $ with radius \textit{r} for all halos with a size $n_{p} $ between [20 40] at \textit{z}=0 (unit: $(km/s)^2$). For all three velocity dispersions, the contribution from $\sigma _{r0}^{2} $ is dominant at small \textit{r}, while contributions from $u_{\varphi }^{2} $ is dominant at large \textit{r}. A good relation $u_{\phi }^{2} =\sigma _{\varphi \varphi }^{2} -\sigma _{\theta \theta }^{2} $ between mean flow and in-plane velocity dispersions can be clearly identified (Eq. \eqref{ZEqnNum521885}).}
\label{fig:4}
\end{figure}

We first focus on small halos with a stable core and slow mass accretion rate. Figure \ref{fig:4} plots the variation of (spherical and group averaged) velocity dispersions and mean azimuthal flow $u_{\varphi }^{2} $ with the radius \textit{r} for all halos with a size $n_{p} $ between [20 40]. For velocity dispersions (Eqs. \eqref{ZEqnNum528783} to \eqref{ZEqnNum892394}), the contribution from $\sigma _{r0}^{2} $ (term 1) is dominant at small \textit{r}, while the contribution from $u_{\varphi }^{2} $ (term 2) can be dominant at large \textit{r}. We also found a good agreement of $u_{\phi }^{2} =\sigma _{\varphi \varphi }^{2} -\sigma _{\theta \theta }^{2} $ for large \textit{x} (Eq. \eqref{ZEqnNum521885}), i.e. a surprisingly simple result that directly connects the mean flow and random motion (turbulence) at halo scale. 

For small halos with a stable core, coupling term $F_{a} \left(r,t\right)=0$ is expected such that $\sigma _{r0}^{2} =\sigma _{r}^{2} $ (Eqs. \eqref{ZEqnNum662332} and \eqref{ZEqnNum401346}). For core region with a small \textit{r,} $\sigma _{rr}^{2} \approx \sigma _{r0}^{2} =\sigma _{r}^{2} $, while $\sigma _{rr}^{2} \gg \sigma _{r0}^{2} =\sigma _{r}^{2} $ for outer region due to a significant contribution from azimuthal flow $u_{\varphi }^{2} $ at large \textit{r} (see Fig. \ref{fig:4}). In addition, the radial flow vanishes with $u_{r} \left(r\right)=0$ (see Eq. \eqref{ZEqnNum773554} small halos are well bound and virialized structure). Small halos are incompressible in (proper) velocity field with $u_{r} =u_{\theta } =0$ and $u_{\varphi } =u_{\varphi } \left(r,\theta ,t\right)$, i.e. $\nabla \cdot \boldsymbol{\mathrm{u}}=0$. While in comoving system, the peculiar velocity \textbf{v} field has constant divergence with 
\begin{equation} 
\label{eq:60} 
\nabla \cdot \boldsymbol{\mathrm{v}}=\frac{1}{r^{2} } \frac{\partial \left(r^{2} v_{r} \right)}{\partial r} =-3Ha,         
\end{equation} 
where peculiar radial flow $v_{r} =u_{r} -Har=-Har$ if $u_{r} =0$ (also from stable clustering hypothesis demonstrated by a two-body collapse model \citep{Xu:2021-A-non-radial-two-body-collapse}). The constant divergence flow in small halos was also supported by the correlation-based statistical analysis, where dark matter flow is shown to be constant divergence on small scale and irrotational on large scale \citep{Xu:2022-The-statistical-theory-of-2nd,Xu:2022-The-statistical-theory-of-3rd}.  

The angular velocity $\omega _{h} $ of small halos is relatively time-invariant (small halos grow slowly with a constant $r_{s} $ and vanishing radial flow $u_{r} =0$). Small halos with a stable core are expected to be relatively isotropic with the anisotropic parameter $\beta _{h1} =0$ (Eq. \eqref{ZEqnNum486654}) (however, the old definition of anisotropic parameter in Eq. \eqref{ZEqnNum597498} $\beta _{h} \ne 0$ for small halos), i.e. 
\begin{equation}
\alpha _{a} =\frac{\alpha _{\varphi } +\beta _{\varphi } +1}{2\gamma _{\varphi } } =1 \quad \textrm{or} \quad 2\sigma _{rr}^{2} =\sigma _{\varphi \varphi }^{2} +\sigma _{\theta \theta }^{2} +u_{\phi }^{2}.
\label{ZEqnNum220251}
\end{equation}
\noindent With $\beta _{h1} =0$ (or equivalently $\alpha _{a} =1$) and $F_{a} \left(r,t\right)=0$, Eq. \eqref{ZEqnNum451086} for $\gamma _{\varphi } $ reduces to,
\begin{equation} 
\label{ZEqnNum477261} 
\frac{\partial \ln \gamma _{\varphi } }{\partial \ln x} +2\frac{\partial \ln u_{\varphi } }{\partial \ln x} +\frac{\partial \ln \rho _{h} }{\partial \ln x} =0.        
\end{equation} 

For outer region (large \textit{x}) of small halos with an isothermal density profile (the logarithmic slope of density is -2) and $u_{\varphi } \sim \omega _{h} r_{z} \sim \omega _{h} r_{s}^{} x\sin \theta $ (as shown in Figs. \ref{fig:2} and \ref{fig:3}), Eq. \eqref{ZEqnNum477261} predicts ${\partial \gamma _{\varphi } /\partial x} =0$, i.e. $\gamma _{\varphi } \left(r,t\right)$ is almost a constant of location \textit{r}. If we also expect $\sigma _{rr}^{2} =\sigma _{\varphi \varphi }^{2}$, i.e. $\gamma _{\varphi } =\beta _{\varphi}$ (as shown in Figs. \ref{fig:4} and \ref{fig:8}) for large \textit{r}, Eq. \eqref{ZEqnNum220251} requires $1+\alpha _{\varphi } =\beta _{\varphi } $ such that 
\begin{equation}
\sigma _{rr}^{2} =\sigma _{\varphi \varphi }^{2} =\sigma _{\theta \theta }^{2} +u_{\phi }^{2} \quad \textrm{and} \quad u_{r} =u_{\theta } =0,    \label{ZEqnNum521885}
\end{equation}
\noindent as shown in both Fig. \ref{fig:4} and Fig. \ref{fig:8} for small halos. Equation \eqref{ZEqnNum521885} may be considered as how energy is partitioned along each direction for isotropic ($\beta _{h1} =0$), incompressible, fully virialized ($u_{r} =0$), and rotating halos with extremely slow mass accretion. As shown in Table \ref{tab:2}, the total kinetic energy (both random motion and mean flow) is partitioned along each coordinate:
\begin{equation}
\begin{split}
&\sigma _{rr}^{2} =\sigma _{\varphi \varphi }^{2}\textrm{ (radial)}, \quad \sigma _{\varphi \varphi }^{2} +u_{\varphi }^{2}\textrm{ (azimuthal)},\\
&\textrm{and}\quad \sigma _{\theta \theta }^{2} =\sigma _{\varphi \varphi }^{2} -u_{\varphi }^{2}\textrm{ (polar)},
\end{split}
\label{ZEqnNum381411}
\end{equation}

\begin{table}
\caption{Dispersions and mean flow for rotating and non-rotating halos}
\label{tab:2}
\begin{tabular}{p{0.45in}p{0.3in}p{0.6in}p{0.65in}p{0.65in}} 
\hline 
 &  & Radial ($r$) & Azimuthal ($\varphi $) & Polar ($\theta $) \\ \hline 
Rotating (Eq. \eqref{ZEqnNum535400}) & Random & $\sigma _{rr}^{2} =\sigma _{r0}^{2} +2u_{\varphi }^{2} $ & $\sigma _{\varphi \varphi }^{2} =\sigma _{r0}^{2} +2u_{\varphi }^{2} $ & $\sigma _{\theta \theta }^{2} =\sigma _{r0}^{2} +u_{\varphi }^{2} $ \\ \hline 
 & Mean flow & 0 & $u_{\varphi }^{2} $ & 0 \\ \hline 
Non-rotating (Eq. \eqref{ZEqnNum662332}) & Random & $\sigma _{rr}^{2} =\sigma _{r}^{2} =\sigma _{r0}^{2} $ & $\sigma _{\varphi \varphi }^{2} =\sigma _{r0}^{2} $ & $\sigma _{\theta \theta }^{2} =\sigma _{r0}^{2} $ \\ \hline 
 & Mean flow & 0 & 0 & 0 \\ \hline 
\end{tabular}
\end{table}
Energy is not equipartitioned along each direction, with the largest kinetic energy in azimuthal direction and the smallest kinetic energy in polar direction. The exponent $\alpha _{\theta } ={1/\alpha _{\varphi } } $ for angular function $K_{\varphi } \left(\theta \right)$ can be obtained from Eq. \eqref{ZEqnNum624087}. For $K_{\varphi } \left(\theta \right)\sim \sin \theta $ such that $\alpha _{\theta } =1$, we should have $\alpha _{\varphi } =1$ and $\beta _{\varphi } =\gamma _{\varphi } =2$ for small halos. This can be confirmed by simulation data in Fig. \ref{fig:8}.

Now let's compare the energy of an initially virialized non-rotating halo that has an isotropic velocity dispersion $\sigma _{r}^{2} $ with the energy of a rotating halo of the same size. The density profile and the potential energy should be the same for both halos. The axial dispersions of rotating halo is always $\sigma _{r0}^{2} =\sigma _{r}^{2}$ (Eq. \eqref{ZEqnNum401346} with $F_{a}(r,t)=0$). For a rotating halo with a rotational kinetic energy of $\bar{K}_{a}$ (see Eq. \eqref{eq:29} for definition), there will be around $5\bar{K}_{a} $ extra kinetic energy in the form of random motion with $\beta _{\varphi } =\gamma _{\varphi } =2\alpha _{\varphi } =2$ when compared with non-rotating halo (see Table \ref{tab:2}). In addition, small halos with a finite spin will have an additional spin-induced pressure $\propto \rho _{h} u_{\varphi }^{2}$ when compared to a non-rotating halo. The spin-induced pressure is independent of \textit{r} for an isothermal density profile ($\rho _{h} \propto r^{-2}$ and $u_{\varphi } \propto r$) such that the gradient of spin-induced pressure vanishes. Total pressure gradient of rotating halos is the same as the non-rotating halo to balance the gravitational force (Eq. \eqref{ZEqnNum535400}). 

\begin{figure}
\includegraphics*[width=\columnwidth]{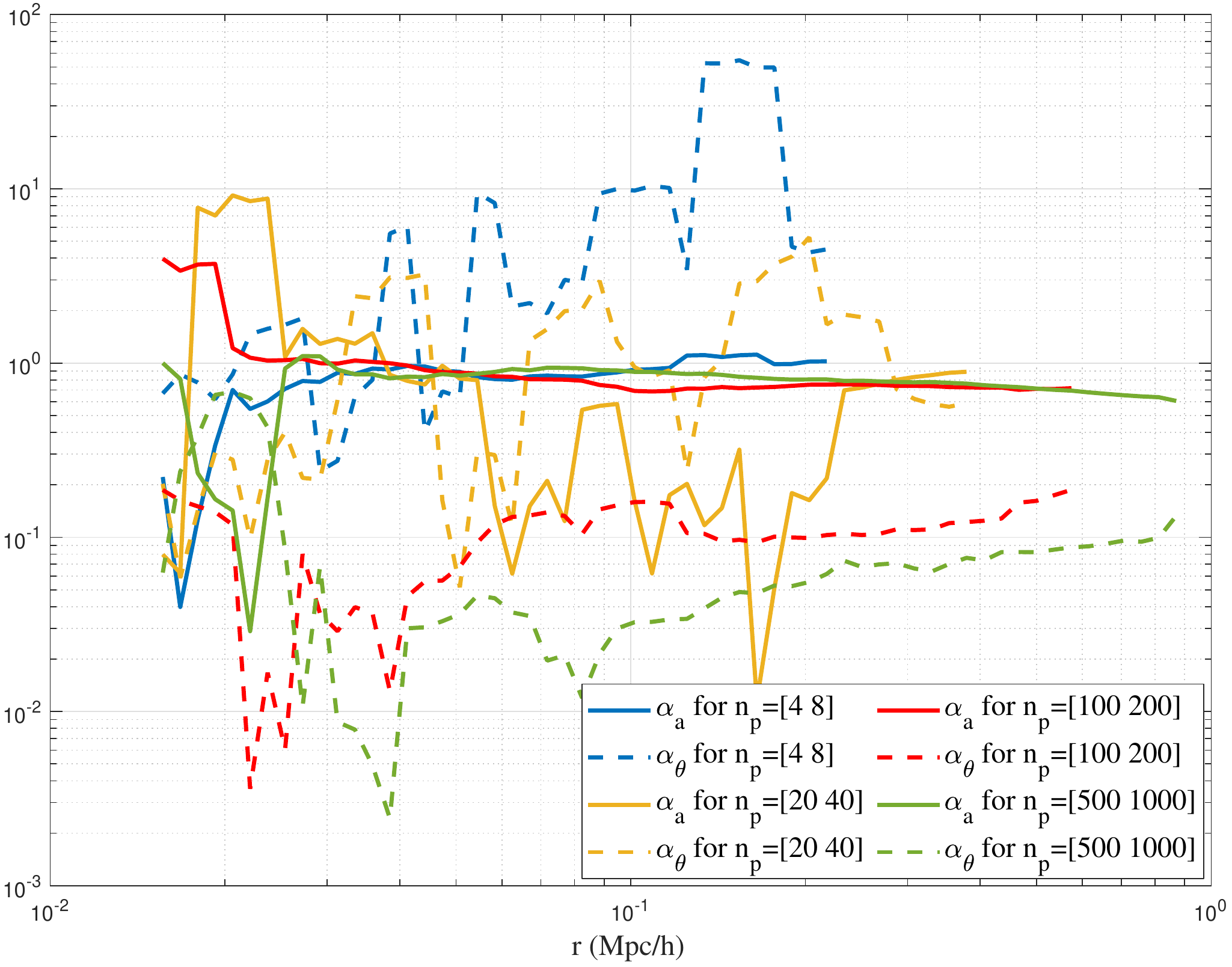}
\caption{The variation of two dimensionless parameters $\alpha _{a} $ (Eq. \eqref{ZEqnNum594242}) and $\alpha _{\theta } $ (Eq. \eqref{ZEqnNum624087}) with radius \textit{r} for halo groups of different sizes $n_{p} $. For small halos with a stable core, $\alpha _{a} =1$ and halo is relatively isotropic with $\beta _{h1} =0$ everywhere. Large halos tend to have an anisotropic outer region with $\alpha _{a} <1$ and an isotropic core with $\alpha _{a} \approx 1$. The azimuthal flow $u_{\varphi }^{} $ tends to have a stronger dependence on polar angle $\theta $ for small halos, while $\alpha _{\theta } \ll 1$ for large halos such that $u_{\varphi }^{} $ is less dependent on angle $\theta $.}
\label{fig:5}
\end{figure}

In contrast to normal object whose temperature is independent of the speed of spin, faster rotating halos (with fixed mass) are expected to be hotter with greater entropy due to the random motion associated with velocity dispersion. 
Figure \ref{fig:5} plots the variation of two parameters $\alpha _{a} $ (Eq. \eqref{ZEqnNum594242}) and $\alpha _{\theta } $ (Eq. \eqref{ZEqnNum624087}) with radius \textit{r} for halo groups of different sizes $n_{p} $. For small halos, $\alpha _{a} =1$ and halo is isotropic with $\beta _{h1} =0$ almost everywhere. Large halos tend to have an anisotropic outer region with $\alpha _{a} <1$ and an isotropic core with $\alpha _{a} \approx 1$. In addition, the azimuthal flow $u_{\varphi }^{} $ tends to strongly depend on the polar angle $\theta $ for small halos. While $\alpha _{\theta } \ll 1$ for large halos, i.e. $u_{\varphi }^{} $ is less dependent on $\theta $ for large halos. More discussion for the solutions of large halos is presented in next section \ref{sec:3.5}.  

\subsection{Solutions for large halos at early stage (high peak height \texorpdfstring{$\nu$}{})}
\label{sec:3.5}
We now turn to solutions for the other limiting situation, i.e. large halos (high peak height $\nu $) with an expanding core, fast mass accretion, and constant halo concentration \textit{c}. We first focus on the solution for azimuthal flow $u_{\varphi }^{} $. For large halos with fast mass accretion, there exists a non-zero radial flow $u_{r}$ (Eq. \eqref{ZEqnNum659883}), where the normalized radial flow $u_{h}$ is  
\begin{equation}
u_{h} \left(x\right)=\frac{u_{r} \left(r\right)t}{r_{s} \left(t\right)} =x-\frac{F\left(x\right)}{F^{'} \left(x\right)} \quad \textrm{and} \quad u_{h} \left(c\right)=c\left(1-\frac{1}{\alpha _{h} } \right).   \label{ZEqnNum176930}
\end{equation}

\noindent A halo deformation parameter is introduced here as
\begin{equation} 
\label{ZEqnNum716050} 
\alpha _{h} ={cF^{'} \left(c\right)/F\left(c\right)}  
\end{equation} 
to quantify the radial deformation at halo surface (no deformation if $\alpha _{h} =1$ for isothermal density profile). The (normalized) peculiar radial flow that excludes the Hubble flow is
\begin{equation} 
\label{ZEqnNum579659} 
\begin{split}
u_{p} \left(x\right) =u_{rp} \left(r\right)\frac{t}{r_{s} \left(t\right)} &=\left[u_{r} \left(r\right)-Hr\right]\frac{t}{r_{s} \left(t\right)}\\
&=u_{h} \left(x\right)-\frac{2}{3} x=\frac{1}{3} x-\frac{F\left(x\right)}{F^{'} \left(x\right)}
\end{split}
\end{equation} 
and 
\begin{equation}
\label{ZEqnNum477267} 
u_{p} \left(x=c\right)=c\left(\frac{1}{3} -\frac{1}{\alpha _{h} } \right). 
\end{equation} 

With radial flow from Eq. \eqref{ZEqnNum176930}, the logarithmic slope of density at halo center can be related to a halo deformation rate parameter $\gamma _{h} $ \citep[see][Eq. (38)]{Xu:2021-Inverse-mass-cascade-halo-density}, 
\begin{equation}
\left. \left(\frac{\partial \ln \rho _{h} }{\partial \ln x} +2\right)\right|_{x=0} =\left. \frac{\partial \ln F^{'} }{\partial \ln x} \right|_{x=0} =\frac{\gamma _{h} }{1-\gamma _{h} } \textrm{and} \left. \frac{\partial \ln F}{\partial \ln x} \right|_{x=0} =\frac{1}{1-\gamma _{h} },   
\label{ZEqnNum698012}
\end{equation}
\noindent where the deformation rate parameter $\gamma _{h} =\left. \left({\partial u_{h} /\partial x} \right)\right|_{x=0} $ quantifies the rate of deformation at the center of halo ($\gamma _{h} $=0, 1/2, and 2/3 for isothermal, NFW and Einasto profiles, respectively). 

The complete solution of the radial function ($F_{\varphi } \left(x\right)$ in Eq. \eqref{ZEqnNum376341}) for azimuthal flow is obtained by substituting $u_{h} \left(x\right)$ from Eq. \eqref{ZEqnNum176930} into Eq. \eqref{ZEqnNum622731} and reads,
\begin{equation}
F_{\varphi } \left(x\right)=\alpha _{f} \frac{F\left(x\right)^{\alpha _{\omega } } }{x} \quad \textrm{with} \quad \alpha _{\omega } =2+\frac{\partial \ln \omega _{h} }{\partial \ln t},
\label{eq:70}
\end{equation}

\noindent where the dimensionless constant $\alpha _{f} $ will be determined later. The angular velocity of large halos is expected to decrease with time as $\omega _{h} \sim a^{-{3/2}} \sim H\sim t^{-1}$ (Eq. \eqref{ZEqnNum501598}) such that $\alpha _{\omega } =1$. From Eq. \eqref{ZEqnNum698012},  
\begin{equation} 
\label{eq:71} 
\left. \frac{\partial \ln F_{\varphi } }{\partial \ln x} \right|_{x=0} =\left. \frac{\partial \ln F^{'} }{\partial \ln x} \right|_{x=0} =\frac{\gamma _{h} }{1-\gamma _{h} } .        
\end{equation} 
The final solution of the mean azimuthal flow $u_{\varphi } $ is
\begin{equation}
\label{ZEqnNum694993} 
u_{\varphi } \left(r,\theta ,t\right)=u_{\varphi } \left(x,\theta \right)=\alpha _{f} \omega _{h} \left(t\right)r_{s}^{} \left(t\right)\left(\sin \theta \right)^{\alpha _{\theta } } \frac{F\left(x\right)}{x} .     
\end{equation} 

Next, we need to determine the dimensionless constant $\alpha _{f} $ and effective angular velocity $\omega _{h} $ for entire halo. For a given halo density profile $\rho _{h} \left(x\right)$ that is determined by function $F\left(x\right)$ (Eq. \eqref{ZEqnNum990243}), the root mean square radius $r_{g}$ is \citep[see][Fig. 13]{Xu:2021-Inverse-and-direct-cascade-of-}
\begin{equation} 
\label{ZEqnNum353891} 
\begin{split}
r_{g}^{2}&=\frac{1}{m_{h} } \int _{0}^{r_{h} }4\pi r^{2} \rho _{h} \left(r\right)r^{2}dr\\
&=r_{h}^{2} \left[1-\frac{2}{c^{2} F\left(c\right)} \int _{0}^{c}xF\left(x\right)dx \right]=\gamma _{g}^{2} r_{h}^{2},
\end{split}
\end{equation} 
where $\gamma _{g}^{} ={r_{g} /r_{h} } $ is a dimensionless ratio of root mean square radius to halo size. The moment of inertia $I_{\omega } $ for that halo is given by,
\begin{equation} 
\label{ZEqnNum963993} 
\begin{split}
I_{\omega }&=\int _{0}^{r_{h} }\int _{0}^{\pi }\int _{0}^{2\pi }\rho _{h} r_{z}^{2} r^{2} \sin \theta d\varphi d\theta dr\\
&=\int _{0}^{r_{h} }2\pi r^{2} \rho _{h} \left(\int _{0}^{\pi }\left(r\sin \theta \right)^{2} \sin \theta d\theta  \right) dr=\frac{2}{3} m_{h} r_{g}^{2}, 
\end{split}
\end{equation} 
where the radius of gyration about axis of rotation is given by $r_{rg}^{2} ={I_{\omega } /m_{h} =} {2r_{g}^{2} /3} $. The halo (specific) angular momentum $H_{h} $ is
\begin{equation} 
\label{ZEqnNum960633} 
\begin{split}
H_{h}&=\omega _{h} r_{rg}^{2} =\frac{\omega _{h} I_{\omega } }{m_{h}}\\ 
&=\frac{2}{3} \omega _{h} r_{h}^{2} \left[1-\frac{2}{c^{2} F\left(c\right)} \int _{0}^{c}xF\left(x\right)dx \right]=\frac{2}{3} \omega _{h} r_{g}^{2}.
\end{split}
\end{equation} 

The specific angular momentum $H_{h} $ can also be derived by a direct integration of azimuthal flow $u_{\varphi } $ using Eq. \eqref{ZEqnNum961855}, where
\begin{equation} 
\label{ZEqnNum295029} 
\begin{split}
H_{h}&=\frac{1}{m_{h} } \int _{0}^{r_{h} }2\pi r^{3} \rho _{h} \left(r\right)\left(\int _{0}^{\pi }u_{\varphi } \sin ^{2} \theta d\theta  \right) dr\\
&=\frac{r_{h}^{} }{2cF\left(c\right)} \int _{0}^{c}xF^{'} \left(x\right)\left(\int _{0}^{\pi }u_{\varphi } \sin ^{2} \theta d\theta  \right) dx.
\end{split}
\end{equation} 

With solution of $u_{\varphi } $ given by Eq. \eqref{ZEqnNum694993}, the dimensionless constant $\alpha _{f} $ can be determined by comparing Eqs. \eqref{ZEqnNum960633} and \eqref{ZEqnNum295029},   
\begin{equation}
\label{ZEqnNum875729} 
\alpha _{f} =\frac{8c^{2} }{3F\left(c\right)} \left[1-\frac{2}{c^{2} F\left(c\right)} \int _{0}^{c}xF\left(x\right)dx \right]\frac{\Gamma \left(2+{\alpha _{\theta } /2} \right)}{\sqrt{\pi } \Gamma \left({3/2} +{\alpha _{\theta } /2} \right)} .
\end{equation} 

The peculiar radial velocity at halo virial radius $r=r_{h}$ is proportional to circular velocity with a proportional constant ${1/3\pi}$ (using Eq. \eqref{ZEqnNum477267})
\begin{equation} 
\label{eq:78} 
\begin{split}
u_{rp} \left(r_{h} \right)&=u_{r} \left(r_{h} \right)-Hr_{h}\\
&=u_{p} \left(c\right)\frac{r_{s} \left(t\right)}{t} =\frac{r_{h} }{t} \left(\frac{1}{3} -\frac{1}{\alpha _{h} } \right)=-\frac{v_{cir} }{3\pi }.
\end{split}
\end{equation} 
This is true for an isothermal density profile with $u_{r} =0$ and $\alpha _{h} =1$, where $v_{cir} $ is the circular velocity at the virial radius. The proportional constant ${1/3\pi}$ is essentially related to the angle of incidence \citep[see][Section 3.4]{Xu:2021-Inverse-mass-cascade-halo-density}, i.e. the angle for single merger merging with halos in mass cascade \citep[see][Fig. 8]{Xu:2021-Inverse-mass-cascade-mass-function}. It is also required to interpret the critical MOND (modified Newtonian dynamics) acceleration $a_0$ by the mass and energy cascade in dark matter flow \citep[see][Eq. (12) and Fig. 8]{Xu:2022-The-origin-of-MOND-acceleratio}.

Specifically, for large halos with an isothermal profile, $F\left(x\right)={x/c}$ and $\alpha _{\theta}=1$, we have $\alpha _{f} ={2c^{2}/3}$ and the mean azimuthal flow
\begin{equation}
u_{\varphi} \left(r,\theta ,t\right)=\frac{2}{3} \omega _{h} \left(t\right)r_{h}^{} \left(t\right)\sin \theta 
\label{eq:79}
\end{equation}
\noindent that is independent of the radius \textit{r}. 

Here if we assume the mean azimuthal flow $u_{\varphi }$ on halo surface with a polar angle of ${\pi/2}$ (halo equator) is equal to the peculiar radial flow (two velocities are equal on the halo equator), from Eq. \eqref{eq:78},
\begin{equation}
u_{\varphi} \left(r_{h} ,\frac{\pi }{2} ,t\right)=-u_{rp} \left(r_{h} ,t\right)=-u_{p} \left(x=c\right)\frac{r_{s} }{t} \approx \frac{v_{cir} }{3\pi }.     
\label{ZEqnNum499805}
\end{equation}
Substitution of expression for $u_{\varphi } $ from Eq. \eqref{ZEqnNum694993} and $u_{p} $ from Eq. \eqref{ZEqnNum477267} into Eq. \eqref{ZEqnNum499805} leads to the expression of halo angular velocity,
\begin{equation}
\label{ZEqnNum501598} 
\omega _{h} =\left(\frac{3}{2\alpha _{h} } -\frac{1}{2} \right)\frac{c^{2} }{F\left(c\right)\alpha _{f} } H,         \end{equation} 
where the angular velocity of large halos $\omega _{h} \sim H\sim t^{-1} \sim a^{{-3/2}}$. 

\begin{figure}
\includegraphics*[width=\columnwidth]{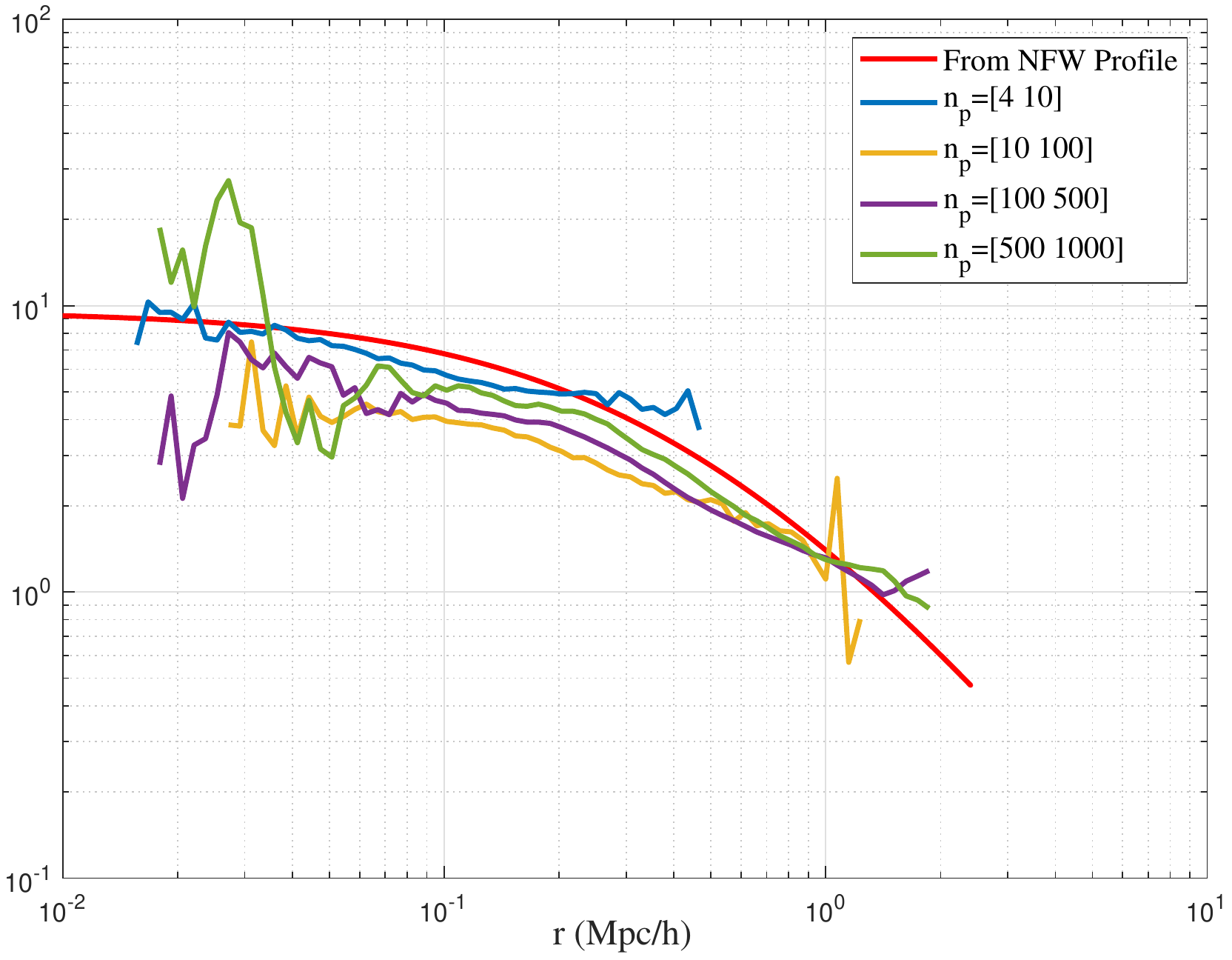}
\caption{The variation of normalized mean azimuthal flow $u_{\varphi}/(Hr)$ with radius \textit{r} for halo groups of different sizes $n_{p}$. The average is taken over all particles in the same spherical shell of radius \textit{r}, and over all halos with a size $n_{p}$ in the range given in figure. The analytical solution (Eq. \eqref{ZEqnNum875988}) is obtained using a NFW profile along with $c=3.5$, $r_{s} =0.34Mpc/h$, and $\alpha _{\theta} ={1/2}$. Solution suggests a faster spinning core and slower spinning outer region.}
\label{fig:6}
\end{figure}

Now we can compare our solution of the mean azimuthal flow $u_{\varphi } $ with \textit{N}-body simulation. The spherical averaged azimuthal flow $u_{n\varphi } $ (normalized by the Hubble flow) can be defined as (with solutions of $u_{\varphi } $ and $\omega _{h} $ from Eqs. \eqref{ZEqnNum694993} and \eqref{ZEqnNum501598}),  
\begin{equation} 
\label{ZEqnNum875988} 
\begin{split}
u_{n\varphi }&=\frac{{1/2} \int _{0}^{\pi }u_{\varphi } \left(r,\theta ,t\right)\sin \theta d\theta  }{Hr} \\
&=\frac{1}{2} \left(\frac{3}{2\alpha _{h} } -\frac{1}{2} \right)\frac{\sqrt{\pi } \Gamma \left(1+{\alpha _{\theta } /2} \right)}{\Gamma \left({3/2} +{\alpha _{\theta } /2} \right)} \frac{c^{2} F\left(x\right)}{x^{2} F\left(c\right)}.
\end{split}
\end{equation} 
Figure \ref{fig:6} presents the variation of normalized (spherical and group averaged) azimuthal flow $u_{n\varphi } $ with radius \textit{r} for different size of halos. Function $F\left(x\right)$ for a NFW density profile \begin{equation} 
\label{ZEqnNum437028} 
F\left(x\right)=\ln \left(1+x\right)-\frac{x}{1+x}  
\end{equation} 
is used for comparison along with other parameters $c=3.5$, $r_{s} =0.34Mpc/h$, and $\alpha _{\theta } ={1/2} $. 

Halos of the same size $n_{p} $ are first aligned by the axis of rotation and assembled into a composite halo containing all particles from the same halo group. The average is taken over the normalized azimuthal flow $u_{n\varphi } $ of all particles in the same spherical shell of radius \textit{r} of composite halos. Next, average is also taken over all halo groups with size $n_{p} $ in the given range as indicated in Fig. \ref{fig:6}. The azimuthal flow $u_{\varphi}$ approaches around 10 times of Hubble flow $Hr$ at the halo core region and is comparable to Hubble flow at halo outer region. This solution also suggests a faster spinning core and slower spinning outer region of halos with $\omega _{r} \sim H$ (Fig. \ref{fig:3}).   

\begin{figure}
\includegraphics*[width=\columnwidth]{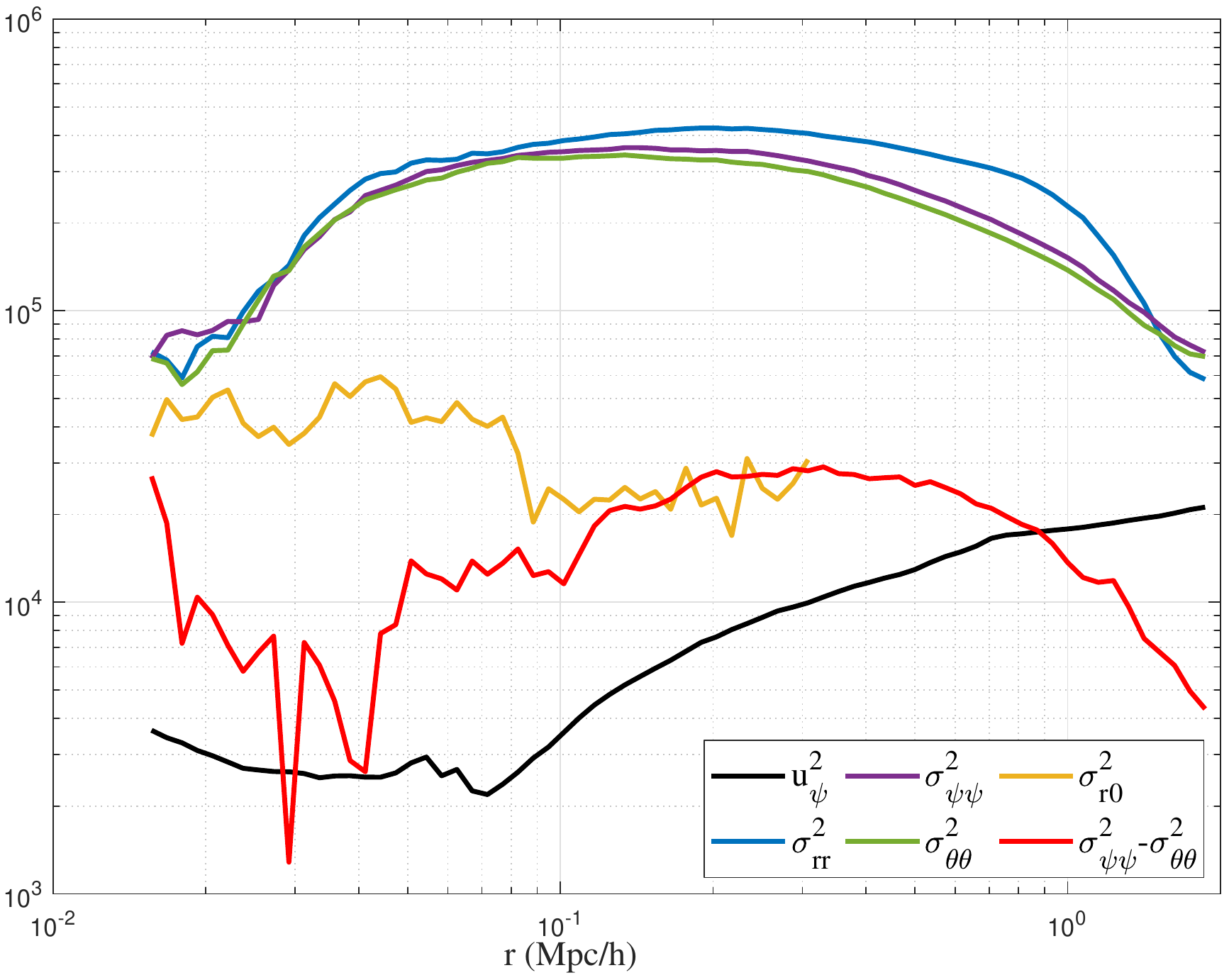}
\caption{The variation of (spherical averaged) velocity dispersion in unit of $(km/s)^2$ and the azimuthal flow $u_{\varphi }^{2}$ with radius \textit{r} for halo groups of size $n_{p}$ between [500 1000] at \textit{z}=0. By contrast to velocity dispersion for small halos in Fig. \ref{fig:4}, the spin-induced dispersion from azimuthal flow $u_{\varphi }^{2}$ is dominant in large halos over the axial dispersion $\sigma_{r0}^2$.}
\label{fig:7}
\end{figure}

\begin{figure}
\includegraphics*[width=\columnwidth]{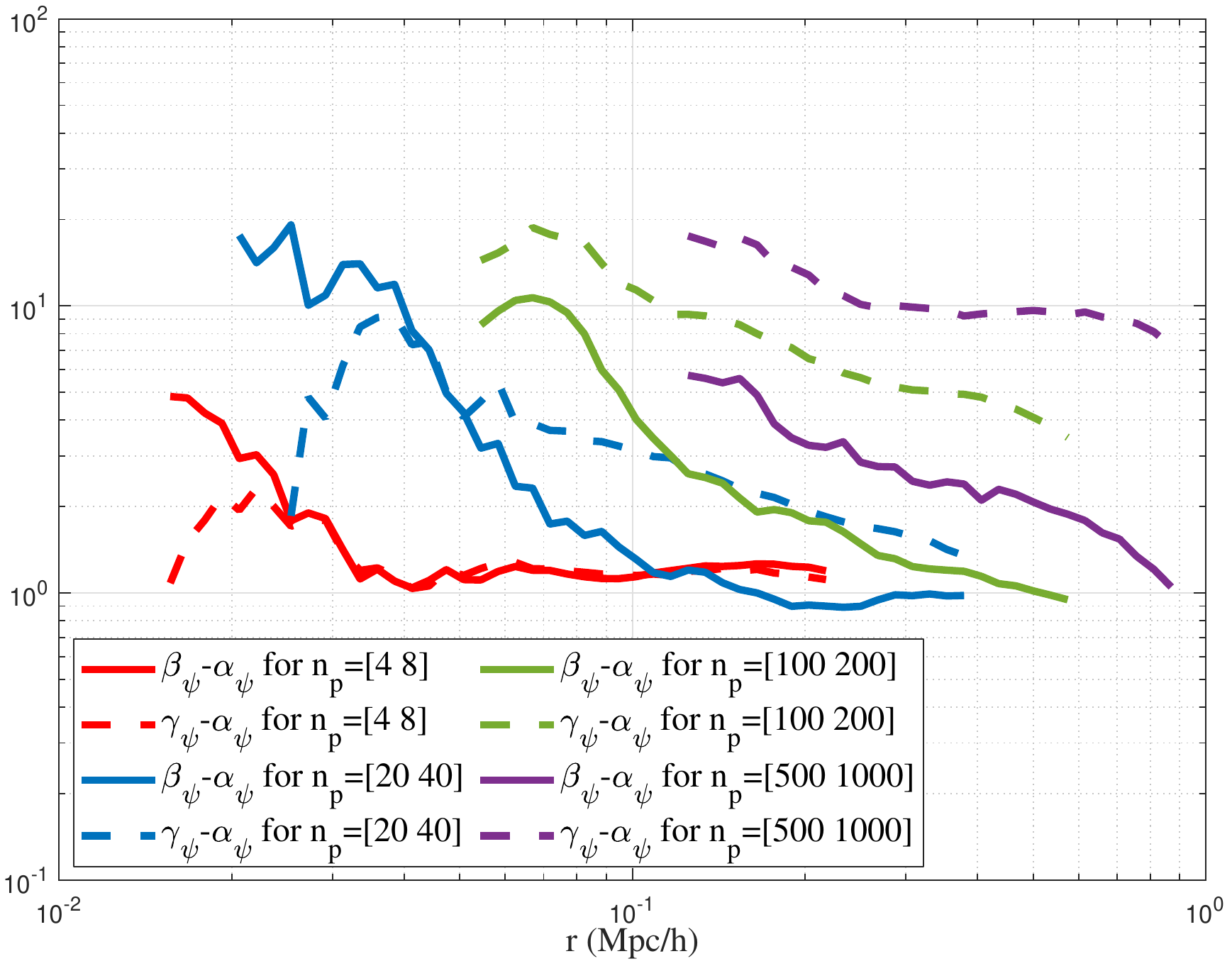}
\caption{The variation of $C_{2} =\beta _{\varphi } -\alpha _{\varphi } $ (solid line) and $C_{1} =\gamma _{\varphi } -\alpha _{\varphi } $ (dash line), i.e. $\sigma _{\varphi \varphi }^{2} -\sigma _{\theta \theta }^{2} =C_{2} u_{\varphi }^{2} $ and $\sigma _{rr}^{2} -\sigma _{\theta \theta }^{2} =C_{1} u_{\varphi }^{2} $, with radius \textit{r} for halo groups of different sizes at \textit{z}=0. Small halos are entirely isotropic with $C_{1} =C_{2} =1$ , i.e. $\sigma _{\varphi \varphi }^{2} -\sigma _{\theta \theta }^{2} =u_{\varphi }^{2} $ and $\sigma _{rr}^{2} =\sigma _{\varphi \varphi }^{2} $ such that the anisotropic parameter $\beta _{h1} =0$. For large halos, $C_{1} $ and $C_{2} $ are more likely to be dependent on \textit{r} with $C_{1} \gg C_{2} $. At halo surface, $C_{1} =\gamma _{\varphi } -\alpha _{\varphi } \approx 10$ and $C_{2} =\beta _{\varphi } -\alpha _{\varphi } \approx 1$.}
\label{fig:8}
\end{figure}

Next let us turn to solutions for velocity dispersions of large halos. Figure \ref{fig:7} plots velocity dispersions and azimuthal flow $u_{\varphi }^{2}$ varying with radius \textit{r} for halos of size $n_{p} $ between [500 1000] at \textit{z}=0. The spin-induced contributions from $u_{\varphi }^{2}$ are dominant for dispersions, where we should have $\sigma _{r}^{2} \approx \sigma _{rr}^{2} \approx \gamma _{\varphi } u_{\varphi }^{2} \gg \sigma _{r0}^{2}$
\noindent , i.e. the term 2 in Eqs. \eqref{ZEqnNum528783}-\eqref{ZEqnNum892394} is dominant over the term 1 with $\alpha _{\varphi } \gg 1$, $\beta _{\varphi } \gg 1$, and $\gamma _{\varphi } \gg 1$ (see Fig. \ref{fig:8} for more details). 

The coupling function $F_{a} \left(r,t\right)<0$ in Eqs. \eqref{ZEqnNum662332} and \eqref{ZEqnNum451086} such that (from Eqs. \eqref{ZEqnNum479172} and \eqref{ZEqnNum401346}),
\begin{equation} 
\label{eq:84}
\begin{split}
\frac{rF_{a} \left(r,t\right)}{\sigma _{r}^{2} } &\approx \frac{\partial \ln \left[\rho _{h} \sigma _{r}^{2} \right]}{\partial \ln r} =\frac{v_{cir}^{2} }{\sigma _{r}^{2} } \left(\frac{x^{2} -xu_{h} }{4\pi ^{2} c^{2} } \frac{\partial u_{h} }{\partial x} -\frac{v_{c}^{2} }{v_{cir}^{2} } \right)\\
&\approx \frac{x^{2} -xu_{h} }{4\pi ^{2} c^{2} } \frac{\partial u_{h} }{\partial x} \frac{v_{cir}^{2} }{\gamma _{\varphi } u_{\varphi }^{2} } -\frac{v_{c}^{2} }{\gamma _{\varphi } u_{\varphi }^{2} }.
\end{split}
\end{equation} 
This can be further reduced to (with $u_{h}$ from Eq. \eqref{ZEqnNum176930}) 
\begin{equation} 
\label{ZEqnNum684840} 
\begin{split}
\frac{rF_{a} \left(r,t\right)}{\gamma _{\varphi } u_{\varphi }^{2} } &\approx \left[\underbrace{\frac{x^{2} }{4\pi ^{2} c^{2} } \left(\frac{\partial \ln F}{\partial \ln x} \right)^{-2} \frac{\partial \ln F^{'} }{\partial \ln x} }_{1}-\underbrace{\frac{cF\left(x\right)}{xF\left(c\right)} }_{2}\right]\frac{v_{cir}^{2} }{\gamma _{\varphi } u_{\varphi }^{2} }\\
&\approx -\frac{cF\left(x\right)}{xF\left(c\right)} \frac{v_{cir}^{2} }{\gamma _{\varphi } u_{\varphi }^{2} } =-\frac{v_{c}^{2} }{\gamma _{\varphi } u_{\varphi }^{2} }
\end{split}
\end{equation} 
that is in terms of the unknown function $F\left(x\right)$. Term 1 in Eq. \eqref{ZEqnNum684840} is the contribution from mean radial flow and is expected to be much smaller when compared to term 2 from the gravitational potential. 

The approximation of coupling function $F_a$ (from Eq. \eqref{ZEqnNum684840})
\begin{equation} 
\label{ZEqnNum832669} 
F_{a} \left(r,t\right)\approx -\frac{F\left(x\right)}{F\left(c\right)} \frac{r_{h} }{r^{2} } v_{cir}^{2} =-\frac{\partial \phi _{r} }{\partial r}  
\end{equation} 
can be obtained and used in Eq. \eqref{ZEqnNum662332} for large halos. 

With $\alpha _{\varphi } $ and $\beta _{\varphi } $ are comparable  and both are much greater than 1 , we will have $\alpha _{\theta } $ (exponent of $\sin \theta $ in Eq. \eqref{ZEqnNum694993} for $u_{\varphi } $), 
\begin{equation}
\alpha _{\theta } =\frac{1+\beta _{\varphi } -\alpha _{\varphi } }{2\alpha _{\varphi } } \ll 1 \quad \textrm{with} \quad \alpha _{\varphi } \gg 1 \quad \textrm{and} \quad \beta _{\varphi } \gg 1, 
\label{ZEqnNum801277}
\end{equation}
\noindent such that the dependence on the coordinate variable $\theta $ can be eliminated, i.e. all variables are only weakly dependent on $\theta$. This is also clearly shown in the plot of $\alpha _{\theta}$ in Fig. \ref{fig:5}, where azimuthal flow $u_{\varphi}$ is weakly dependent on $\theta$ for large halos. 

With approximation of coupling function $F_{a} \left(r,t\right)$ in Eq. \eqref{ZEqnNum832669}, Eq. \eqref{ZEqnNum662332} for axial velocity dispersion $\sigma _{r0}^{2}$ reduces to
\begin{equation} 
\label{ZEqnNum356649} 
\frac{\partial u_{r} }{\partial t} +u_{r} \frac{\partial u_{r} }{\partial r} +\frac{1}{\rho _{h} } \frac{\partial \left(\rho _{h} \sigma _{r0}^{2} \right)}{\partial r} =0,         
\end{equation} 
where $\sigma _{r0}^{2}$ is entirely determined by the mean radial flow $u_{r}$. Using solution of $u_{r}$ in Eq. \eqref{ZEqnNum176930}, the solution of $\sigma _{r0}^{2}$ reads
\begin{equation} 
\label{ZEqnNum103136} 
\begin{split}
\sigma _{r0}^{2} \left(x\right)=\frac{v_{cir}^{2} x^{2} }{4\pi ^{2} c^{2} F^{'} \left(x\right)}&\left\{\left. \frac{F^{2} \left(x\right)}{x^{2} F^{'} \left(x\right)} \right|_{x}^{\infty }\right.\\
&\left.-\int _{x}^{\infty }\left[\frac{2F\left(x\right)}{x^{2} } -\frac{2F^{2} \left(x\right)}{F^{'} \left(x\right)x^{3} } \right] dx\right\},
\end{split}
\end{equation} 
which is the first term in the solution for radial dispersion $\sigma _{r}^{2}$ of isotropic and non-rotating halos \citep[see][Eq. (68)]{Xu:2021-Inverse-mass-cascade-halo-density}. 

Next, Eqs. \eqref{ZEqnNum451086} and \eqref{ZEqnNum684840} are now used to solve for the in-plane and radial velocity dispersions. The equation for $\gamma _{\varphi}$ now reads,  
\begin{equation} 
\label{ZEqnNum185329} 
\frac{\partial \ln \gamma _{\varphi } }{\partial \ln x} +2\frac{\partial \ln u_{\varphi } }{\partial \ln x} +\frac{\partial \ln \rho _{h} }{\partial \ln x} +2-\frac{\left(1+\alpha _{\varphi } +\beta _{\varphi } \right)}{\gamma _{\varphi } } =-\frac{cF\left(x\right)}{xF\left(c\right)} \frac{v_{cir}^{2} }{\gamma _{\varphi } u_{\varphi }^{2} }  .     
\end{equation} 
Substitution of the solution of $u_{\varphi } $ (Eq. \eqref{ZEqnNum694993}) into Eq. \eqref{ZEqnNum185329} leads to  
\begin{equation}
\label{ZEqnNum150487} 
\frac{\partial \ln \gamma _{\varphi } }{\partial \ln x} +2\frac{\partial \ln F}{\partial \ln x} -2+\frac{\partial \ln F^{'} }{\partial \ln x} +\frac{\lambda _{f} x}{F\left(x\right)\gamma _{\varphi } } =\frac{\left(1+\alpha _{\varphi } +\beta _{\varphi } \right)}{\gamma _{\varphi } } =2\alpha _{a} .    
\end{equation} 
With $v_{cir}^{2} $ from Eq. \eqref{ZEqnNum477002} and $\omega _{h}^{} $ from Eq. \eqref{ZEqnNum501598}, the dimensionless constant $\lambda _{f} $ is defined as
\begin{equation} 
\label{ZEqnNum684593} 
\lambda _{f} =\frac{cv_{cir}^{2} }{\alpha _{f}^{2} \omega _{h}^{2} r_{s}^{2} F\left(c\right)} =\frac{9\pi ^{2} F\left(c\right)}{\left({3/\left(2\alpha _{h} \right)} -{1/2} \right)^{2} c}.         
\end{equation} 

To obtain a solution of $\gamma _{\varphi } $ and hence the solution of velocity dispersions, we need to introduce some additional constraints between three expansion coefficients, 
\begin{equation}
\beta _{\varphi } =\alpha _{\varphi } +C_{2} \left(x\right) \quad \textrm{and} \quad \gamma _{\varphi } =\alpha _{\varphi } +C_{1} \left(x\right),     
\label{eq:93}
\end{equation}
\noindent where $C_{1} $ and $C_{2} $ are two functions of \textit{x} that can be determined from simulation. This requires
\begin{equation}
C_{2} u_{\varphi }^{2} =\sigma _{\varphi \varphi }^{2} -\sigma _{\theta \theta }^{2} \quad \textrm{and} \quad C_{1} u_{\varphi }^{2} =\sigma _{rr}^{2} -\sigma _{\theta \theta }^{2},     
\label{eq:94}
\end{equation}
\noindent i.e. the difference between velocity dispersions is always proportional to $u_{\varphi }^{2} $. Figure \ref{fig:8} presents the variation of $C_{1} $ and $C_{2} $ with radius \textit{r} for halo groups of different sizes. Clearly, $C_{1} =C_{2} =1$ for small halos, as predicted in the previous section since small halos are relatively isotropic with anisotropic parameter $\beta _{h1} =0$. However, large halos are anisotropic with $\beta _{h1} >0$, where $C_{1} $ and $C_{2} $ are \textit{r}-dependent with $C_{1} \gg C_{2} $. At halo surface, $C_{1} =\gamma _{\varphi } -\alpha _{\varphi } \approx 10$ and $C_{2} =\beta _{\varphi } -\alpha _{\varphi } \approx 1$.  

We first look at a special case: large halos with extremely fast mass accretion and infinitesimal halo lifespan, where the radial flow $u_r$ vanishes \citep[see][Fig. 3]{Xu:2021-Inverse-mass-cascade-halo-density} and axial velocity dispersion $\sigma_{r0}^2=0$ from Eq. \eqref{ZEqnNum356649}). These halos should have an isothermal density profile with $F(x)=x/c$ \citep[see][Section 3.7]{Xu:2021-A-non-radial-two-body-collapse}. Therefore, from Eq. \eqref{ZEqnNum150487}, the expansion coefficients for large halos with isothermal density profile should be,
\begin{equation}
\label{eq:95}
    \alpha_{\phi} = \frac{9\pi^2-C_2-1}{2}, \beta_{\phi} = \frac{9\pi^2+C_2-1}{2},
    \gamma_{\phi} = \frac{9\pi^2-C_2-1+2C_1}{2}
\end{equation}

For a general density profile, with these relations, the final equation for the expansion coefficient $\gamma _{\varphi } $ reads (from Eq. \eqref{ZEqnNum150487}) 
\begin{equation} 
\label{ZEqnNum978381} 
\frac{\partial \gamma _{\varphi } }{\partial x} +\frac{\gamma _{\varphi } }{x} \left[\frac{\partial \ln \left({F^{2} F^{'} /x^{4} } \right)}{\partial \ln x} \right]+\frac{\lambda _{f} }{F\left(x\right)} =\underbrace{\frac{C_{2} }{x} }_{1}+\frac{1-2C_{1} }{x} .      
\end{equation} 
Exact solution of $\gamma _{\varphi } $ will depend on the model of $C_{1} $ and $C_{2} $. One reasonable simplification  is to neglect term 1 in Eq. \eqref{ZEqnNum978381} because of $C_{1} \gg C_{2} \approx 1$ and assume a constant $C_{1} \left(x\right)=C_{1} =10$. The corresponding solution for $\gamma _{\varphi } $ can be obtained in terms of $F\left(x\right)$,  
\begin{equation} 
\label{ZEqnNum206423}
\begin{split}
\gamma _{\varphi } \left(x\right)=\frac{x^{4} }{F^{2} \left(x\right)F^{'} \left(x\right)} &\left(\left(2C_{1} -1-C_{2} \right)\underbrace{\int _{x}^{\infty }\frac{F^{2} \left(y\right)F^{'} \left(y\right)}{y^{5} } dy }_{1}\right.\\
&\left.+\lambda _{f} \underbrace{\int _{x}^{\infty }\frac{F\left(y\right)F^{'} \left(y\right)}{y^{4} } dy }_{2}\right).
\end{split}
\end{equation} 
With $F\left(x\right)\sim x^{2} $ for small \textit{x} (NFW profile), we should expect $\gamma _{\varphi } \sim x^{-1} $ from Eq. \eqref{ZEqnNum978381}. For any given density profile (or function $F\left(x\right)$), the velocity dispersions (Eqs. \eqref{ZEqnNum528783} to \eqref{ZEqnNum892394}) can be eventually obtained with solution of $\sigma _{r0}^{2} $ from Eq. \eqref{ZEqnNum103136} and solutions of $u_{\varphi }^{2} $ and $\gamma _{\varphi } $ from Eqs. \eqref{ZEqnNum694993} and \eqref{ZEqnNum206423}, respectively. For NFW profile, the two terms in Eq. \eqref{ZEqnNum206423} can be obtained analytically,
\begin{equation} 
\label{eq:97} 
\begin{split}
&\textrm{term1}=\frac{2+83x+147x^{2} +68x^{3} }{6x\left(1+x\right)^{3} } +\frac{35}{12} \ln x\\
&+\frac{\ln \left(-x\right)}{12} \left[-35+8\ln \left(1+x\right)\left(5+6\ln \left(1+x\right)\right)\right]\\ 
&+\frac{\ln \left(1+x\right)}{3x^{3} \left(1+x\right)^{2} } \left[-2x+6x^{2} +45x^{3} +34x^{4} \right]\\
&-\frac{\ln ^{2} \left(1+x\right)}{3x^{3} \left(1+x\right)} \left[-1+2x+x^{2} \left(x-2\right)\left(3+5x\right)\right]\\ 
&+\frac{2}{3} \left[5+12\ln \left(1+x\right)\right]poly\log \left(2,1+x\right)\\
&-8poly\log \left(3,1+x\right)+\frac{5\pi \left(21i-8\pi \right)}{36} -\frac{4}{3} \ln ^{3} \left(1+x\right)\\
& \textrm{and}\\
&\textrm{term2}=\frac{\ln \left(1+x\right)}{2x^{2} \left(1+x\right)^{2} } +\frac{1}{2x\left(1+x\right)^{2} } \left\{-1-9x-7x^{2}\right.\\ 
&+\left[-2-8x-4x^{2} +x^{3} \right]\ln\left(1+x\right)\\
&+\left. \left[\pi ^{2} +6\textrm{polylog}\left(2,-x\right)-\ln x+3\left(\ln \left(1+x\right)\right)^{2} \right]x\left(1+x\right)^{2} \right\},
\end{split}
\end{equation} 

For large halos with $\sigma _{r0}^{2} \ll \gamma _{\varphi } u_{\varphi }^{2} $, the anisotropy parameters $\beta _{h1} $ and $\beta _{h} $ (Eq. \eqref{ZEqnNum486654}) are equal and reduced to the same expression in terms of $\gamma _{\varphi } $, 
\begin{equation} 
\label{ZEqnNum895234} 
\beta _{h1} \approx 1-\frac{1+\alpha _{\varphi } +\beta _{\varphi } }{2\gamma _{\varphi } } \approx \frac{2C_{1} -1-C_{2} }{2\gamma _{\varphi } } \approx \beta _{h} ,       
\end{equation} 
which is inversely proportional to coefficient $\gamma _{\varphi }$ in Eq. \eqref{ZEqnNum206423}. 
\begin{figure}
\includegraphics*[width=\columnwidth]{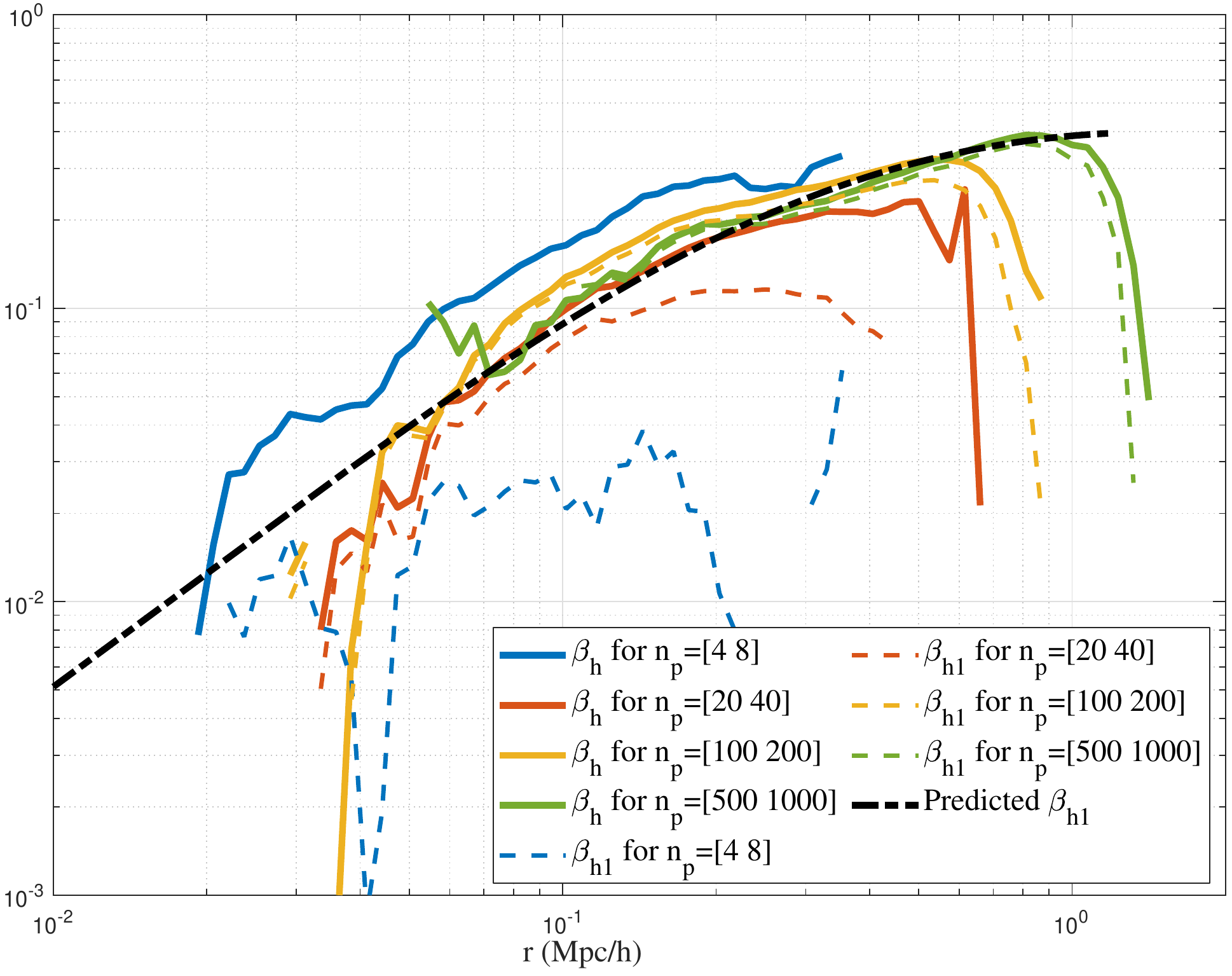}
\caption{The variation of new anisotropy parameters $\beta _{h1} $ (dash lines) and conventional anisotropy parameters $\beta _{h} $ (solid lines) with radius \textit{r} for halo groups of different sizes $n_{p} $. Small halos are isotropic with $\beta _{h1} \approx 0$, while $\beta _{h} \ne 0$ since $\beta _{h} $ does not include the effect of azimuthal flow $u_{\varphi }^{2} $ (Eqs. \eqref{ZEqnNum865565} and \eqref{ZEqnNum597498}). However, $\beta _{h1} \approx \beta _{h} $ for large halos and increases with radius \textit{r}. The predicted $\beta _{h1} $ (dash-dot line) from Eq. \eqref{ZEqnNum895234} is also presented showing good agreement with simulation results.}
\label{fig:9}
\end{figure}

Figure \ref{fig:9} plots the variation of anisotropy parameters $\beta _{h1} $ (dash lines) and $\beta _{h} $ (solid lines) with radius \textit{r} for halo groups of different sizes $n_{p} $. For small halos that are isotropic, $\beta _{h1} \approx 0$ while $\beta _{h} \ne 0$ since $\beta _{h} $ does not include the effect of azimuthal flow $u_{\varphi }^{2} $. However, $\beta _{h1} \approx \beta _{h} $ for large halos and increases with radius \textit{r}. The predicted $\beta _{h1} $ (dash-dot line) from Eqs. \eqref{ZEqnNum206423} to \eqref{ZEqnNum895234} is also presented showing good agreement with simulation results. This prediction is made with function $F\left(x\right)$ for a NFW profile given in Eq. \eqref{ZEqnNum437028} and $c=3.5$ such that $\lambda _{f} =10.89$ from Eq. \eqref{ZEqnNum684593}. Other relevant parameters used to make this prediction are $C_{1} =10$, $C_{2} =1$, and $r_{s} =0.34{Mpc/h} $. 

\section{Momentum and energy of axisymmetric rotating growing halos}
\label{sec:4}
\subsection{Momentum and energy solutions}
\label{sec:4.1}
With full solutions developed for large halos in the previous section, this section summarize the momentum and energy solutions for large halos. With radial flow $u_r$ from Eq. \eqref{ZEqnNum176930}, the physical and peculiar radial linear momentum (zeroth order moment) read
\begin{equation}
\label{ZEqnNum170799} 
L_{h} =\frac{1}{m_{h} } \int _{0}^{r_{h} }4\pi r^{2} \rho _{h} u_{r} dr =\frac{3}{2} \left(1-\frac{2}{cF\left(c\right)} \int _{0}^{c}F\left(x\right)dx \right)Hr_{h} ,     
\end{equation} 
\begin{equation}
\label{ZEqnNum186975} 
L_{hp} =\frac{1}{m_{h} } \int _{0}^{r_{h} }4\pi r^{2} \rho _{h} u_{rp} dr =\frac{1}{2} \left(1-\frac{4}{cF\left(c\right)} \int _{0}^{c}F\left(x\right)dx \right)Hr_{h} .     
\end{equation} 
The virial quantity (the first order moment of mean radial flow) is
\begin{equation}
\label{ZEqnNum458282} 
G_{h} =\frac{1}{m_{h} } \int _{0}^{r_{h} }4\pi r^{3} \rho _{h} u_{r} dr =\frac{3}{2} \left[1-\frac{3}{c^{2} F\left(c\right)} \int _{0}^{c}xF\left(x\right)dx \right]Hr_{h}^{2} .     
\end{equation} 
The peculiar virial quantity (excluding Hubble flow) is (Eq. \eqref{ZEqnNum579659}),
\begin{equation} 
\label{ZEqnNum978532} 
\begin{split}
G_{hp} &=\frac{1}{m_{h} } \int _{0}^{r_{h} }4\pi r^{3} \rho _{h} u_{rp} dr \\
&=\frac{1}{2} \left[1-\frac{5}{c^{2} F\left(c\right)} \int _{0}^{c}xF\left(x\right)dx \right]Hr_{h}^{2}. 
\end{split}
\end{equation} 
For any density profiles, the specific halo angular momentum reads
\begin{equation} 
\label{ZEqnNum696358} 
H_{h} =\left(\frac{1}{\alpha _{h} } -\frac{1}{3} \right)\frac{c^{2} }{F\left(c\right)\alpha _{f} } \left(G_{h} -G_{hp} \right)=\left(\frac{1}{\alpha _{h} } -\frac{1}{3} \right)\frac{c^{2} }{F\left(c\right)\alpha _{f} } Hr_{g}^{2}  
\end{equation} 
from Eqs. \eqref{ZEqnNum960633}, \eqref{ZEqnNum458282}, \eqref{ZEqnNum978532}, and \eqref{ZEqnNum501598}. 

With Eq. \eqref{ZEqnNum353891} for relations between $r_{g}^{2} $ and $r_{h}^{2} $, the halo angular momentum from Eq. \eqref{ZEqnNum696358} can be finally written in terms of $r_{h}$, 
\begin{equation} 
\label{ZEqnNum546367} 
H_{h} =\gamma _{H} Hr_{h}^{2} =\frac{1}{8} \left(\frac{3}{\alpha _{h} } -1\right)\frac{\sqrt{\pi } \Gamma \left({3/2} +{\alpha _{\theta } /2} \right)}{\Gamma \left(2+{\alpha _{\theta } /2} \right)} Hr_{h}^{2} ,      
\end{equation} 
where the coefficient $\gamma _{H} $ for angular momentum is 
\begin{equation} 
\label{ZEqnNum775083} 
\gamma _{H} =\frac{1}{8} \left(\frac{3}{\alpha _{h} } -1\right)\frac{\sqrt{\pi } \Gamma \left({3/2} +{\alpha _{\theta } /2} \right)}{\Gamma \left(2+{\alpha _{\theta } /2} \right)}.        
\end{equation} 
The specific momentum tensor of a spherical halo reads (from Eqs. \eqref{ZEqnNum295029} and \eqref{ZEqnNum978532}),
\begin{equation} 
\label{eq:107} 
\frac{1}{m_{h} } \int _{V}\boldsymbol{\mathrm{x}}\otimes \boldsymbol{\mathrm{u}}_{p}  \rho _{h} dV=\left[\begin{array}{ccc} {{G_{hp} /3} } & {-{H_{h} /2} } & {0} \\ {{H_{h} /2} } & {{G_{hp} /3} } & {0} \\ {0} & {0} & {{G_{hp} /3} }
\end{array}\right].      
\end{equation} 
It can be found the diagonal terms of halo momentum tensor are the virial quantity in Eq.\eqref{ZEqnNum978532}, while the off-diagonal terms are the angular momentum in Eq. \eqref{ZEqnNum546367}. The evolution of momentum tensor on both halo and large scales is extensively studied in a separate paper \citep[see][Section 5]{Xu:2022-The-evolution-of-energy--momen}. 

Finally, the halo specific radial kinetic energy is derived as (with Eq. \eqref{ZEqnNum176930} for $u_{r}$) \citep[also see][Eq. (54)]{Xu:2021-Inverse-mass-cascade-halo-density},
\begin{equation}
\label{ZEqnNum903694} 
\begin{split}
K_{r}&=\frac{1}{2m_{h} } \int _{0}^{r_{h} }u_{r}^{2} \left(r,a\right)4\pi r^{2} \rho _{h} \left(r,a\right)dr\\
&=\frac{9}{8} \left(1-\frac{4}{c^{2} F\left(c\right)} \int _{0}^{c}xF\left(x\right)dx +\frac{1}{c^{2} F\left(c\right)} \int _{0}^{c}\frac{F^{2} \left(x\right)}{F^{'} \left(x\right)} dx \right)H^{2} r_{h}^{2}.
\end{split}
\end{equation} 
The halo (specific) peculiar radial kinetic energy (excluding Hubble flow) can be obtained as (with Eq. \eqref{ZEqnNum477267} for $u_{rp}$),
\begin{equation} 
\label{ZEqnNum100420}
\begin{split}
K_{rp}&=\frac{1}{2m_{h} } \int _{0}^{r_{h} }u_{rp}^{2} 4\pi r^{2} \rho _{h} \left(r,a\right)dr\\ 
&=\left(\frac{1}{8} -\frac{1}{c^{2} F\left(c\right)} \int _{0}^{c}xF\left(x\right)dx +\frac{9}{8c^{2} F\left(c\right)} \int _{0}^{c}\frac{F^{2} \left(x\right)}{F^{'} \left(x\right)} dx \right)H^{2} r_{h}^{2}.
\end{split}
\end{equation} 
The halo (specific) rotational kinetic energy is derived as (with Eq. \eqref{ZEqnNum694993} for $u_{\varphi } $),
\begin{equation} 
\label{ZEqnNum155330} 
\begin{split}
K_{a}&=\frac{1}{m_{h} } \int _{0}^{r_{h} }2\pi r^{3} \rho _{h} \left(r\right)\left(\int _{0}^{\pi }\frac{1}{2} u_{\varphi }^{2} \sin \theta d\theta  \right) dr\\ 
&=\frac{1}{4} \left(\frac{3}{2\alpha _{h} } -\frac{1}{2} \right)^{2} \frac{c^{2} }{F\left(c\right)^{3} } \frac{\sqrt{\pi } \Gamma \left(1+\alpha _{\theta } \right)}{\Gamma \left({3/2} +\alpha _{\theta } \right)} \int _{0}^{c}\frac{F^{2} \left(x\right)F^{'} \left(x\right)}{x^{2} } dx H^{2} r_{h}^{2}.
\end{split}
\end{equation} 

All these momentum and energy quantities are derived in terms of function $F(x)$ (Eq. \eqref{ZEqnNum990243}) that is dependent on halo density profile \citep{Xu:2021-Inverse-mass-cascade-halo-density} and summarized in Table \ref{tab:3} for isothermal and NFW profiles.

\subsection{Calculation of halo spin parameter}
\label{sec:4.2}
The halo spin parameter $\lambda _{p} $ is commonly used to characterize the importance of angular momentum to the random motion. The energy solutions obtained can be used to estimate the value of $\lambda _{p} $ for large halos with fast mass accretion. With angular momentum explicitly derived in Eq. \eqref{ZEqnNum546367}, the two usual definitions of dimensionless spin parameter can be defined as \citep{Peebles:1969-Origin-of-the-Angular-Momentum,Bullock:2001-Profiles-of-dark-haloes--evolu}, 
\begin{equation}
\lambda _{p} =\frac{H_{h} \left|E_{h} \right|^{{1/2} } }{Gm_{h}} \quad \textrm{and} \quad \lambda _{p}^{'} =\frac{H_{h} }{\sqrt{2} v_{cir} r_{h} },       
\label{ZEqnNum475450}
\end{equation}
\noindent where $E_{h} =\Phi _{h} +K_{h} $ is the total specific energy.  The halo specific potential energy
\begin{equation} 
\label{ZEqnNum658729} 
\begin{split}
\Phi _{h}&=-\gamma _{\Phi } \frac{Gm_{h} }{r_{h} }\\
&=-\frac{1}{m_{h} } \int _{0}^{r_{h} }4\pi r^{2} \rho _{h} \left(r,a\right)\frac{Gm_{r} }{r}  dr=-\frac{1}{2} \gamma _{\Phi } \Delta _{c} H^{2} r_{h}^{2} ,   
\end{split}
\end{equation} 
where the coefficient $\gamma _{\Phi } $ for potential energy is
\begin{equation}
\label{ZEqnNum430481} 
\gamma _{\Phi } =\left(\frac{c}{F{}^{2} \left(c\right)} \int _{0}^{c}\frac{F\left(x\right)F^{'} \left(x\right)}{x}  dx\right)\approx 1.        
\end{equation} 
The critical density ratio $\Delta _{c} =18\pi ^{2} $ can be obtained from spherical collapse model or two-body collapse model \citep{Xu:2021-A-non-radial-two-body-collapse}. The halo specific kinetic energy $K_{h} ={3/2} \sigma _{v}^{2} =\left({n_{e} /2} \right)\Phi _{h} $, with $n_{e} \approx -1.3$ for large halos is the effective potential exponent for virial theorem that considers surface energy due to non-zero radial flow and velocity dispersion \citep[see][Eq. (96)]{Xu:2021-Inverse-mass-cascade-halo-density}.

It should be noted that Eq. \eqref{ZEqnNum658729} can be used to derive the relation for virial kinetic energy $\sigma _{v}^{2}$. Halo size $r_{g}=\gamma _{g} r_{h}$ can be written as \citep[see][Eq. (61)]{Xu:2021-Inverse-and-direct-cascade-of-} ,
\begin{equation} 
\label{eq:114} 
r_{g}^{} =\gamma _{g} a\left(\frac{2Gm_{h} }{\Delta _{c} H_{0}^{2} } \right)^{{1/3} }  
\end{equation} 
such that (with Eq. \eqref{ZEqnNum658729} for $\Phi _{h} $)
\begin{equation}
\label{ZEqnNum676335} 
\sigma _{v}^{2} =-\Phi _{h} \frac{\gamma _{v} }{3} =\frac{1}{3} \gamma _{\Phi } \gamma _{v} \left(\frac{\Delta _{c} }{2} \right)^{{1/3} } \left(Gm_{h} H_{0} \right)^{{2/3} } a^{-1} ,      
\end{equation} 
where $\gamma _{v} \approx -n_{e}$ is the virial ratio and $\Delta _{c} $ is the critical density ratio. Here $\gamma _{v} \approx 1.3$ for NFW profile and $\gamma _{v} =1.5$ for isothermal profile, \citep[Eq. (96)]{Xu:2021-Inverse-mass-cascade-halo-density}. Combining Eq. \eqref{ZEqnNum676335} with the model of $\sigma _{v}^{2}$ from N-body simulation \citep[Eq. (19)]{Xu:2021-Inverse-and-direct-cascade-of-} leads to a good equation for velocity dispersion $u^{2}$ of entire N-body system \citep[see][Fig. 1a]{Xu:2022-The-evolution-of-energy--momen},
\begin{equation} 
\label{eq:116} 
u^{2} =\gamma _{\Phi } \gamma _{v} \left(\Delta _{c} \right)^{{1/3} } \left(GH_{0} \cdot 5.8\times 10^{12} \frac{M_{\odot } }{h} \right)^{{2/3} } \frac{t}{t_{0} } .       
\end{equation} 

Rotational kinetic energy $K_{a} $ can be approximated as (Eq. \eqref{ZEqnNum960633})
\begin{equation} 
\label{eq:117} 
K_{a} \approx \frac{1}{2} \left|\boldsymbol{\mathrm{H}}_{h} \right|\omega _{h} =\frac{3}{4} \left({\left|\boldsymbol{\mathrm{H}}_{h} \right|/r_{g} } \right)^{2} .         
\end{equation} 
With Eq. \eqref{ZEqnNum353891} for root mean square radius $r_{g}$ and Eq. \eqref{ZEqnNum475450}, the two halo spin parameters read
\begin{equation} 
\label{eq:118} 
\begin{split}
&\lambda _{p} =\gamma _{\Phi } \gamma _{g} \sqrt{\frac{4}{3} \left(1+\frac{n_{e} }{2} \right)\frac{K_{a} }{\left|\Phi _{h} \right|} } =\frac{2}{3} \gamma _{\Phi } \gamma _{g} \sqrt{\gamma _{v} \left(1-\frac{\gamma _{v} }{2} \right)\frac{K_{a} }{\sigma _{v}^{2} } }\\ 
&\textrm{and}\\   
&\lambda _{p}^{'} =\gamma _{g} \sqrt{\frac{2\gamma _{\Phi } K_{a} }{3\left|\Phi _{h} \right|} } =\frac{1}{3} \gamma _{g} \sqrt{2\gamma _{\Phi } \gamma _{v} \frac{K_{a} }{\sigma _{v}^{2} } },
\end{split}
\end{equation} 
where both definitions reflect the ratio of rotational kinetic energy $K_a$ to virial kinetic energy $\sigma _{v}^{2}$. 

With Eq. \eqref{ZEqnNum546367} for $H_{h} $, circular velocity $v_{cir} =\sqrt{{\Delta _{c} /2} } Hr_{h} =3\pi Hr_{h} $ ($\Delta _{c}$ is the critical density ratio), and Eq. \eqref{ZEqnNum475450}, spin parameters $\lambda _{p}$ and  $\lambda _{p}^{'} $ finally read  (for NFW profile in Table \ref{tab:3})
\begin{equation}
\lambda _{p} =\frac{\gamma_{H} }{3\pi } \sqrt{\gamma_{\Phi } \left(1+\frac{n_{e} }{2} \right)} \approx 0.03, \quad \lambda _{p}^{'} =\frac{\gamma_{H} }{3\pi \sqrt{2} } \approx 0.038.
\label{ZEqnNum951104}
\end{equation}
Results for halo spin parameter agrees well with other simulations \citep{Hetznecker:2006-The-evolution-of-the-dark-halo}. In addition, the halo mass dependence of spin parameter that decreases with halo size is discussed in a separate paper \citep{Xu:2022-The-evolution-of-energy--momen}. All relevant parameters are summarized in Table \ref{tab:3} for two density profiles.

\begin{table}
\caption{Relevant parameters for two different density profiles} 
\label{tab:3}
\begin{tabular}{p{0.2in}p{0.7in}p{0.45in}p{0.55in}p{0.7in}} 
\hline 
Symbol & Physical meaning & Equation & Isothermal \newline profile\newline with $\alpha _{\theta } =1$ & NFW profile\newline with $\alpha _{\theta } =0$ \newline and $c=3.5$ \\ \hline 
$F\left(x\right)$ & Function for density $\rho _{h} $ & Eq. \eqref{ZEqnNum990243} & ${x/c} $ & $\ln \left(1+x\right)-{x/\left(1+x\right)} $ \\ \hline 
$\alpha _{h} $ & Deformation parameter & Eq. \eqref{ZEqnNum716050}  & $1.0$ & $0.833$  \\ \hline 
$\gamma _{h} $ & Deformation rate parameter & Eq. \eqref{ZEqnNum698012}\textit{} & 0 & ${1/2} $ \\ \hline 
$\alpha _{f} $ & Constant for function$F_{\varphi } \left(x\right)$  & Eq. \eqref{ZEqnNum875729} & ${2c^{2} /3} $  & $9.20$  \\ \hline 
$\lambda _{f} $ & Constant for equation for $\gamma _{\varphi } $ & Eq. \eqref{ZEqnNum684593} & ${9\pi ^{2} /c} $ & 10.895 \\ \hline 
$\gamma _{H} $ & Coefficient for $H_{h} $ & Eq. \eqref{ZEqnNum775083} & ${1/3} $ & $0.511$  \\ \hline 
$\gamma _{\Phi } $ & Coefficient for potential $\Phi _{h} $  & Eq. \eqref{ZEqnNum430481} & 1 & 0.936 \\ \hline 
$\gamma _{v} $ & Virial ratio  & Eq. \eqref{ZEqnNum676335} & 1.5 & 1.3 \\ \hline 
$\gamma _{g}^{2}$ & Ratio of two halo sizes & Eq. \eqref{ZEqnNum353891} & ${1/3} $ & 0.3214 \\ \hline 
$L_{h} $ & Specific radial momentum & Eq. \eqref{ZEqnNum170799} & 0 & 0 \\ \hline 
$L_{hp} $ & Peculiar radial momentum & Eq. \eqref{ZEqnNum186975} & $-{Hr_{h} /2} $ & $-0.501Hr_{h} $ \\ 
\hline 
$G_{h} $ & Specific virial quantity & Eq. \eqref{ZEqnNum458282} & 0 & $-0.027Hr_{h}^{2} $ \\ \hline 
$G_{hp} $ & Peculiar virial quantity & Eq. \eqref{ZEqnNum978532} & $-{Hr_{h}^{2} /3} $ & $-0.348Hr_{h}^{2} $ \\ \hline 
$H_{h} $ & Specific angular momentum & Eq. \eqref{ZEqnNum546367} & ${Hr_{h}^{2} /3} $  & $0.511Hr_{h}^{2} $ \\ \hline 
$\omega _{h} $ & Angular velocity & Eq. \eqref{ZEqnNum501598} & $1.5H$  & $2.38H$  \\ \hline 
$K_{r} $ & Radial kinetic energy & Eq. \eqref{ZEqnNum903694} & 0 & $0.0062H^{2} r_{h}^{2} $ \\ \hline 
$K_{rp} $ & Peculiar radial kinetic energy & Eq. \eqref{ZEqnNum100420} & ${H^{2} r_{h}^{2} /6} $ & $0.1937H^{2} r_{h}^{2} $ \\ \hline 
$K_{a} $ & Rotational kinetic energy & Eq. \eqref{ZEqnNum155330} & ${H^{2} r_{h}^{2} /3} $ & $0.7658H^{2} r_{h}^{2} $ \\ \hline 
$\Phi _{h} $  & Halo potential energy & Eq. \eqref{ZEqnNum658729} & $-9\pi ^{2} H^{2} r_{h}^{2} $ & $-8.424\pi ^{2} H^{2} r_{h}^{2} $  \\ \hline 
$\lambda _{p} $ & First halo spin parameter & Eq. \eqref{ZEqnNum951104} & 0.018 & 0.031 \\ \hline 
$\lambda _{p}^{'} $ & Second halo spin parameter & Eq. \eqref{ZEqnNum951104} & 0.025 & 0.038 \\ \hline 
\end{tabular}
\end{table}

\section{Energy transfer between mean flow and random motion}
\label{sec:5}
The energy exchange between mean flow and random motion is the key to understand how the turbulence initiates, propagates and evolves in dark matter flow to maximize system entropy.

\subsection{General formulation for energy transfer}
\label{sec:5.1}
First, we present a generalized formulation for the evolution of an arbitrary scalar quantity $S\left(r,\theta ,t\right)$. Using continuity Eq. \eqref{ZEqnNum321851}, it is easy to write down the evolution of an arbitrary quantities $\rho _{h} S^{n} $ ($n=1$ for momentum and $n=2$ for kinetic energy if \textit{S} is velocity), 
\begin{equation}
\label{ZEqnNum545276} 
\frac{\partial \left(\rho _{h} S^{n} \right)}{\partial t} +\frac{1}{r^{2} } \frac{\partial \left[\left(\rho _{h} S^{n} \right)u_{r} r^{2} \right]}{\partial r} \underbrace{-n\rho _{h} S^{n-1} \left(\frac{\partial S}{\partial t} +u_{r} \frac{\partial S}{\partial r} \right)}_{P_{S} }=0,     
\end{equation} 
where the term $P_{S} $ stands for the production or consumption of scalar \textit{S.} Integrating Eq. \eqref{ZEqnNum545276} with $\int _{0}^{r_{h} }2\pi r^{2} r^{k} \int _{0}^{\pi }\left(\bullet \right)\sin ^{m} \theta d\theta dr  $ leads to the time evolution of the \textit{k}th moment of $S^{n}$ in entire halo,
\begin{equation} 
\label{ZEqnNum711069} 
\begin{split}
&\frac{\partial }{\partial t} \left[\int _{0}^{r_{h} }2\pi \rho _{h} r^{2+k} \left(\int _{0}^{\pi }S^{n} \left(r,\theta \right) \sin ^{m} \theta d\theta \right)dr \right]\\
&=\underbrace{2\pi \rho _{h} \left(r_{h} \right)r_{h}^{2+k} \left(\int _{0}^{\pi }S^{n} \left(r_{h} ,\theta \right) \sin ^{m} \theta d\theta \right)\left(\frac{\partial r_{h} }{\partial t} -u_{r} \left(r_{h} \right)\right)}_{S_{2} }\\ 
&+\underbrace{\int _{0}^{r_{h} }2\pi \rho _{h} r^{k+1} \left[ku_{r} \left(\int _{0}^{\pi }S^{n} \left(r,\theta \right) \sin ^{m} \theta d\theta \right)\right.}_{S_{1}}\\
&+\underbrace{\left.nr\left(\int _{0}^{\pi }S^{n-1}  \left(\frac{\partial S}{\partial t} +u_{r} \frac{\partial S}{\partial r} \right)\sin ^{m} \theta d\theta \right)\right] dr}_{S_{1} }. 
\end{split}
\end{equation} 
In general, the rate of change of scalar $S$ includes two parts: the surface contribution from mass accretion ($S_{2}$) and the bulk contribution from exchange between mean flow and random motion ($S_{1}$). By replacing \textit{S} with the mean flow $u_{r} $ and $u_{\varphi } $, we can choose appropriate values for \textit{k}, \textit{m} and \textit{n} to recover the equations for radial and rotational momentum and energy evolution (Eqs. \eqref{ZEqnNum898930} to \eqref{ZEqnNum463094}). 

The evolution of radial \& peculiar radial kinetic energy, and rotational kinetic energy can be easily obtained by applying Eq. \eqref{ZEqnNum545276} with \textit{S} replaced by $u_{r} $, $u_{rp} $, and $u_{\varphi } $, respectively, 
\begin{equation} 
\label{ZEqnNum644755} 
\frac{\partial \left(\rho _{h} u_{r}^{2} \right)}{\partial t} +\frac{1}{r^{2} } \frac{\partial \left[\left(\rho _{h} u_{r}^{2} \right)u_{r} r^{2} \right]}{\partial r} \underbrace{-2\rho _{h} u_{r}^{} \left[\frac{\partial u_{r} }{\partial t} +u_{r} \frac{\partial u_{r} }{\partial r} \right]}_{P_{ur} }=0,     
\end{equation} 
\begin{equation} 
\label{eq:123} 
\frac{\partial \left(\rho _{h} u_{rp}^{2} \right)}{\partial t} +\frac{1}{r^{2} } \frac{\partial \left[\left(\rho _{h} u_{rp}^{2} \right)u_{r} r^{2} \right]}{\partial r} \underbrace{-2\rho _{h} u_{rp}^{} \left[\frac{\partial u_{rp} }{\partial t} +u_{r} \frac{\partial u_{rp} }{\partial r} \right]}_{P_{urp} }=0,     
\end{equation} 
\begin{equation} 
\label{ZEqnNum808825} 
\frac{\partial \left(\rho _{h} u_{\varphi }^{2} \right)}{\partial t} +\frac{1}{r^{2} } \frac{\partial \left[\left(\rho _{h} u_{\varphi }^{2} \right)u_{r} r^{2} \right]}{\partial r} \underbrace{-2\rho _{h} u_{\varphi }^{} \left[\frac{\partial u_{\varphi } }{\partial t} +u_{r} \frac{\partial u_{\varphi } }{\partial r} \right]}_{P_{u\varphi } }=0. 
\end{equation} 
These equations can be used to illustrate the energy transfer between mean flow and the random motion. 

\subsection{Energy transfer between mean flow and random motion}
\label{sec:5.2}
We first substitute Eq. \eqref{ZEqnNum356649} for large halos into Eq. \eqref{ZEqnNum644755} to show that 
\begin{equation} 
\label{eq:125} 
\frac{\partial \left(\rho _{h} u_{r}^{2} \right)}{\partial t} +\frac{1}{r^{2} } \frac{\partial \left[\left(\rho _{h} u_{r}^{2} \right)u_{r} r^{2} \right]}{\partial r} +\underbrace{2\frac{\partial \left(\rho _{h} \sigma _{r0}^{2} \right)}{\partial \ln r} \frac{u_{r} }{r} }_{P_{ur} }=0,      
\end{equation} 
i.e. the mean radial flow $u_{r}^{2} $ exchanges kinetic energy with the radial velocity dispersion $\sigma _{r0}^{2} $ (the axial velocity dispersion). 

The momentum Eq. \eqref{ZEqnNum720202} and Eq. \eqref{ZEqnNum808825} can be used to derived the energy exchange between azimuthal flow and velocity dispersions, 
\begin{equation} 
\label{ZEqnNum156930} 
\frac{\partial \left(\rho _{h} u_{\varphi }^{2} \right)}{\partial t} +\frac{1}{r^{2} } \frac{\partial \left[\left(\rho _{h} u_{\varphi }^{2} \right)u_{r} r^{2} \right]}{\partial r} +\underbrace{2\rho _{h} u_{\varphi }^{2} \frac{u_{r} }{r} }_{P_{u\varphi } }=0,      
\end{equation} 
which is essentially the same as Eq. \eqref{ZEqnNum749450} that has been directly derived from the continuity and momentum equations. Here since $u_{\varphi }^{2} $ in production term $P_{u\varphi } $ is actually related to in-plane velocity dispersions as shown in Eq. \eqref{ZEqnNum618520}, Eq. \eqref{ZEqnNum156930} describes the energy transfer between azimuthal flow and random motion (the in-plane velocity dispersions) of SG-CFD via a fictitious stress $\rho _{h} u_{\varphi }^{2}$ (similar to Reynolds stress) acting on the mean flow gradient ${u_{r}/r}$.

The production terms in Eqs. \eqref{ZEqnNum644755}-\eqref{ZEqnNum808825} can be explicitly written in terms of function $F\left(x\right)$ (using Eqs. \eqref{ZEqnNum990243}, \eqref{ZEqnNum176930}, \eqref{ZEqnNum477267} and \eqref{ZEqnNum694993}),
\begin{equation} 
\label{ZEqnNum278480}
\begin{split}
P_{ur}&=\frac{2\rho _{h} v_{cir}^{2} }{t} \frac{\left(x-u_{h} \right)}{4\pi ^{2} c^{2} } u_{h} \frac{\partial u_{h} }{\partial x}\\ &=\frac{2\bar{\rho }_{h} v_{cir}^{2} }{t} \frac{\left(x-u_{h} \right)}{12\pi ^{2} x^{2} } \frac{cF^{'} \left(x\right)}{F\left(c\right)} u_{h} \frac{\partial u_{h} }{\partial x},
\end{split}
\end{equation} 
\begin{equation} 
\label{eq:128} 
\begin{split}
P_{urp}&=\frac{2\rho _{h} v_{cir}^{2} }{t} \frac{\left(x-u_{h} \right)}{4\pi ^{2} c^{2} } u_{p} \frac{\partial u_{p} }{\partial x}\\ &=\frac{2\bar{\rho }_{h} v_{cir}^{2} }{t} \frac{\left(x-u_{h} \right)}{12\pi ^{2} x^{2} } \frac{cF^{'} \left(x\right)}{F\left(c\right)} u_{p} \frac{\partial u_{p} }{\partial x},
\end{split}
\end{equation} 
\begin{equation} 
\label{ZEqnNum824164} 
P_{u\varphi } =\frac{2\bar{\rho }_{h} v_{cir}^{2} }{t} \frac{c^{4} F^{'} \left(x\right)}{3\lambda _{f} F\left(c\right)^{2} } \left(\sin \theta \right)^{2\alpha _{\theta } } \frac{u_{h} \left(x\right)}{x^{5} } F^{2} \left(x\right),      
\end{equation} 
where $\bar{\rho }_{h} $ is the mean halo density. The $P_{2} $ contribution of $P_{ur} $ in Eq. \eqref{ZEqnNum185081} reads,
\begin{equation} 
\label{eq:130} 
P_{2} =\frac{2\bar{\rho }_{h} v_{cir}^{2} }{t} \frac{c^{4} F\left(x\right)F^{'} \left(x\right)}{3x^{4} F\left(c\right)^{2} } u_{h} ,        
\end{equation} 
while the contributions $P_{1} $ and $P_{3} $ (from radial velocity dispersions in Eq. \eqref{ZEqnNum185081}) of $P_{ur} $ can be obtained as $P_{1} +P_{3} =P_{ur} -P_{2} $. 
\begin{figure}
\includegraphics*[width=\columnwidth]{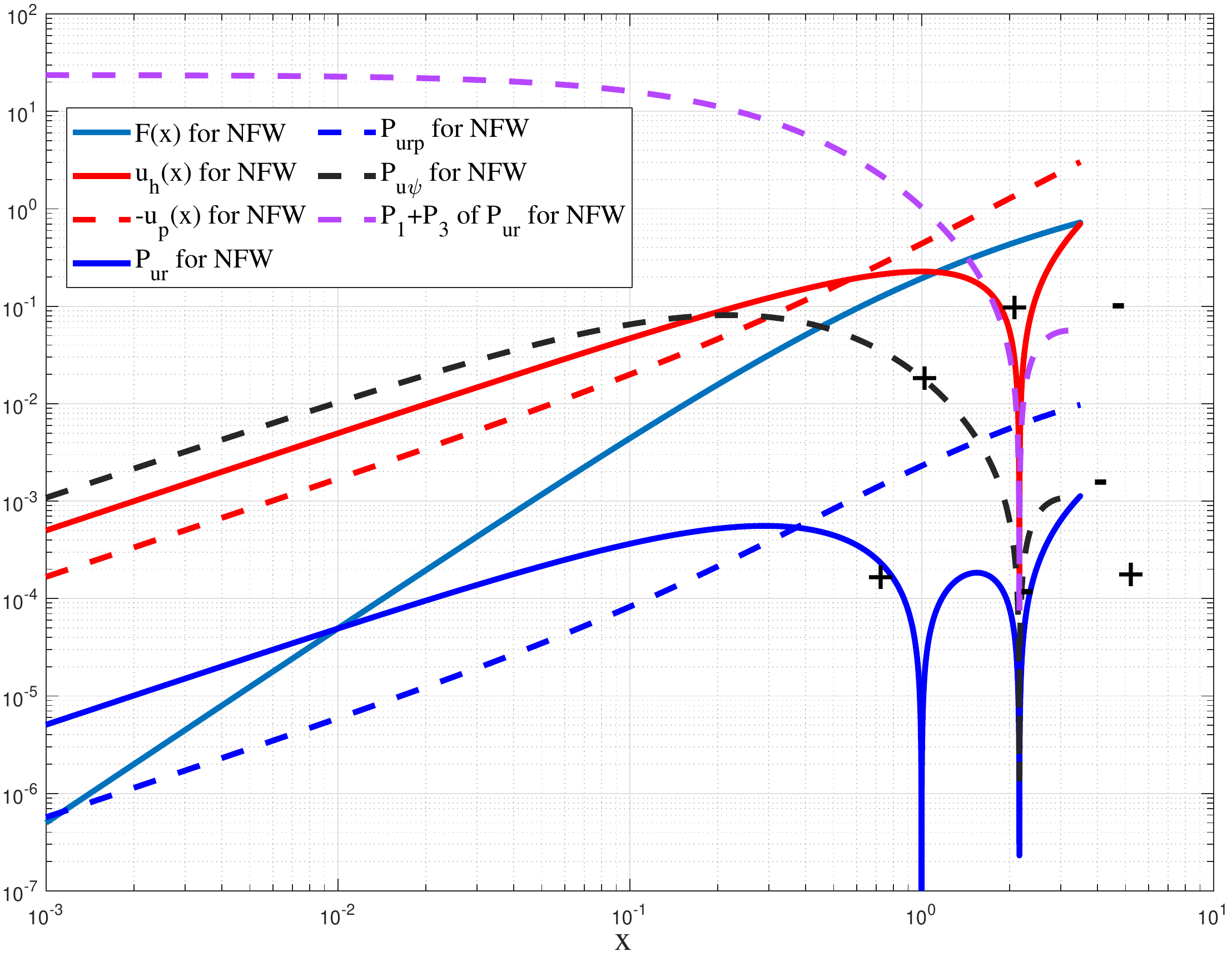}
\caption{The variation of function $F\left(x\right)$, normalized radial flow $u_{h} \left(x\right)$, peculiar radial flow $u_{p} \left(x\right)$, and production terms $P_{ur} \left(x\right)$, $P_{urp} \left(x\right)$ and $P_{u\varphi } \left(x\right)$ (Eqs. \eqref{ZEqnNum278480}-\eqref{ZEqnNum824164}) with a reduced coordinate $x={r/r_{s} } \left(t\right)$ for NFW profile. The radial flow $u_{h} \left(x\right)$ is positive (out-flow) at halo core region, reaching its maximum at $x=1$, and become negative (in-flow) at around $x=x_{0} $ for outer region with $x_{0} \approx 2.15$, as indicated by the '+' and '-' signs in figure. The peculiar radial flow $u_{p} \left(x\right)$ is always negative. The production terms for radial and rotational kinetic energies are normalized by ${2\bar{\rho }_{h} v_{cir}^{2} /t} $. A positive production term means the energy transfer from mean flow to random motion, and vice versa. The radial flow loses its energy to random motion in halo core region $x<1$ and gains energy for $x=\left[1,2\right]$, and loses it energy again for outer region with $x>x_{0} $. The production term $P_{u\varphi } \left(x\right)>0$ for $x<x_{0} $ means the rotational flow loses its energy to random motion in core region, while gains energy from random motion in outer region.}
\label{fig:10}
\end{figure}

Figure \ref{fig:10} presents the variation of function $F\left(x\right)$, normalized radial flow $u_{h} \left(x\right)$, peculiar radial flow $u_{p} \left(x\right)$, and production terms $P_{ur} \left(x\right)$, $P_{urp} \left(x\right)$ and $P_{u\varphi } \left(x\right)$ (Eqs. \eqref{ZEqnNum278480}-\eqref{ZEqnNum824164} ) with a reduced coordinate $x={r/r_{s} } \left(t\right)$ for NFW profile. The mean radial flow $u_{h} \left(x\right)$ is positive (out-flow) at halo core region, reaching its maximum at $x=1$, and become negative (in-flow) at $x=2$ for outer region. The peculiar radial flow $u_{p} \left(x\right)$ is always negative. A positive production term means the energy transfer from mean flow to random motion in SG-CFD, and vice versa. The radial flow loses its energy to random motion in core region $x<1$, gains energy for $x=\left[1,x_{0} \right]$ with $x_{0} \approx 2.15$, and loses it energy again at outer region with $x>x_{0}$. The term $P_{u\varphi } \left(x\right)>0$ (with $\alpha _{\theta } =0$) for $x<x_{0} $ means the azimuthal flow loses its energy to random motion in core region, while gains energy from random motion in outer region for $x>x_{0}$. 

The net rate of change of quantity $\rho_h S^{n} $ for the entire halo has two contributions (as shown in the general Eq. \eqref{ZEqnNum711069}), i.e. term $S_{1} $ due to the energy transfer with the random motion inside halo, and the term $S_{2} $ from the halo surface due to the halo mass accretion (growth) and mass cascade. One example is for radial momentum $\bar{L}_{h}$ by replacing \textit{S} with $u_{r} $ in Eq. \eqref{ZEqnNum711069} and $n=1$, $m=1$, and $k=0$,
\begin{equation} 
\label{eq:131} 
\begin{split}
\frac{\partial \bar{L}_{h} }{\partial t}&=4\pi \rho _{h} \left(r_{h} \right)r_{h}^{2} u_{r} \left(r_{h} \right)\left(\frac{\partial r_{h} }{\partial t} -u_{r} \left(r_{h} \right)\right)\\
&+\int _{0}^{r_{h} }4\pi \rho _{h} r^{2} \left(\frac{\partial u_{r} }{\partial t} +u_{r} \frac{\partial u_{r} }{\partial r} \right)dr.
\end{split}
\end{equation} 
With Eq. \eqref{ZEqnNum176930} for $u_{r} $ and Eq. \eqref{ZEqnNum253473} for derivatives, the final expression can be expressed as,   
\begin{equation} 
\label{eq:132} 
\frac{\partial \bar{L}_{h} }{\partial t} =\frac{m_{h} r_{h} }{t^{2} } \left[\underbrace{\left(1-\frac{1}{\alpha _{h} } \right)}_{S_{2} }+\underbrace{\left(\frac{1}{\alpha _{h} } -\frac{2}{cF\left(c\right)} \int _{0}^{c}F\left(x\right)dx \right)}_{S_{1} }\right],      
\end{equation} 
with two contributions, i.e. $S_{2} $ from halo surface and $S_{1} $ from bulk respectively. For different kinetic energy, i.e. $\bar{K}_{r}$ (radial), $\bar{K}_{rp}$ (peculiar radial), and $\bar{K}_{a}$ (rotational), two terms $S_{1} $ and $S_{2} $ can all be computed with given density profiles and listed in Table \ref{tab:4}. 

Finally, this section describes the evolution of momentum and energies and the energy transfer between coherent (mean) flow and random motion for large halos with fast mass accretion. For radial momentum $\bar{L}_{h} $, $S_{1} =-S_{2} >0$ and the total $\bar{L}_{h} =0$ is time-invariant for large halos \citep[see][Eq. (51)]{Xu:2021-Inverse-mass-cascade-halo-density}. For angular momentum $\bar{H}_{h} $, $S_{1} =0$ and $S_{2} >0$ such that the total angular momentum $\bar{H}_{h} $ increases as $\bar{H}_{h} \propto t^{2} $ (specific angular momentum $H_{h} \propto t$ and angular velocity $\omega _{h} \propto t^{-1} $) with all contributions from $S_{2} $ due to the mass accretion ($m_{h} \left(t\right)\sim t$ and $r_{h} \left(t\right)\sim t$ in Eq. \eqref{ZEqnNum681799}). The radial and rotational kinetic energies of mean flow ($\bar{K}_{r} $ and $\bar{K}_{a} $) increase proportional to \textit{t }with $S_{1} <0$ and $S_{2} >0$ , i.e. the mean flow kinetic energy of entire halo is increasing mainly due to the mass accretion ($S_{2} >0$). The energy transfer between mean flow and random motion is described by Eqs. \eqref{ZEqnNum644755} to \eqref{ZEqnNum824164} and Fig. \ref{fig:10}. The local energy transfer can be two-way between coherent and random motion. For entire halo, a net kinetic energy is transferred from mean flow to random motion in SG-CFD (the bulk contribution is always negative $S_{1} <0$). 

\begin{table}
\caption{The rate of change of halo momentum and energy}
\label{tab:4}
\centering
\begin{tabular}{p{0.3in}p{1.0in}p{0.6in}p{0.9in}} 
\hline 
Symbol & Physical meaning & Isothermal profile with $\alpha _{\theta } =0$ & NFW profile with $\alpha _{\theta } =0$ and $c=3.5$ \\ 
\hline 
$\frac{\partial \bar{L}_{h}}{\partial t} $ &Radial momentum& 0 & 0 \\ \hline 
$S_{1} $ &Bulk contribution& 0 & ${0.2r_{h}\frac{m_{h}}{t^{2}}} $ \\ \hline 
$S_{2} $ &Surface contribution& 0 & ${-0.2r_{h}\frac{m_{h}}{t^{2}}} $ \\ \hline 
$\frac{\partial {\bar{H}_{h}}}{ \partial t} $ &Angular momentum& $\frac{\pi }{4} \frac{m_{h} Hr_{h}^{2} }{t} $  & $\frac{\pi }{4} \frac{m_{h} Hr_{h}^{2} }{t} \left(\frac{3}{2\alpha _{h} } -\frac{1}{2} \right)$ \\ \hline 
$S_{1} $ &Bulk contribution& 0 & 0 \\ \hline 
$S_{2} $ &Surface contribution& $\frac{\pi }{4} \frac{m_{h} Hr_{h}^{2} }{t} $  & $\frac{\pi }{4} \frac{m_{h} Hr_{h}^{2} }{t} \left(\frac{3}{2\alpha _{h} } -\frac{1}{2} \right)$ \\ \hline 
$\frac{\partial \bar{K}_{r}}{ \partial t} $ &Radial kinetic energy & 0 & $0.0062H^{2} r_{h}^{2}\frac{m_{h}}{t} $ \\ \hline 
$S_{1} $ &Bulk contribution& 0 & $-0.0391H^{2} r_{h}^{2}\frac{m_{h}}{t} $ \\ \hline 
$S_{2} $ &Surface contribution & 0 & $0.0453H^{2} r_{h}^{2}\frac{m_{h}}{t} $ \\ \hline 
$\frac{\partial \bar{K}_{rp}}{\partial t} $ &Peculiar radial kinetic energy& $\frac{H^{2} r_{h}^{2} m_{h}} {\left(6t\right)} $ & $0.1937H^{2} r_{h}^{2}\frac{m_{h}}{t} $ \\ \hline 
$S_{1} $ &Bulk contribution& $\frac{-H^{2} r_{h}^{2} m_{h}}{\left(3t\right)} $ & $-0.6525H^{2} r_{h}^{2}\frac{m_{h}}{t} $ \\ \hline 
$S_{2} $ &Surface contribution& $\frac{H^{2} r_{h}^{2} m_{h}}{\left(2t\right)} $ & $0.8462H^{2} r_{h}^{2}\frac{{m_{h}}}{t} $ \\ \hline 
$\frac{\partial \bar{K}_{a}}{\partial t} $ &Rotational kinetic energy & $\frac{H^{2} r_{h}^{2} m_{h}}{\left(2t\right)} $ & $0.7661H^{2} r_{h}^{2} \frac{m_{h}}{t} $ \\ \hline 
$S_{1} $ &Bulk contribution& 0 & \makecell{$-0.0801H^{2} r_{h}^{2}\frac{m_{h}}{t} $} \\ \hline 
$S_{2} $ &Surface contribution& $\frac{H^{2} r_{h}^{2}}{2} $ & $0.8462H^{2} r_{h}^{2}\frac{m_{h}}{t} $ \\ \hline 
\end{tabular}
\end{table}

\section{Halo relaxation from early to late stage}
\label{sec:6}
Previous sections provide the mean flow and velocity dispersion solutions for large halos (high $\nu $ at the early stage of halo life with fast mass accretion and constant concentration) with a non-zero radial flow (Eq. \eqref{ZEqnNum659883}). The other limiting situation consists of halos with a stable core, low mass accretion and vanishing radial flow (low peak height $\nu $ at the late stage of halo life with a constant core mass, scale radius and a time-varying concentration). This section focuses on the transition (relaxation) of halos from their early to late stages. 

Let's assume a typical large halo of mass $m_{h}^{L} \left(t\right)$ that is constantly growing with the waiting time exactly to be $\tau _{g} \sim am_{h}^{-{2/3} } $ for every single merging event during its entire mass accretion history \citep[see][Eq. (45)]{Xu:2021-Inverse-mass-cascade-mass-function}. With $m_{h}^{L} \left(t\right)\sim a^{{3/2} } $, the life span of that typical halo $\tau _{g}^{L} \equiv \tau _{g} \sim t^{0} $ should be time-invariant. The actual halo lifespan $\tau _{gr} $ can be random in nature and either less or greater than $\tau _{g} $. If for any merging event, the random waiting time $\tau _{gr} >\tau _{g} $ such that the actual halo mass $m_{h} \left(t\right)<m_{h}^{L} \left(t\right)$ after that merging. A positive feedback process is established since the waiting time $\tau _{g} \propto am_{h}^{-{2/3}}$ in the propagation range such that $m_{h} \left(t\right)$ will increase slower and slower with longer and longer waiting time or lifespan $\tau _{g} $. On the other hand, if the random waiting time $\tau _{gr} <\tau_{g}$ for a merging event such that halo mass $m_{h} \left(t\right)>m_{h}^{L} \left(t\right)$ is in the deposition range after that merging, where the average waiting time for a single merging is significantly longer. A negative feedback will be established to self-limit and slower down the further growth of $m_{h} \left(t\right)$ such that rare halos can have mass much greater than $m_{h}^{L} \left(t\right)$. 

The feedback process leads to the transition (relaxation) from high $\nu $ to low $\nu $ halos with slower mass accretion. During halo relaxation, there is a continuous variation of halo shape, density profile, mean flow, momentum, and energies. We will start from the general solution for mean radial flow, which facilitates the mass and momentum exchange between different spherical shells and the energy transfer between random motion and mean flow (Eqs. \eqref{ZEqnNum185081} and \eqref{ZEqnNum749450}). 

\subsection{Evolution of mean radial flow from early to late stage}
\label{sec:6.1}
To discuss the halo relaxation, the starting point is to extend the key function $F\equiv F\left(x\right)$ (Eq. \eqref{ZEqnNum990243}) to a more general form of $F\equiv F\left(x,\alpha \right)$, where an additional shape parameter $\alpha \equiv \alpha \left(t\right)$ is introduced. A good example is the function $F\left(x,\alpha \right)$ of an Einasto profile in Eq. \eqref{ZEqnNum339041}. During halo relaxation, we assume a continues variation of function $F\left(x,\alpha \right)$ with time-dependent shape parameter $\alpha$ and concentration \textit{c}. Like Eq. \eqref{ZEqnNum990243}, the halo density and mass $m_{r}$ within radius \textit{r} is,
\begin{equation} 
\label{ZEqnNum203770} 
\rho _{h} \left(r,t\right)=\frac{1}{4\pi r^{2} } \frac{\partial m_{r} \left(r,a\right)}{\partial r} =\frac{m_{h} \left(t\right)}{4\pi r_{s}^{3} } \frac{F^{'} \left(x,\alpha \right)}{x^{2} F\left(c,\alpha \right)}  
\end{equation} 
and 
\begin{equation} 
\label{ZEqnNum729266} 
m_{r} \left(r,t\right)=m_{h} \left(t\right)\frac{F\left(x,\alpha \right)}{F\left(c,\alpha \right)} .          
\end{equation} 

The time derivative of halo density is obtained from Eq. \eqref{ZEqnNum203770},
\begin{equation} 
\label{ZEqnNum302488} 
\frac{\partial \rho _{h} \left(r,a\right)}{\partial t} =\frac{1}{4\pi r^{2} } \frac{\partial ^{2} m_{r} \left(r,a\right)}{\partial r\partial t} .        
\end{equation} 
Using the continuity Eq. \eqref{ZEqnNum321851} and Eq. \eqref{ZEqnNum302488}, the time derivative of $m_{r} \left(r,a\right)$ reads 
\begin{equation} 
\label{eq:136} 
\frac{\partial m_{r} \left(r,a\right)}{\partial t} =-4\pi r^{2} u_{r} \left(r,a\right)\rho _{h} \left(r,a\right).        
\end{equation} 
A general expression of the mean radial flow reads,
\begin{equation} 
\label{ZEqnNum715816} 
u_{r}=-\frac{1}{4\pi r^{2} } \frac{\partial \ln m_{r} }{\partial \ln t} \frac{m_{r} \left(r,a\right)}{\rho _{h} \left(r,a\right)t} =-\frac{r_{s} }{t} \frac{\partial \ln m_{r} }{\partial \ln t} \frac{F\left(x,\alpha \right)}{F^{'} \left(x,\alpha \right)} .     
\end{equation} 

From the definition of $m_{r} \left(r,a\right)$ in Eq. \eqref{ZEqnNum729266}, the logarithmic derivative of $m_{r} \left(r,a\right)$ reads,
\begin{equation} 
\label{ZEqnNum922772} 
\begin{split}
\frac{\partial \ln m_{r} }{\partial \ln t}&=\frac{\partial \ln m_{h} }{\partial \ln t} -\frac{\partial \ln F\left(x,\alpha \right)}{\partial \ln x} \frac{\partial \ln r_{s} }{\partial \ln t}\\ &-\frac{\partial \ln F\left(c,\alpha \right)}{\partial \ln c} \frac{\partial \ln c}{\partial \ln t}+\frac{\partial \ln \frac{F\left(x,\alpha \right)}{F\left(c,\alpha \right)} }{\partial \ln \alpha } \frac{\partial \ln \alpha }{\partial \ln t}.
\end{split}
\end{equation} 
Substitution of Eq. \eqref{ZEqnNum922772} into Eq. \eqref{ZEqnNum715816} leads to the dimensionless radial flow $u_{h} =u_{r} {t/r_{s}}$, 
\begin{equation} 
\label{ZEqnNum645205} 
\begin{split}
&u_{h} =\underbrace{x\frac{\partial \ln r_{s} }{\partial \ln t} -\frac{F\left(x,\alpha \right)}{F^{'} \left(x,\alpha \right)} \frac{\partial \ln m_{h} }{\partial \ln t} }_{1}\\
&+\frac{F\left(x,\alpha \right)}{F^{'} \left(x,\alpha \right)} \left[\underbrace{\frac{\partial \ln F\left(c,\alpha \right)}{\partial \ln c} \frac{\partial \ln c}{\partial \ln t} }_{2}-\underbrace{\frac{\partial \ln \frac{F\left(x,\alpha \right)}{F\left(c,\alpha \right)} }{\partial \ln \alpha } \frac{\partial \ln \alpha }{\partial \ln t} }_{3}\right],
\end{split}
\end{equation} 
where terms 1, 2 and 3 represent the contributions from mass accretion, change of concentration, and change of the shape of halo density profile, respectively. Here $F^{'} \left(x,\alpha \right)$ stands for the derivative with respect to \textit{x}, not $\alpha $. For constant $\alpha $ and $c$, Eq. \eqref{ZEqnNum645205} reduces to Eq. \eqref{ZEqnNum659883} for large halos (high $\nu $) with fast mass accretion. The mean radial flow is given by $u_{h} =u_{hm} +u_{hc} +u_{h\alpha } $ from Eq. \eqref{ZEqnNum645205} with contributions from mass accretion ($u_{hm} $), concentration ($u_{hc} $) and shape parameter ($u_{h\alpha }$), 
\begin{equation} 
\label{ZEqnNum730470} 
u_{hm}(x,t) =x\frac{\partial \ln r_{s} }{\partial \ln t} -\frac{F\left(x,\alpha \right)}{F^{'} \left(x,\alpha \right)} \frac{\partial \ln m_{h} }{\partial \ln t} ,         
\end{equation} 
\begin{equation} 
\label{ZEqnNum998321} 
u_{hc}(x,t) =\frac{\partial \ln c}{\partial \ln t} \frac{F\left(x,\alpha \right)}{F^{'} \left(x,\alpha \right)} \frac{\partial \ln F\left(c,\alpha \right)}{\partial \ln c} ,        
\end{equation} 
\begin{equation} 
\label{ZEqnNum133334} 
u_{h\alpha}(x,t) =\frac{\partial \ln \alpha }{\partial \ln t} \frac{F\left(x,\alpha \right)}{F^{'} \left(x,\alpha \right)} \left[\frac{\partial \ln F\left(c,\alpha \right)}{\partial \ln \alpha } -\frac{\partial \ln F\left(x,\alpha \right)}{\partial \ln \alpha } \right],      
\end{equation} 
and the relevant boundary conditions are
\begin{equation}
\begin{split}
&u_{h} \left(0,t\right)=0, \quad \left. \frac{\partial u_{hm} }{\partial x} \right|_{x=1} =0,\\
&u_{hc} \left(c,t\right)=c\frac{\partial \ln c}{\partial \ln t},\quad \textrm{and} \quad u_{h\alpha } \left(c,t\right)=0.
\end{split}
\label{ZEqnNum366493}
\end{equation}
\noindent It can be easily verified that
\begin{equation} 
\label{ZEqnNum691867} 
u_{hc}\left(c,t\right)\frac{r_{s} }{t} =\frac{\partial \ln c}{\partial \ln t} \frac{r_{h} }{t} =\left(\frac{\partial \ln r_{h} }{\partial \ln t} -\frac{\partial \ln r_{s} }{\partial \ln t} \right)\frac{r_{h} }{t} =\frac{\partial r_{h} }{\partial t} -\frac{r_{h} }{t} \frac{\partial \ln r_{s} }{\partial \ln t} .    
\end{equation} 

For large halos with high peak height $\nu$ and constant concentration $c$, $u_{hc} =u_{h\alpha } =0$ and only radial flow $u_{hm}<0$ is dominant such that halo angular momentum increases with time (Eq. \eqref{ZEqnNum861207}). However, halo relaxation involves an increasing concentration with a fixed scale radius $r_{s}$, i.e. an isotropic "halo stretching" along all directions with increasing c and halo size $r_h$. The radial flow $u_{hc}$ on halo surface is (Eq. \eqref{ZEqnNum691867})
\begin{equation} 
\label{ZEqnNum374119} 
u_{hc} \left(c,t\right)\frac{r_{s} }{t} =\frac{\partial r_{h} }{\partial t}  .           
\end{equation} 
From Eq. \eqref{ZEqnNum861207}, the concentration flow $u_{hc} $ by itself does not change the angular momentum of halos. Since $u_{hc} >0$ for all \textit{x} and using Eq. \eqref{ZEqnNum374119}, the radial flow $u_{hc} $ leads to decreasing rotational kinetic energy (Eq. \eqref{ZEqnNum463094}). This can be understood as the increase of moment of inertia from halo stretching (Eq. \eqref{ZEqnNum325642}). The shape induced radial flow $u_{h\alpha } $ vanishes on halo surface ($u_{h\alpha } \left(c,t\right)=0$) and can be neglected for small change in $\alpha $. Hence, both radial flows $u_{hc} $ and $u_{h\alpha } $ does not lead to the change of halo angular momentum (Eq. \eqref{ZEqnNum861207}). This is important as the halo angular momentum should be conserved if no mass accretion ($u_{hm} =0$) if the halo mass $m_h$ is also fixed with no mass accretion (see Eq. \eqref{ZEqnNum730470}). 

However, we do expect a slow but nonzero mass accretion during halo relaxation. Halo angular momentum slowly increases with time and should not be conserved. Instead, during halo stretching, a vanishing total radial flow $u_{h} =0$ is expected in Eq. \eqref{ZEqnNum645205} that requires the radial flow $u_{hm}$ from mass accretion to cancel the concentration flow $u_{hc}$, i.e. $m_{h} \propto F\left(c,\alpha \right)$ from Eqs. \eqref{ZEqnNum730470} and \eqref{ZEqnNum998321}. It turns out a conserved rotational kinetic energy during halo relaxation (Section \ref{sec:6.4}). 

For Einasto profile, $F\left(x,\alpha \right)$ reads,
\begin{equation} 
\label{ZEqnNum339041} 
F\left(x,\alpha \right)=\Gamma \left({3/\alpha } \right)-\Gamma \left({3/\alpha } ,{2x^{\alpha } /\alpha } \right),
\end{equation} 
where $u_{hc} $ can be explicitly obtained from Eq. \eqref{ZEqnNum998321}
\begin{equation} 
\label{eq:147} 
u_{hc} =\frac{\partial \ln c}{\partial \ln t} \frac{\Gamma \left({3/\alpha } \right)-\Gamma \left({3/\alpha } ,{2x^{\alpha } /\alpha } \right)}{\Gamma \left({3/\alpha } \right)-\Gamma \left({3/\alpha } ,{2c^{\alpha } /\alpha } \right)} \frac{c^{3} }{x^{2} } \exp \left[-\frac{2}{\alpha } \left(c^{\alpha } -x^{\alpha } \right)\right].    
\end{equation} 

\subsection{Path of evolution in (c, \texorpdfstring{$\alpha$}{}) space from early to late stage}
\label{sec:6.2}
To better describe the halo evolution from early stage (high $\nu $) to late stage (low $\nu $), a relation between shape parameter $\alpha $ and concentration $c$ can be identified from Eq. \eqref{ZEqnNum729266},
\begin{equation} 
\label{ZEqnNum154489} 
\frac{F\left(1,\alpha \right)}{F\left(c,\alpha \right)} =\frac{m_{r} \left(r_{s} \right)}{m_{h} } =C_{F} \left(t\right),
\end{equation} 
where $C_{F} <1$ is the ratio of core mass to total mass of halo. The ratio $C_{F} $ should approach a constant for small halos ($\nu \to 0$) with extremely slow mass accretion, where the scale radius $r_{s} $, core mass $m_{r} \left(r_{s} \right)$ and halo mass $m_{h} $ are all relatively time-invariant. 

 Shape parameter $\alpha $ and concentration $c$ for halos of different sizes at different redshifts can be conveniently expressed in terms of the peak height $\nu ={\delta _{c} /\sigma \left(m_{h} ,z\right)} $ of density fluctuation \citep{Klypin:2016-MultiDark-simulations--the-sto}. The relevant expressions read,
\begin{equation}
\alpha =0.115+0.0165\nu ^{2} \quad \textrm{and} \quad c=6.5\nu ^{-1.6} \left(1+0.21\nu ^{2} \right),   
\label{ZEqnNum147169}
\end{equation}

\noindent where $\delta _{c} \approx 1.68$ is the critical overdensity from spherical collapse model and $\sigma \left(m_{h} ,z\right)$ is the root mean square fluctuation of the smoothed density field. This equation gives minimum values of $\min \left(\alpha \right)=0.115$ and $\min \left(c\right)=3.08$ for arbitrary peak height $\nu $.

\begin{figure}
\includegraphics*[width=\columnwidth]{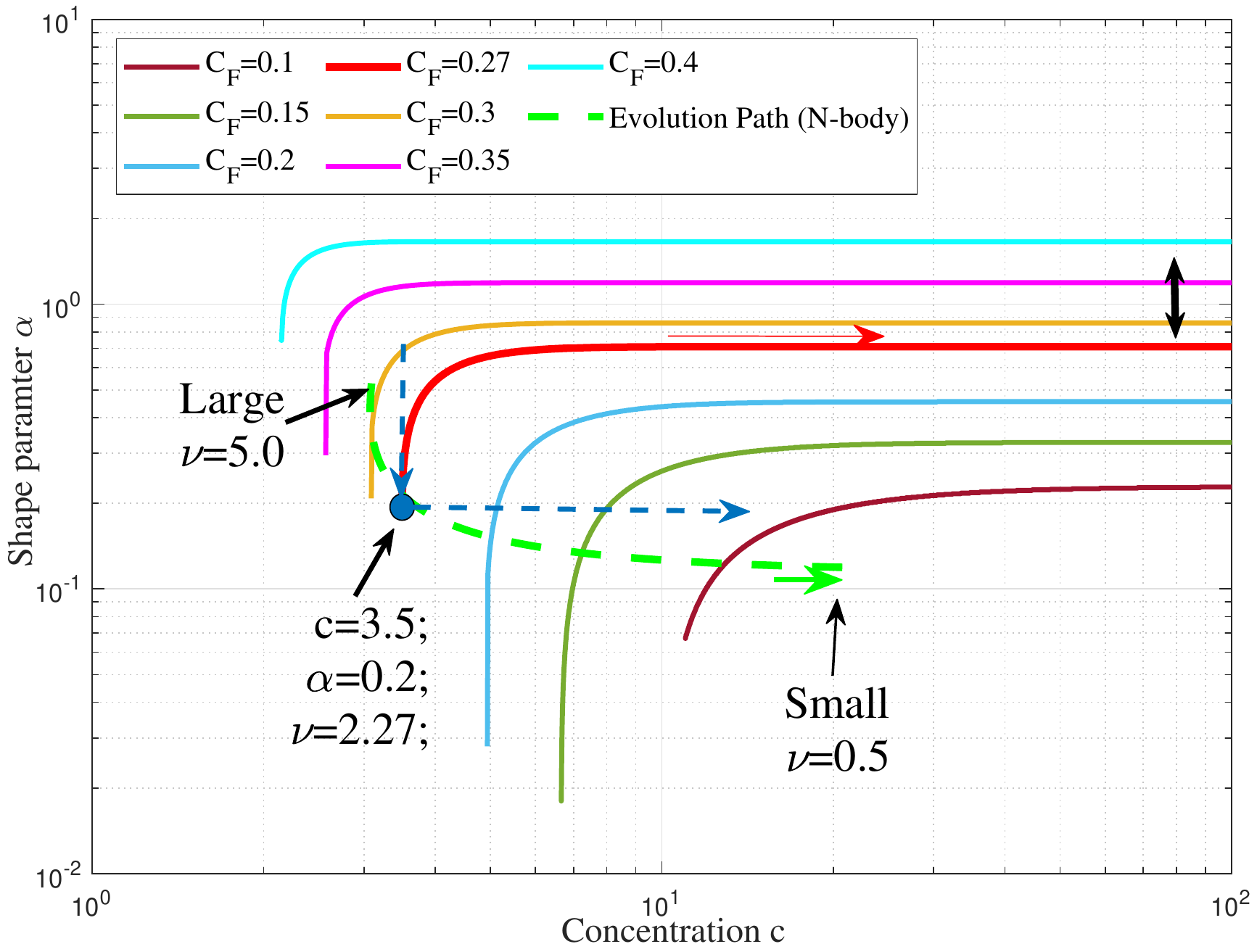}
\caption{The path of halo evolution in the space of shape parameter $\alpha$ and concentration $c$ for different ratio $C_{F}$ of core mass to halo mass. An Einasto profile is used for the calculation. Each curve describes the variation of $\alpha$ with respect to $c$ for a constant $C_{F}$. The blue dot in figure gives $\alpha $ and $c$ of typical halos (high $\nu $ and early stage) with fast mass accretion where $C_{F} =0.27$ and $\alpha =0.2$. Three paths of evolution can be identified: 1) Constant halo mass $m_{h} $ and constant $C_{F} $ where halos evolve along the thick red curve until a constant $\alpha \approx 0.7$; 2) The green dash line for the evolution of halos from \textit{N}-body simulations with a decreasing $\alpha $ and increasing $c$; 3) The blue dash line with two segments as a simplified path for green dash line: constant \textit{c} before blue dot (high $\nu $) and constant $\alpha $after blue dot (low $\nu $). First segment does not present for a NFW profile ($\alpha $is not present). Double arrow indicates the range of $\alpha $ from the distribution of all particles in the same halo group (halos of same mass) from \textit{N}-body simulation \citep[see][Fig. 9]{Xu:2021-Inverse-mass-cascade-halo-density}.} 
\label{fig:11}
\end{figure}

Figure \ref{fig:11} plots different paths of halo evolution in the space of shape parameter $\alpha$ and concentration $c$. The thick red line gives a path of evolution in ($c$, $\alpha$) space that follows a constant ratio $C_{F} =0.27$. Other solid lines plot different paths along different ratio $C_{F}$ using Eq. \eqref{ZEqnNum154489}. All paths end with a limiting shape parameter $\alpha $ when concentration $c\to \infty $ and ${\partial \alpha /\partial c} \to 0$. The corresponding evolution path in ($c$, $\alpha $) space (for halos in N-body simulations from Eq. \eqref{ZEqnNum147169}) is presented as the green dash line with peak height $\nu $ between [0.5 5.0]. Halos with fast mass accretion and vanishing radial momentum should have a constant $\alpha =0.2$, a limiting concentration $c=3.5$ \citep[see][Eq. (53)]{Xu:2021-Inverse-mass-cascade-halo-density} and $C_{F} =0.27$ (Eq. \eqref{ZEqnNum154489}) that is denoted by the blue dot in Fig. \ref{fig:11}. Halos at their early stage of life (high $\nu $) will gradually evolving to the low $\nu$ (late stage) along the green dash line in N-body simulation. Both the shape parameter $\alpha$ and ratio $C_{F}$ are decreasing along that path, while $c$ is increasing along that path. With $\nu \to 0$ along the green line, we have limiting $\alpha =0.115$ and $C_{F} \approx 0.03$ for halos reaching their final stage. For blue dash line with constant $\alpha =0.2$, the limiting ratio $C_{F} \approx 0.083$.

\subsection{Evolution of density profile and moment of inertia}
\label{sec:6.3}
\begin{figure}
\includegraphics*[width=\columnwidth]{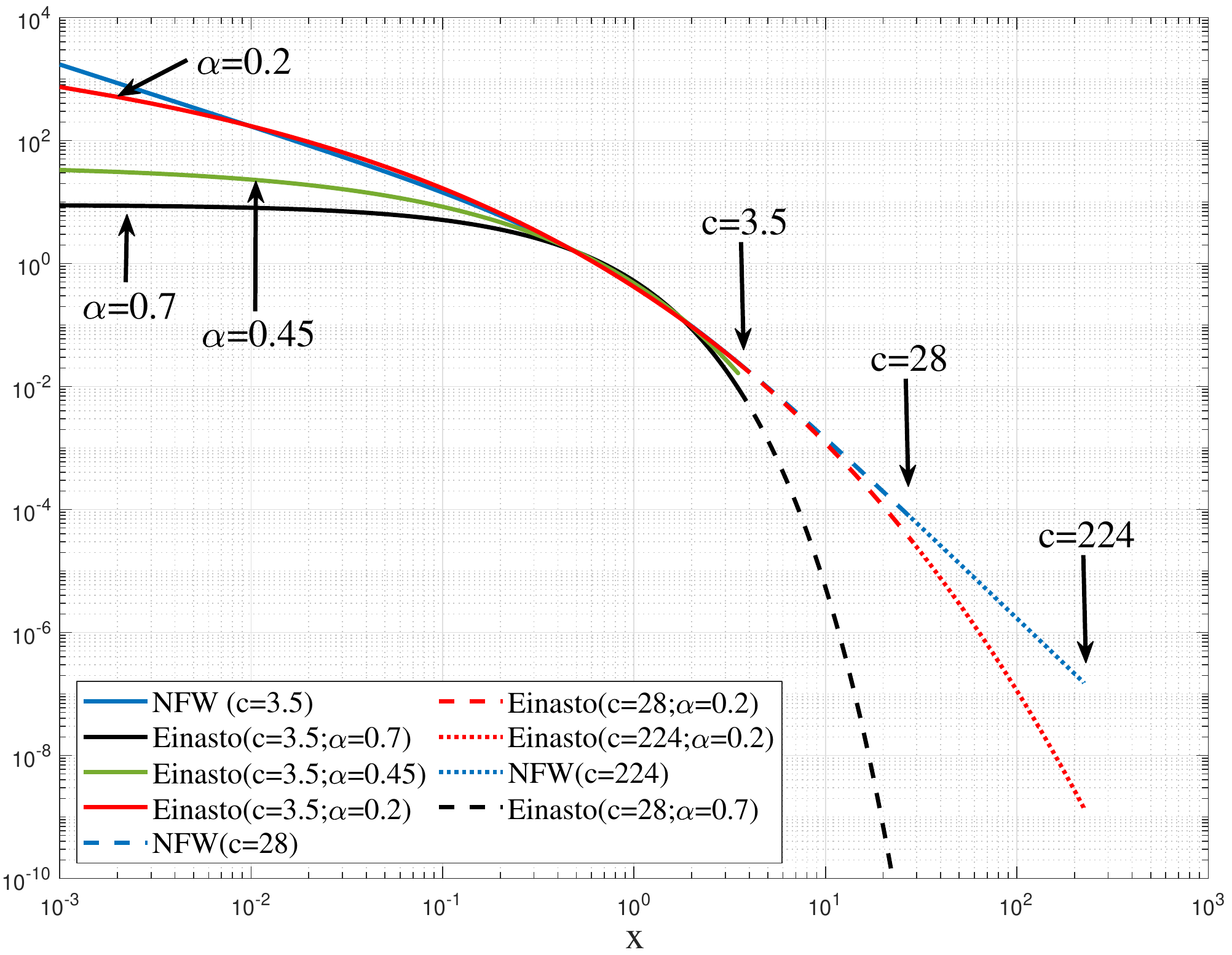}
\caption{The variation of normalized density $\rho _{h}^{*} \left(x\right)$ with time along the relaxation path 3) in Fig. \ref{fig:11}. An Einasto profile is used for the calculation. For high \textit{v} halos with fast mass accretion, the evolution is along a constant \textit{c} and $C_{F} $ path in Fig. \ref{fig:11}. The decreasing shape parameter $\alpha $ leads to a steeper density of inner region and core mass increases proportional to the total halo mass (constant $C_{F} $). For low \textit{v} halos with slower mass accretion, the evolution is along a constant $\alpha $ path. The density profile is simply stretching to larger \textit{c} with inner density fixed ($\rho _{c} $ is fixed in Eq. \eqref{ZEqnNum788345}). A NFW profile (Blue) is also plotted for comparison that is quite different from Einasto profile for high \textit{v} halos.}
\label{fig:12}
\end{figure}

Now let us look at the density profile variation during halo relaxation. Halo density profile reads (from Eq. \eqref{ZEqnNum990243})
\begin{equation} 
\label{ZEqnNum788345} 
\rho _{h} =\frac{m_{h} F\left(1,\alpha \right)}{\left({4/3} \right)\pi r_{s}^{3} F\left(c,\alpha \right)} \cdot \frac{F^{'} \left(x,\alpha \right)}{3F\left(1,\alpha \right)x^{2} } =\rho _{c} \cdot \frac{F^{'} \left(x,\alpha \right)}{3F\left(1,\alpha \right)x^{2} } ,     
\end{equation} 
where $\rho _{c} $ is the mean density of core region with $r<r_{s} $. Figure \ref{fig:12} plots the variation of normalized density profile of $\rho _{h}^{*} ={\rho _{h} /\rho _{c} } $ along the path 3) (blue dash line in Fig. \ref{fig:11}) using an Einasto (red lines) and a NFW model (blue lines). 
The first segment for high \textit{v} halos with constant $c$ (before blue dot) is also almost along a constant $C_{F} $ path (see the red line in Fig.  \ref{fig:11}). The change of density profile is from black solid line to green, and to red solid lines in Fig. \ref{fig:12}. With decreasing $\alpha$ and constant \textit{c}, fast mass accretion leads to an increasing core mass that is proportional to the total halo mass $m_{h} $. 

During fast mass accretion stage (high \textit{v}), the mass accretion induced radial flow ($u_{hm}$) is dominant and core structure is changing significantly. The NFW and Einasto profiles are different in inner region during this stage. It was shown that Einasto profile is a better choice for massive (high \textit{v}) halos \citep{Klypin:2016-MultiDark-simulations--the-sto}. The reason is that NFW is a single parameter profile and cannot reflect the change in shape parameter $\alpha$ during this stage of evolution. In addition, the mean core density $\rho_{c} \sim t^{-2} \sim a^{-3} $ with $m_{h} \sim t$ and $r_{s}^{} \sim t$. Note that for NFW profile with only one parameter \textit{c}, the first segment simply reduces to the blue dot for high \textit{v} halos (full solutions are discussed in Sections \ref{sec:3.5} and \ref{sec:4.1}). 

The second segment for evolving toward low \textit{v} halos with a constant $\alpha$ (after blue dot) should have decreasing $C_{F}$ with time. During this slower mass accretion stage (low \textit{v}), the radial flow is negligible with contributions from both $u_{hm}$ and $u_{hc}$ canceling each other such that $m_{h} \propto F\left(c,\alpha \right)$ (Eqs. \eqref{ZEqnNum730470} and \eqref{ZEqnNum998321}). This means a constant core mass $m_{r} \left(r_{s} \right)$ (Eq. \eqref{ZEqnNum154489}) and core density $\rho _{c}$ during this stage. The density profile during this stage simply stretches to larger \textit{c} with inner density $\rho _{c} $ fixed ("halo stretching"). Similar observations were also discussed in \citep{Zhao:2009-Accurate-Universal-Models-for-}, i.e. the slower mass accretion during this stage simply adds more mass to the outer region with core structure fixed. Full solutions for low \textit{v} halos with fully vanishing radial flow are presented in Section \ref{sec:3.4}.

To better understand the halo relaxation ("stretching"), the variation of momentum of inertia should also be checked. For any density profile, the \textit{k}th order moment of inertia can be obtained as (with density $\rho _{h} $ from Eq. \eqref{ZEqnNum203770}),
\begin{equation} 
\label{eq:151} 
\left(r_{k} \right)^{k} =\frac{1}{m_{h} } \int _{0}^{r_{h} }4\pi r^{2} \rho _{h}  \left(r\right)r^{k} dr=r_{s}^{k} \int _{0}^{c}\frac{F^{'} \left(x,\alpha \right)}{F\left(c,\alpha \right)} x^{k} dx .      
\end{equation} 

Specifically, the moment of inertial for isothermal, NFW ($2nd$ order), and Einasto profiles are,
\begin{equation} 
\label{ZEqnNum851638} 
\begin{split}
&\left(r_{k} \right)^{k} =r_{s}^{k} {c^{k} /\left(1+k\right)},\\ 
&\left(r_{2} \right)^{2} =\frac{r_{s}^{2} }{2} \frac{c\left(c^{2} -3c-6\right)+6\left(1+c\right)\ln \left(1+c\right)}{\left(1+c\right)\ln \left(1+c\right)-c},\\ 
&\left(r_{k} \right)^{k} =r_{s}^{k} \left(\frac{\alpha }{2} \right)^{\frac{k}{\alpha } } \frac{\Gamma \left({\left(3+k\right)/\alpha } \right)-\Gamma \left({\left(3+k\right)/\alpha } ,{2c^{\alpha } /\alpha } \right)}{\Gamma \left({3/\alpha } \right)-\Gamma \left({3/\alpha } ,{2c^{\alpha } /\alpha } \right)}.
\end{split}
\end{equation} 

Halo moment of inertia (Eq. \eqref{ZEqnNum963993}) can be related to the root mean square radius $r_{g}^{2} =r_{2}^{2} $ (i.e. $k=2$), 
\begin{equation}
\label{ZEqnNum414741} 
I_{\omega } =\frac{2}{3} m_{h} r_{g}^{2} =\frac{2}{3} m_{h} r_{s}^{2} F_{\omega } \left(\alpha ,c\right).         
\end{equation} 
Dimensionless moments of inertia for NFW and Einasto profiles are
\begin{equation} 
\label{ZEqnNum325642}
\begin{split}
&F_{\omega } \left(c\right)=\frac{c\left(c^{2} -3c-6\right)+6\left(1+c\right)\ln \left(1+c\right)}{2\left(1+c\right)\ln \left(1+c\right)-2c}\\ 
&\textrm{and}\\
&F_{\omega } \left(\alpha ,c\right)=\left(\frac{\alpha }{2} \right)^{\frac{2}{\alpha } } \frac{\Gamma \left({5/\alpha } \right)-\Gamma \left({5/\alpha } ,{2c^{\alpha } /\alpha } \right)}{\Gamma \left({3/\alpha } \right)-\Gamma \left({3/\alpha } ,{2c^{\alpha } /\alpha } \right)} \\
&\quad\quad\quad\quad=\left(\frac{\alpha }{2} \right)^{\frac{2}{\alpha } } \frac{\gamma \left({5/\alpha } ,{2c^{\alpha } /\alpha } \right)}{\gamma \left({3/\alpha } ,{2c^{\alpha } /\alpha } \right)},
\end{split}
\end{equation} 
where $\gamma \left(x,y\right)$ is a lower incomplete Gamma function. For Einasto profile with a constant shape parameter $\alpha $ during halo stretching,
\begin{equation}
\begin{split}
&F_{\omega } \left(\alpha ,c\right)\approx \frac{3}{5} c^{2} \quad \textrm{for} \quad c\to 0\\
&\textrm{and}\\
&F_{\omega } \left(\alpha ,c\right)=\left(\frac{\alpha }{2} \right)^{{2/\alpha } } \frac{\Gamma \left({5/\alpha } \right)}{\Gamma \left({3/\alpha } \right)} \quad \textrm{for} \quad c\to \infty.  
\end{split}
\label{ZEqnNum248210}
\end{equation}

\begin{figure}
\includegraphics*[width=\columnwidth]{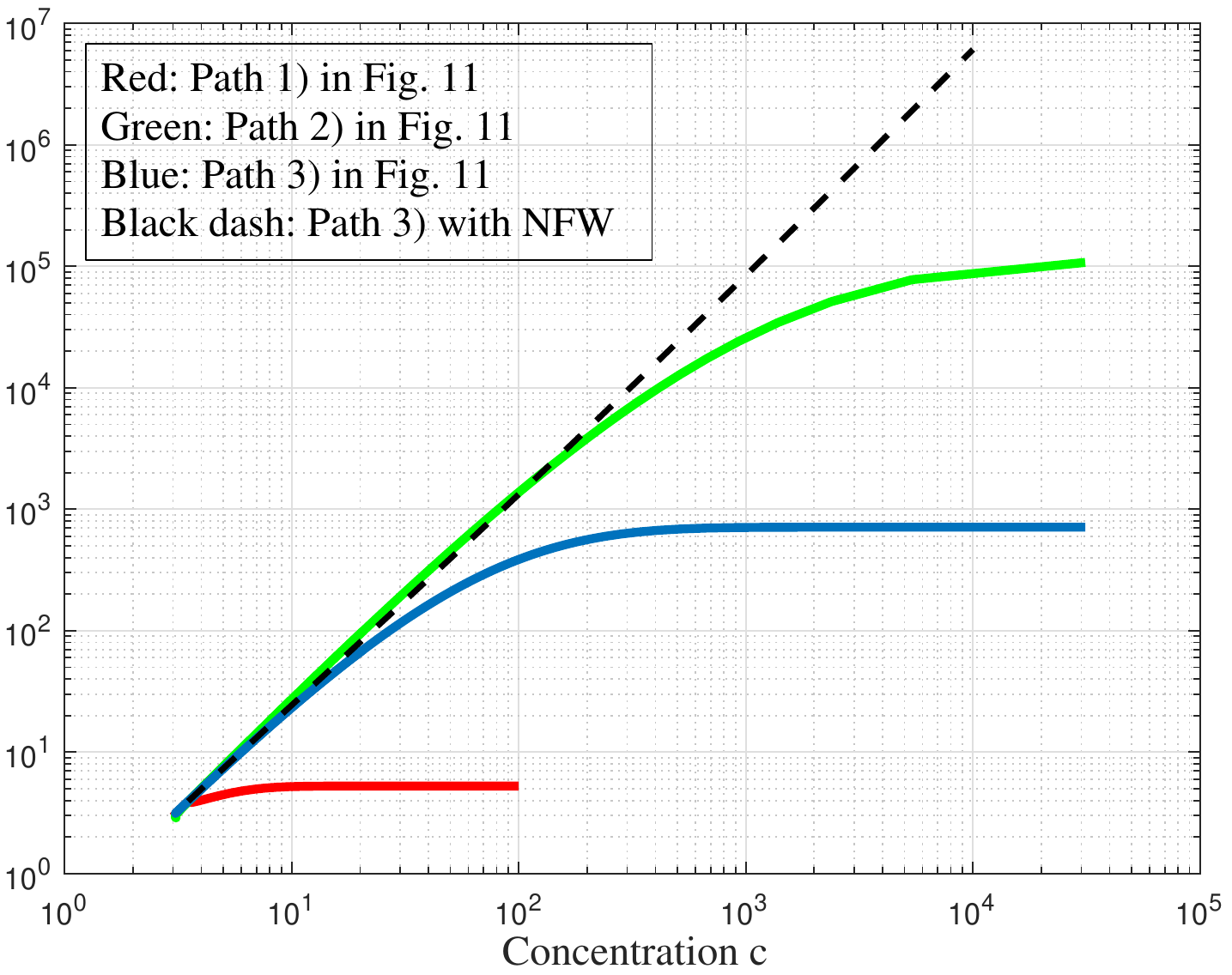}
\caption{The variation of dimensionless moment of inertia $F_{\omega } \left(\alpha ,c\right)$ with concentration $c$ along three different paths of evolution in Fig. \ref{fig:11}. An Einasto profile is used for the calculation. The moment of inertia is constant along path 1) with a constant $C_{F} $ and halo mass, which indicates a constant angular velocity and rotational kinetic energy. Path 1) leads to the equilibrium distribution of all particles in the same halo group. The moment of inertia increases significantly along paths 2) and 3). The variation of $F_{\omega } \left(\alpha ,c\right)$ along path 3) is also plotted for a NFW profile that diverges at large \textit{c}.}
\label{fig:13}
\end{figure}

Figure \ref{fig:13} plots the variation of dimensionless momentum of inertia $F_{\omega} \left(\alpha ,c\right)$ along three different paths in Fig. \ref{fig:11}. The scale radius $r_{s} $ is assumed to be relatively constant along all paths. Along path 1) with a constant $C_{F} =0.27$ and a constant halo mass $m_{h} $, the radial flow $u_{hm} =0$ ($u_{hc} $ and $u_{h\alpha } $ may not be zero, see Eqs. \eqref{ZEqnNum730470} to \eqref{ZEqnNum374119}). The angular momentum is conserved. The moment of inertia is relatively constant in Fig. \ref{fig:13}. This means that both angular velocity and rotational kinetic energy are also constant along path 1), i.e. no energy transfer between mean flow and random motion. The limiting $\alpha \approx 0.7$ along this path is the $\alpha $ for the density distribution from all particles in the same halo group (red arrow in Fig. \ref{fig:11}). Therefore, the final stage along that path is the equilibrium distribution of all particles in the same halo group. That equilibrium particle distribution (Black line in Fig. \ref{fig:12}) was studied via random walk of particles in a halo with varying size \citep[see][Section 4]{Xu:2021-Inverse-mass-cascade-halo-density}. The double arrow in Fig. \ref{fig:11} denotes the range of $\alpha $ [0.7 1.2] for equilibrium particle distribution in halo groups of different size from a N-body simulation \citep[see][Fig. 9]{Xu:2021-Inverse-mass-cascade-halo-density}. 

Along path 2) or 3) (dash green and blue lines in Fig. \ref{fig:11}), the normalized moment of inertial $F_{\omega } \left(\alpha ,c\right)$ increases significantly during halo stretching and plateaus with $c\to \infty $ for Einasto profile, while $F_{\omega } \left(\alpha ,c\right)$ diverges for NFW (Eq. \eqref{ZEqnNum851638}), a well-known problem of NFW profile. 

\subsection{Evolution of momentum and energy from early to late stage}
\label{sec:6.4}
To simplify the calculation, the path 3) (second segment of dash blue line in Fig. \ref{fig:11}) with a constant $\alpha $ can be used to represent the path from N-body simulations (Green line) and studied in detail. 

Along this path, the scale radius $r_{s}$ is constant and the concentration \textit{c} is increasing with time. The radial flow vanishes and we expect $u_{r} \left(r_{h} \right)\approx 0$ such that halo mass $m_{h} \sim F\left(c\right)$ (Eq. \eqref{ZEqnNum773554}) with a constant core mass $m_{r} \left(x=1\right)={m_{h} \left(c\right)F\left(1\right)/F\left(c\right)} $. The specific angular momentum and rotational kinetic energy are generally related to the effective angular velocity $\omega _{h} $ as
\begin{equation}
\left|\boldsymbol{\mathrm{H}}_{h} \right|=\frac{2}{3} \omega _{h} r_{g}^{2} \quad \textrm{and} \quad K_{a} =\frac{1}{2} \left|\boldsymbol{\mathrm{H}}_{h} \right|\omega _{h} =\frac{3}{4} \left({\left|\boldsymbol{\mathrm{H}}_{h} \right|/r_{g} } \right)^{2}.    
\label{ZEqnNum308719}
\end{equation}

For halos in their early stage (high \textit{v}), $\left|\boldsymbol{\mathrm{H}}_{h} \right|\sim r_{g} $ and both are proportional to time \textit{t} ($r_{g} \sim t$ and $\left|\boldsymbol{\mathrm{H}}_{h} \right|\sim t$ in Table \ref{tab:3}) such that the specific rotational kinetic energy $K_{a} $ is always conserved. During halo "stretching" (second segment of blue line in Fig. \ref{fig:11}), the root mean square radius $r_{g} \left(c\right)=r_{s} \sqrt{F_{\omega } \left(c\right)} $ (Eq. \eqref{ZEqnNum414741}) that can be different from scaling of $r_{g} \sim t$ for high \textit{v} halos. However, a reasonable estimate is that the scaling $\left|\boldsymbol{\mathrm{H}}_{h} \right|\sim r_{g} $ continuously extends beyond early stage during halo stretching such that rotational kinetic energy $K_{a}$ is still conserved and the angular velocity $\omega _{h} \sim r_{g}^{-1} $ (Eq. \eqref{ZEqnNum308719}). At least, the scaling $\left|\boldsymbol{\mathrm{H}}_{h} \right|\sim r_{g} $ should be a good approximation at the beginning of halo stretching.

To summarize, along path 3) in Fig. \ref{fig:11} with constant $r_{s} $ and core mass, the increasing concentration \textit{c} leads to a decreasing core mass ratio $C_{F} $. The halo stretching with inner density fixed (Fig. \ref{fig:12}) leads to the increasing moment of inertial (Eq. \eqref{ZEqnNum325642}) and angular momentum $H_h$, while halo angular velocity $\omega _{h} $ and azimuthal flow $u_{\varphi }^{2} $ decreases along that path. With the coupling term $F_{a} $ (Eqs. \eqref{ZEqnNum662332} and \eqref{ZEqnNum451086}) approaching zero for low $\nu $ halos, there is a net transfer of spin-induced velocity dispersion to axial dispersion ($\sigma _{r0}^{2}$ dispersion due to gravity) (from part 2 to part 1 in Eq. \eqref{ZEqnNum528783}), i.e. an increasing in $\sigma _{r0}^{2} $. Coefficients $\alpha _{\varphi }^{} $, $\beta _{\varphi }^{} $ and $\gamma _{\varphi }^{} $ also decreases with time (Fig. \ref{fig:8}) such that halos become more isotropic with $\beta _{h1}^{} \to 0$ (Fig. \ref{fig:9}). 

The halo specific potential energy \citep[see][Eq. (90)]{Xu:2021-Inverse-mass-cascade-halo-density} reads
\begin{equation} 
\label{eq:157} 
\Phi _{h} \frac{Gm_{h} }{r_{h} } =-\frac{1}{m_{h} } \int _{0}^{r_{h} }4\pi r^{2} \rho _{h} \frac{Gm_{r} \left(r\right)}{r}  dr=-\frac{Gm_{h} F\left(1\right)}{r_{s} F\left(c\right)} \Phi _{h}^{*}, 
\end{equation} 
where the dimensionless number $\Phi _{h}^{*}$ reads (due to constant $r_{s}$ and core mass ${m_{h} F(1)/F(c)}$),
\begin{equation} 
\label{eq:158} 
\Phi _{h}^{*} =\frac{1}{F\left(1\right)F\left(c\right)} \int _{0}^{c}\frac{F\left(x\right)F^{'} \left(x\right)}{x}  dx.        
\end{equation} 
\begin{figure}
\includegraphics*[width=\columnwidth]{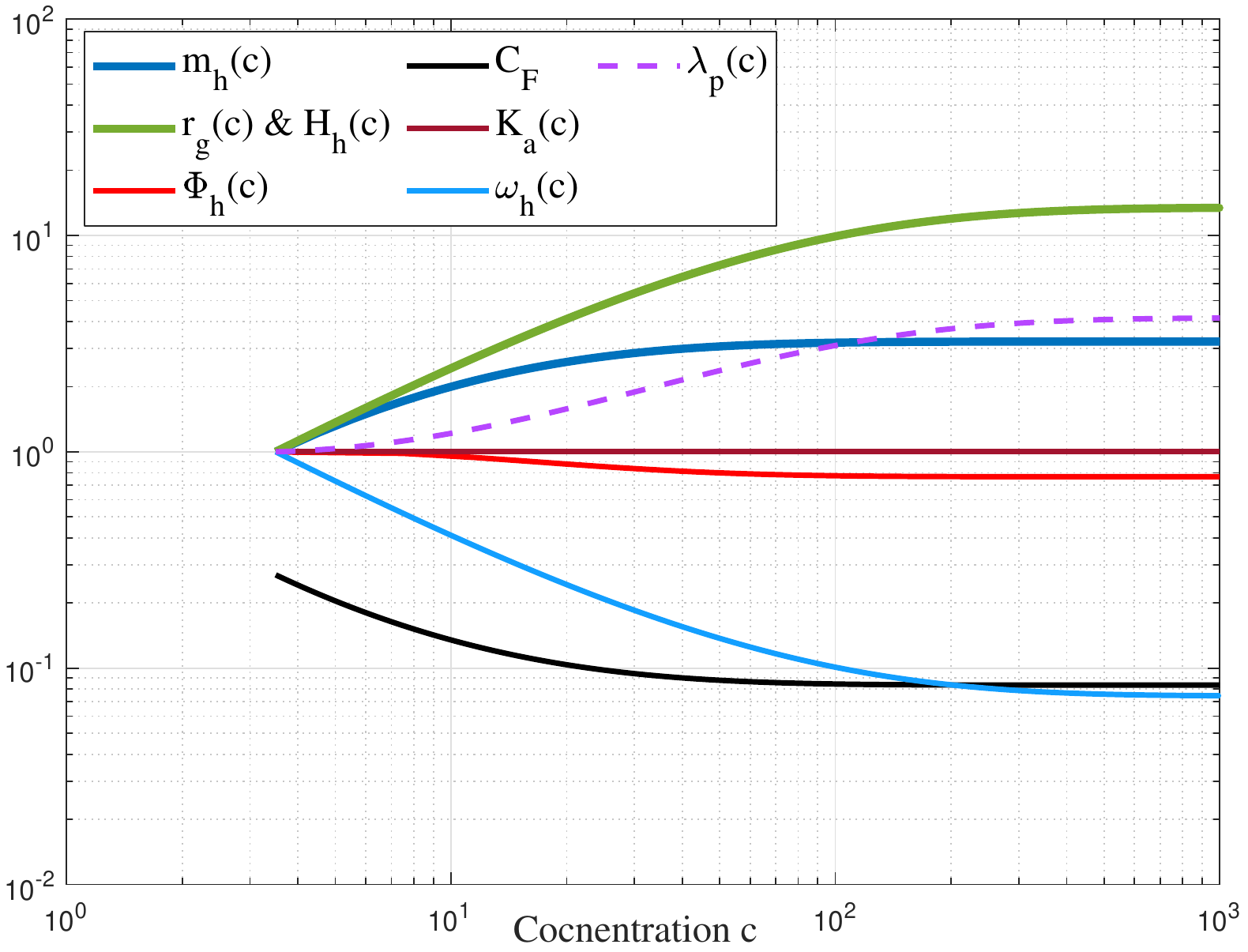}
\caption{The variation of halo mass $m_{h} $, potential $\Phi _{h}^{} $, root mean square radius $r_{g} $, rotational kinetic energy $K_{a} $, and angular velocity $\omega _{h} $ with concentration $c$ during halo stretching (path 3) in Fig. \ref{fig:11}). An Einasto profile is used for the calculation. The scale radius $r_{s} $ and core mass are constant and specific rotational kinetic energy is conserved during halo stretching. The specific potential $\Phi _{h}^{} $ is almost constant. The angular momentum $\left|\boldsymbol{\mathrm{H}}_{h} \right|\sim r_{g} $ and angular velocity $\omega _{h} \sim r_{g}^{-1} $, while the halo spin parameter $\lambda _{p} $ increases due to faster increase in $\left|\boldsymbol{\mathrm{H}}_{h} \right|$ than halo mass $m_{h}^{}$.}
\label{fig:14}
\end{figure}

With $m_{h} \propto F(c)$, constant scale radius $r_{s}$, and conserved rotational kinetic energy $K_{a} $ along path 3) in Fig. \ref{fig:11}, the variation of all relevant quantities can be summarized in Fig. \ref{fig:14} for an Einasto profile. The mass ratio $C_{F} $ decreases from 0.27 to 0.08. Other quantities are normalized by their initial values at $c=3.5$, i.e. the values for halos in their early stage (blue dot in Fig. \ref{fig:11} and shown in Table \ref{tab:3}). Halo spin parameter $\lambda _{p} =0.031$ when $c=3.5$ and increases with time during halo stretching due to the faster increase in angular momentum than halo mass (Eq. \eqref{ZEqnNum475450} and Fig. \ref{fig:14}). This is consistent with simulation results \citep{Ahn:2014-Halo-Spin-Parameter-in-Cosmolo}, where $\lambda _{p} $ increases with time. In addition, $\lambda _{p} $ for halos of different size should converge to a limiting value of $\lambda _{p}$ of low $\nu$ halos (late stage) with $c\to \infty$. 

\section{Conclusions}
\label{sec:7}
By revisiting fundamental ideas of energy transfer and cascade in hydrodynamic turbulence, self-gravitating collisionless dark matter flow (SG-CFD) shares many similarities, but also exhibits some unique features.  In hydrodynamic turbulence, Reynolds stress arising from velocity fluctuations acts as a conduit to continuously transfer energy from mean flow to turbulence and sustain the continuous energy cascade. To quantitatively describe the energy transfer between mean flow and random motion in SG-CFD, general solutions of mean flow and velocity dispersions are derived for axisymmetric, growing, and rotating halos in spherical coordinate. The polar flow can be neglected (Fig. \ref{fig:2}). The azimuthal flow is directly related to in-plane velocity dispersions (Eq. \eqref{ZEqnNum618520}). The radial flow facilitates the exchange of momentum and energy across different spherical shells (Eqs. \eqref{ZEqnNum185081}, \eqref{ZEqnNum701076} and \eqref{ZEqnNum749450}). 

Evolution of halo momentum and kinetic energy are extensively studied (Eqs. \eqref{ZEqnNum898930} to \eqref{ZEqnNum463094}) based on the continuity and momentum equations (Eqs. \eqref{ZEqnNum535400} to \eqref{ZEqnNum720202}). A growing halo may obtain its momentum through a continuous mass acquisition as quantitatively described by Eq. \eqref{ZEqnNum861207}. For large halo at the early stage of its life (Table \ref{tab:3}), the specific angular momentum $H_{h} $ increases linearly with time \textit{t}, while the specific halo angular kinetic energy $K_{a} $ is a constant. Halo angular momentum can only be changed from mass accretion and radial flow at halo surface (Eq. \eqref{ZEqnNum861207}). Halo rotational kinetic energy can be generated from both mass accretion and the energy transfer with random motion (Eq. \eqref{ZEqnNum463094}). The fictitious stress $\rho _{h} u_{\varphi }^{2} $ (equivalent to ``Reynolds stress'') acts on the gradient of mean flow (${u_{r} /r} $) to facilitate the energy transfer between mean flow and random motion (Eq. \eqref{ZEqnNum749450}). While the energy transfer in turbulence is always one-way from mean flow to random motion, the local energy transfer can be two-way in SG-CFD depending on the sign of radial flow $u_{r} $.

By assuming that velocity anisotropy is due to finite halo spin, velocity dispersions can be decomposed into a gravity induced non-spin axial dispersion ($\sigma_{r0}^{2}$) and a spin-induced dispersion that is dependent on the azimuthal flow $u_{\varphi }^{2}$ (Eqs. \eqref{ZEqnNum528783} to \eqref{ZEqnNum892394}). A new definition of halo anisotropic parameter $\beta _{h1} $ is proposed to include the effect of azimuthal flow $u_{\varphi }^{2}$ on anisotropy (Eq. \eqref{ZEqnNum865565}).  Parameter $\beta _{h1} $ reduces to the usual definition $\beta _{h}$ (Eq. \eqref{ZEqnNum597498}) if $u_{\varphi }^{2} $ can be neglected. General solutions of mean flow and velocity dispersion are obtained in Section \ref{sec:3.3} (Eqs. \eqref{ZEqnNum376341}, \eqref{ZEqnNum622731}, \eqref{ZEqnNum624087}, \eqref{ZEqnNum662332} and \eqref{ZEqnNum451086}) and subsequently applied to two limiting situations in Sections \ref{sec:3.4} and \ref{sec:3.5}. 

For "large" halos (high peak height $\nu $ at the early stage of halo life) with fast mass accretion and constant concentration, there exists a non-zero self-similar radial flow induced by fast halo growth (Eq. \eqref{ZEqnNum176930}). The radial flow drives outward mass flow in the core region and inward mass flow in the outer region (the gravitational infall). The halo surface energy can be significant due to the non-zero radial flow and low halo concentration such that the halo virial ratio $\gamma _{v} \approx 1.3>1$ \citep[see][Fig. 9]{Xu:2021-Inverse-and-direct-cascade-of-}. Angular momentum and rotational kinetic energy are transported by the radial flow (Eqs. \eqref{ZEqnNum701076} and \eqref{ZEqnNum749450}). The random motion draws kinetic energy from mean flow in core region, and vice versa in the outer region (Eq. \eqref{ZEqnNum749450} and Fig. \ref{fig:10}). There is a net transfer from mean flow to random motion for the entire halo to maximize system entropy (negative $S_1$ in Table \ref{tab:4}). A growing halo (the early stage of halo life) obtains its angular momentum through continuous mass acquisition (Eq. \eqref{ZEqnNum861207}) that predicts a linear increase of specific angular momentum $H_{h} $ with time \textit{t} (Eq. \eqref{ZEqnNum546367} and Table \ref{tab:3}). The self-similar azimuthal flow is only dependent on radius \textit{r} and not significantly dependent on the polar angle $\theta $ with $\alpha _{\theta } \ll 1$ (Eqs. \eqref{ZEqnNum694993}, \eqref{ZEqnNum801277} and Fig. \ref{fig:5}). The effective halo angular velocity $\omega _{h} $ is proportional to the Hubble parameter \textit{H} and decreases with time (Eq. \eqref{ZEqnNum501598}). Large halos rotate with a faster spinning core and slower outer region. For large halos, spin-induced dispersions are dominant ( $\sigma _{r0}^{2} \ll \gamma _{\varphi } u_{\varphi }^{2} $) and two anisotropy parameters are equal, i.e. $\beta _{h1} \approx \beta _{h} $ (Fig. \ref{fig:9}). The radial velocity momentum vanishes for large halos leads to a limiting concentration $c=3.5$ \citep[see][Eq. (53)]{Xu:2021-Inverse-mass-cascade-halo-density}. Halo mass $m_{h} $, size $r_{h} $, and specific angular momentum $H_{h} $ all increase linearly with time \textit{t}. All specific energies (radial/rotational/kinetic/potential) are time invariant for large halos (Table \ref{tab:3}). The halo spin parameter $\lambda _{p} =0.031$ and the variation of anisotropic parameter $\beta _{h1}$ in halo can be obtained analytically (Eq. \eqref{ZEqnNum951104}, Eqs. \eqref{ZEqnNum206423} to \eqref{ZEqnNum895234} and Fig. \ref{fig:9}).  

The other limiting situation consists of "small" halos with a stable core (well bound and virialized) and low mass accretion (low peak height $\nu $ and the late stage of halo life with an almost constant halo mass, core mass, scale radius and an increasing halo concentration). The radial flow vanishes for small halos (Eq. \eqref{ZEqnNum773554}) without mass, momentum, and energy exchange between different spherical shells. Halo surface energy can be negligible due to the vanishing radial flow and high halo concentration (extremely low density at halo surface). Small halos rotate more like a rigid body. The halo angular velocity $\omega _{h} $ is relatively time-invariant. For small halos, non-spin axial dispersion is dominant ($\sigma _{r0}^{2} \gg \gamma _{\varphi } u_{\varphi }^{2}$) and the anisotropy parameters $\beta _{h1} \approx 0$ (Fig. \ref{fig:9}). Small halos are more spherical in shape, incompressible for proper velocity, and isotropic ($\beta _{h1} \approx 0$). The radial and azimuthal dispersions are comparable for small halos and greater than the polar dispersion, i.e. $\sigma _{rr}^{2} =\sigma _{\varphi \varphi }^{2} =\sigma _{\theta \theta }^{2} +u_{\phi }^{2} $ (Eq. \eqref{ZEqnNum521885}) that reflects a direct connection between mean flow and random motion in SG-CFD. The total kinetic energy including both random motion and mean flow is not equipartitioned along each direction with the greatest kinetic energy along azimuthal direction and the smallest along polar direction, i.e. $\sigma _{\varphi \varphi }^{2} +u_{\varphi }^{2} >\sigma _{rr}^{2} =\sigma _{\varphi \varphi }^{2} >\sigma _{\theta \theta }^{2} =\sigma _{\varphi \varphi }^{2} -u_{\varphi }^{2} $. In short, small halos are isotropic ($\beta _{h1} =0$), incompressible ($u_{r} =u_{\theta } =0$), well bound and virialized structures.

Finally, the halo relaxation from high $\nu $ (early stage) to low $\nu $ (late stage) is studied with a continuous variation of halo shape, density profile, mean flow, momentum, and energy (dash lines in Fig. \ref{fig:11}). Overall, shape parameter $\alpha $ decreases and concentration \textit{c} increases during relaxation (Eq. \eqref{ZEqnNum147169} and Fig. \ref{fig:11}). The "vortex stretching" plays an important role for the energy cascade from large to small scales in turbulence. Due to the conservation of angular momentum, the stretching of vortex along the axis of rotation decreases the moment of inertial and increases the rotational kinetic energy. In SG-CFD, A isotropic "halo stretching" is proposed with increasing concentration and constant inner density (Fig. \ref{fig:12}) and core mass. Halo stretching leads to increasing halo mass, moment of inertial (Eq. \eqref{ZEqnNum325642} and Fig. \ref{fig:13}). In contrast to "vortex stretching", the halo angular momentum is not conserved and increasing with time (Fig. \ref{fig:14}). The specific rotational kinetic energy is relatively conserved during halo stretching such that angular velocity $\omega _{h} $ decreases with time (Eq. \eqref{ZEqnNum308719}). With the coupling term $F_{a}$ (Eqs. \eqref{ZEqnNum662332} and \eqref{ZEqnNum451086}) approaching zero for low $\nu $ halos, there is a net transfer of spin-induced velocity dispersion to the non-spin axial dispersion ($\sigma _{r0}^{2}$) (from part 2 to part 1 in Eq. \eqref{ZEqnNum528783}), i.e. an increasing in $\sigma _{r0}^{2} $ and decreasing in $u_{\varphi}^{2}$, and coefficients $\alpha _{\varphi }$, $\beta _{\varphi}$ and $\gamma _{\varphi }$. Halo becomes more isotropic with $\beta _{h1}\to 0$ during relaxation. The halo spin parameter increases with time due to faster increasing angular momentum than halo mass. 
 

\section*{Data Availability}
Two datasets underlying this article, i.e. a halo-based and correlation-based statistics of dark matter flow, are available on Zenodo \citep{Xu:2022-Dark_matter-flow-dataset-part1,Xu:2022-Dark_matter-flow-dataset-part2}, along with the accompanying presentation slides "A comparative study of dark matter flow \& hydrodynamic turbulence and its applications" \citep{Xu:2022-Dark_matter-flow-and-hydrodynamic-turbulence-presentation}. All data files are also available on GitHub \citep{Xu:Dark_matter_flow_dataset_2022_all_files}.

\bibliographystyle{mnras}
\bibliography{Papers}


\label{lastpage}
\end{document}